\documentclass[%
 reprint,
superscriptaddress,
nofootinbib,
 amsmath,amssymb,
 aps,floatfix,
]{revtex4-1}
\usepackage{color}
\usepackage{graphicx}
\usepackage{graphics}
\usepackage{caption}
\usepackage{bigints}
\usepackage{multirow}
\usepackage{subcaption}
\usepackage{dcolumn}
\usepackage{bm}
\RequirePackage{mathptmx}      
\RequirePackage{amsmath,amssymb}
\RequirePackage{bm}
\usepackage{adjustbox}
\RequirePackage{bbm}
\usepackage[ruled]{algorithm2e}
\RequirePackage{flushend}
\RequirePackage{natbib}
\usepackage[colorlinks,citecolor=blue,urlcolor=blue,linkcolor=blue]{hyperref}
\RequirePackage[usenames, dvipsnames]{xcolor}
\RequirePackage{tikz}
\RequirePackage[percent]{overpic}
\RequirePackage{setspace}
\usetikzlibrary{shapes.geometric, arrows}
\usepackage{natbib}
\newcommand{\rpm}{\sbox0{$1$}\sbox2{$\scriptstyle\pm$}
  \raise\dimexpr(\ht0-\ht2)/2\relax\box2 }
\newcommand\black[1]{\textcolor{black}{#1}}

\begin{document}

\widetext
\leftline{Version  of \today}

\title{Detecting new signals under background mismodelling}
\thanks{The author declares no conflict of interest.}%

\author{Sara Algeri}
\email{salgeri@umn.edu}
\affiliation{School of Statistics, University of Minnesota, Minneapolis (MN), 55455, USA }
\date{\today}

\begin{abstract}
Searches for  new astrophysical phenomena often involve several 
sources of non-random uncertainties which can  lead to highly misleading results.
Among these, model-uncertainty arising from background mismodelling  can dramatically compromise 
the sensitivity of the experiment under study. Specifically, overestimating the background distribution 
in the signal region increases the chances of missing new physics. Conversely, underestimating the background outside the signal region leads to an artificially enhanced sensitivity and a higher likelihood of claiming false discoveries.
The aim of this work is to provide a unified statistical strategy to perform modelling, estimation, inference, and signal characterization under background mismodelling. The method proposed allows to incorporate  the (partial) scientific knowledge available on the background distribution and provides a data-updated version of it in a purely nonparametric fashion without requiring the specification of prior distributions \black{on the unknown parameters}.  Applications in the context of dark matter searches and radio surveys show how the tools presented in this article can be used to incorporate non-stochastic uncertainty due to instrumental noise  and  to overcome  violations of  classical distributional assumptions in stacking experiments.
\end{abstract}


\maketitle
\section{Introduction}

When searching for new physics, a discovery claim is made if the data collected by the experiment provides sufficient statistical evidence in favor of the new phenomenon.  If the background and signal distributions are specified correctly, this can be done by means of statistical tests of hypothesis, upper limits and confidence intervals.

\textbf{\emph{The problem.}} In practice, even if a reliable description of the signal distribution is available,  providing accurate background models may be   challenging,  as the behavior of the sources which contribute to it is often poorly understood. Some examples include searches for nuclear recoils of weakly interacting massive particles   over electron recoils backgrounds \cite{aprile18, agnese18},  searches for gravitational-wave signals over non-Gaussian backgrounds from stellar-mass binary black holes \cite{smith18}, and  searches for a standard model-like Higgs boson over prompt diphoton production \cite{CMS18}.

Unfortunately, model uncertainty due to background mismodelling   can significantly compromise 
the sensitivity of the experiment under study. Specifically, overestimating the background distribution 
in the signal region increases the chances of missing new physics. Conversely, underestimating the background outside the signal region leads to an artificially enhanced sensitivity, which can easily result in false discovery claims. Several methods have been proposed in literature to address this problem  \cite[e.g.,][]{yellin, Priel,dauncey}. However, to the best of the author's  knowledge, none of the methods available \black{provides a unified strategy to} (i) assess the validity of existing models for the background, (ii) fully characterize the background distribution, (iii) perform signal detection even if the signal distribution is not available, (iv) characterize the signal distribution, and (v) detect additional signals of new unexpected sources.

\emph{\textbf{Goal.}} \black{The aim of this work is to integrate   modelling, estimation, and inference  under background mismodelling and provide a general statistical methodology to perform of (i)-(v).}
As a brief overview,  given   a source-free sample and the (partial) scientific knowledge available
on the background distribution,  a data-updated version of it is obtained in a purely nonparametric fashion without requiring the specification of prior distributions \black{on the unknown parameters}. At this stage, a graphical tool is provided in order to assess if and where significant deviations between the true and the postulated background distributions occur.
The ``updated'' background distribution is then used to assess if the distribution of  the data collected by the experiment deviates significantly from the background model. Also in this case, it is possible to assess graphically how the data distribution deviates from the expected background model. If a source-free sample is available, or if control regions can be identified, the solution proposed does not  require the specification  of a model for the signal; however, if the signal distribution is known (up to some free parameters),  the latter can be used to further improve the accuracy  of the analysis and to detect the signals of unexpected new sources. \black{Finally, the method  can be easily  adjusted to cover situations in which a source-free sample or control regions are not available, the background is unknown, or incorrectly specified, but a functional form of the signal distribution is known}.

\emph{\textbf{The key of the solution.}}  The statistical methodologies involved rely on the novel \emph{LP approach to statistical modelling} first introduced by  Mukhopadhyay and Parzen in 2014 \cite{LPapproach}.  \black{As it will become clearer later on in the paper, the letter \emph{L} typically denotes robust nonparametric
methods based on quantiles, whereas \emph{P} stands for polynomials \cite[Supp S1]{ksamples}.}
This approach allows the unification of many of the standard results of classical  statistics by expressing them in terms of
quantiles and comparison distributions and provides a simple and powerful framework for statistical learning and data analysis. The interested reader is directed to \cite{LPmode, LPBayes, LPtime, LPFdr, LPdec}  and references therein, for recent advancements  in  mode detection, nonparametric time series, goodness-of-fit on prior distributions,  and  large-scale inference  using an LP approach.

\emph{\textbf{Organization.}} Section \ref{LPmodelling} is dedicated to a review of the main constructs of LP modelling.  Section \ref{cali} highlights the practical advantages  offered by modelling background distributions using an LP approach. Section \ref{inference}  introduces a novel LP-based framework for statistical inference. 
 Section \ref{DS} outlines the main steps of a \black{data-scientific approach} for signal detection and characterization. 
 \black{In Section \ref{PSDMsec}, the methods proposed are applied in the context of dark matter searches where the goal is to distinguish $\gamma$-ray emissions due to dark matter from those due to pulsars.
In Section \ref{instrument},   the tools discussed are applied to a simulation of the Fermi Large Area Telescope  $\gamma$-ray telescope and it is shown how upper limits and Brazil plots can be constructed by means of comparison distributions}.
Section \ref{denoising} is dedicated to model-denoising. Section \ref{stacking} presents an application to data from the NVSS astronomical survey  and discusses a simple approach to assess the validity of distributional assumptions on the polarized intensity in stacking experiments. 
A  discussion of the main results  and extensions is proposed in Section \ref{discussion}.

\section{LP Approach to Statistical Modelling}
\label{LPmodelling}
The \emph{LP Approach to Statistical Modelling} \citep{LPapproach} is a novel statistical approach which  provides an ideal framework to simultaneously  assess the validity of the scientific knowledge available and  fill the gap between the initial scientific belief and the evidence provided by the data. \black{Sections \ref{skewGsec}, \ref{LegPoly} and \ref{LPestimate} below introduce the LP modelling framework, whereas Section \ref{cali} discusses how the problem of background mismodelling can be formulated under this paradigm.}

\black{
\subsection{The skew-G density model} 
\label{skewGsec}
 Let $X$ be a continuous random variable with cumulative distribution function (cdf) and probability density  function (pdf) $F(x)$.
 Since $F$ is the true distribution of the data, it is typically unknown. However, suppose a suitable cdf $G(x)$ is available, and let $g(x)$  be the respective pdf. In order to understand  if $G$ is a good candidate  for $F$, it is convenient to express the relationship among the two in a concise manner.}

 \black{
The \emph{skew-G density model} \citep{LPapproach,LPmode} is a universal representation scheme which allows to express any 
pdf $f(x)$ as 
\begin{equation}
\label{skewG}
f(x)=g(x)d(G(x);G,F)
\end{equation}
where $d(u;G,F)$ is called \emph{comparison density} \cite{manny2} and it is such that
\begin{equation}
\label{cd1}
d(u;G,F)=\frac{f(G^{-1}(u))}{g(G^{-1}(u))}\qquad\text{with $0\leq u\leq 1$},
\end{equation}
with $u=G(x)$ and $G^{-1}(u)=\inf\{x: G(x)\geq u\}$ denoting the ``postulated'' quantile function of  $X$. The comparison density is the pdf of the   random variable $U=G(X)$; whereas, its cdf is given by 
\begin{equation}
\label{cd2}
D(u)=F\bigl(G^{-1}(u)\bigl)=\int_0^u d(v;G,F)\partial v,
\end{equation}
and it is called \emph{comparison distribution}.
}

 \black{
\noindent\emph{\underline{Practical remarks.}} Equations \eqref{cd1} and  \eqref{cd2}  are of fundamental importance to understand the power of a statistical modelling approach based on the comparison density. Specifically, $d(u;G,F)$ 
 allows to ``connect'' any given pdf $g$ to the true pdf $f$ through the quantile transformation $G^{-1}$ of $u$. Furthermore,   $g\equiv f$ if and only if $d(u;G,F)=1$ for all $u \in [0,1]$, i.e., $U$ is uniformly distributed over the interval $[0,1]$.
Whereas, if $g\not\equiv f$, $d(u;G,F)$ models the departure of the true density $f(x)$ from the postulated model $g(x)$. 
Consequently, an adequate estimate of $d(u;G,F)$, not only  leads  to an estimate of the true $f(x)$ based on \eqref{skewG}, but   it also allows to identify the regions where  $f(x)$ deviates substantially from $g(x)$. }

 \black{
\subsection{LP skew-G series representation}
\label{LegPoly}
Denote with $L_2[0,1]$ the Hilbert space of square integrable functions on the unit interval with respect to the measure $G$. A complete, orthonormal basis of functions in $L_2[0,1]$ can be constructed considering  powers of $G(x)$, i.e., $G(x),G^2(x),G^3(x),\dots$ and
adequately orthonormalized via Gram-Schmidt procedure \cite{LPmode}. The resulting bases can equivalently be expressed as normalized shifted \textbf{L}egendre \textbf{P}olynomials,\footnote{Classical Legendre polynomials are defined over $[-1,1]$; here,   their ``shifted''  counterpart  over the range $[0,1]$ is considered. 
The first  three normalized shifted Legendre polynomials are: $Leg_0(u)=1$, $Leg_1(u) =\sqrt{12}(u-0.5)$, $Leg_2(u)=\sqrt{5}(6u^2-6u+1)$, etc.} namely $Leg_j(u)$, with $u=G(x)$. 
}

\black{
Under  the assumption that \eqref{cd1} is a square integrable function on $[0,1]$, i.e., $d\in L_2[0,1]$, 
we can then represent $d(u;G,F)$ via a series of $\{Leg_j(u)\}_{j\geq0}$ polynomials, i.e.,
\begin{equation}
\label{cd}
d(u;G,F)=1+\sum_{j>0}LP_jLeg_j(u)
\end{equation}
with coefficients $LP_j=\int_0^1Leg_j(u)d(u;G,F)\partial u$. The representation in \eqref{cd} is called \emph{LP skew-G series representation} \citep{LPmode}.
}

\black{
\subsection{LP density estimate}
\label{LPestimate}
Let $x_1,\dots,x_n$ be a sample of independent and identically distributed (i.i.d.) observations from  $X$. Observations from $U$ are given by $u_1=G(x_1),\dots,u_n=G(x_n)$.
The $LP_j$ coefficients in \eqref{cd} can then be estimated via
\begin{equation}
\label{LPest}
\widehat{LP}_j=\frac{1}{n}\sum_{i=1}^n Leg_j(u_i).
\end{equation}
Aternatively, in virtue of \eqref{cd2}, the estimates $\widehat{LP}_j$ can also be specified as 
\begin{equation}
\label{ecdf}
\widehat{LP}_j=\int_0^1 Leg_j(u) \partial \tilde{D}(u)=\int_0^1 Leg_j(u) \partial \tilde{F}(G^{-1}(u))
\end{equation}
where $\tilde{F}$ and $\tilde{D}$ denote the empirical distribution of the samples $x_1,\dots,x_n$ and $u_1,\dots,u_n$, respectively.
 }

\black{
The moments of the $\widehat{LP}_j$ are
\begin{equation}
\label{moments}
E[\widehat{LP}_j]=LP_j, \quad V(\widehat{LP}_j)=\frac{\sigma^2_j}{n}\quad\text{and}\quad Cov(\widehat{LP}_j, \widehat{LP}_k)=\frac{\sigma_{jk}}{n}
\end{equation}
where $\sigma^2_j=\int_0^1(Leg_j(u)-LP_j)^2d(u;G,F)\partial u$ and $\sigma^2_{jk}=\int_0^1(Leg_j(u)-LP_j)(Leg_k(u)-LP_k)d(u;G,F)\partial u$. 
When $f\equiv g$,  the equalities in \eqref{moments} reduce to
\begin{equation}
\label{momentsH0}
E[\widehat{LP}_j]=0,\quad V(\widehat{LP}_j)=\frac{1}{n}\quad\text{and}\quad Cov(\widehat{LP}_j, \widehat{LP}_k)=0
\end{equation} 
for all $j\neq k$. Derivations of  \eqref{moments} and \eqref{momentsH0} are discussed  in Appendix \ref{appA}. }

If   \eqref{cd} is approximated  by the first $M+1$ terms,\footnote{Recall that the first normalized shifted Legendre polynomial is $Leg_0(u)=1$.} an estimate of the comparison density is given by
\begin{equation}
\label{dhat}
\widehat{d}(u;G,F)=1+\sum_{j=1}^{M} \widehat{LP}_{j} Leg_{j}(u),
\end{equation}
\black{with variance
\begin{equation}
\label{variancedhat}
V\bigl[\widehat{d}(u;G,F)\bigl]=\sum_{j=1}^{M}\frac{\sigma^2_j}{n}Leg^2_j(u)+2\sum_{j<k}\frac{\sigma_{jk}}{n}Leg_j(u)Leg_k(u). 
\end{equation}
See Appendix \ref{appB} for more details on the derivation of \eqref{variancedhat}.
Finally, the standard error of $\widehat{d}(u;G,F)$ corresponds the square root of \eqref{variancedhat}, with $\sigma^2_{j}$ and $\sigma_{jk}$ estimated by their sample counterpart, i.e., 
\begin{equation*}
\begin{split}
\widehat{\sigma}^2_{j}&=\frac{1}{n}\sum_{i=1}^n (Leg_j(u_i)-\widehat{LP}_j)^2\\
\widehat{\sigma}_{jk}&=\frac{1}{n}\sum_{i=1}^nLeg_j(u_i)Leg_k(u_i)-\widehat{LP}_j\widehat{LP}_k.
\end{split}
\end{equation*}}

\begin{figure}[htb]
\centering
 \begin{adjustbox}{center}
\includegraphics[width=1\columnwidth]{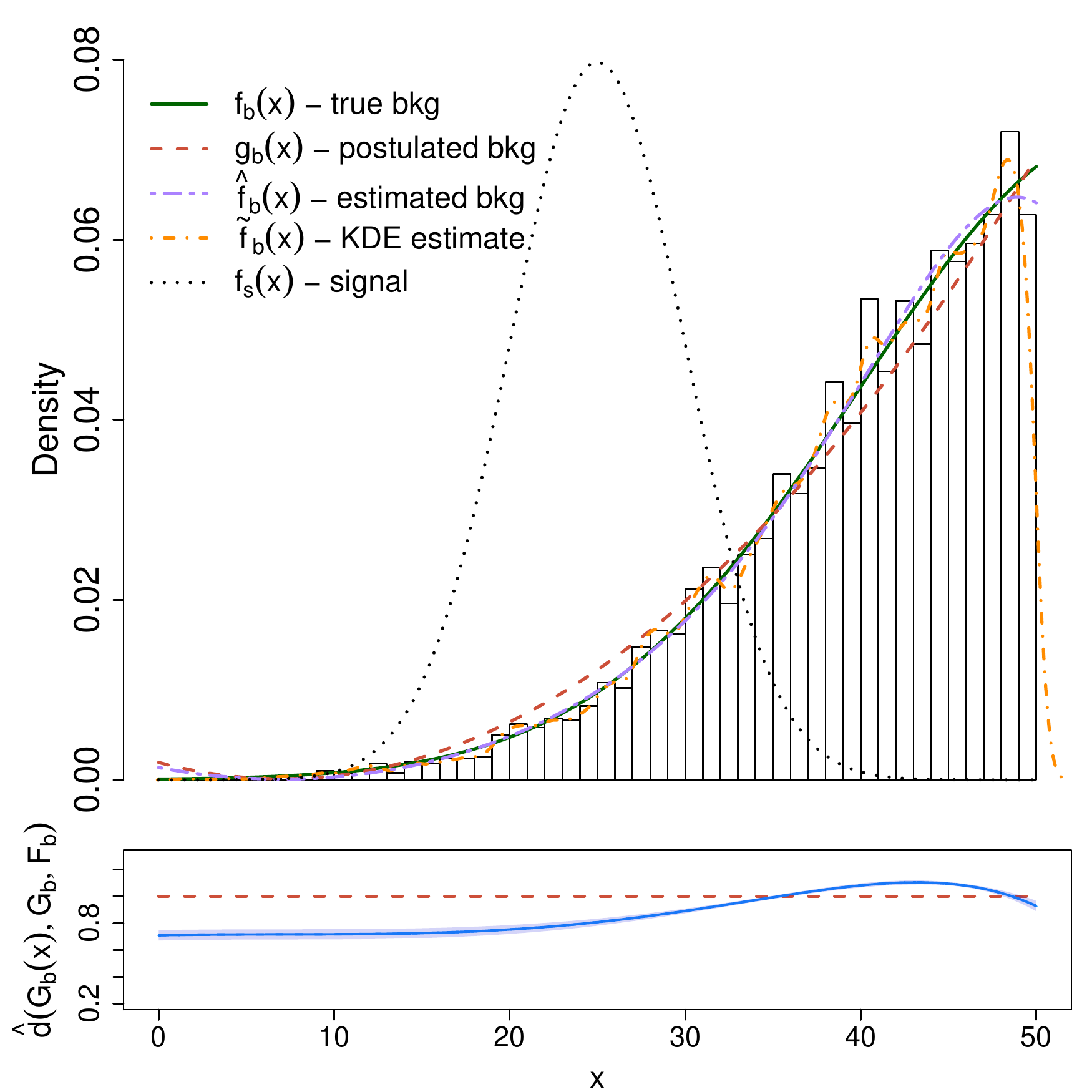} 
 \end{adjustbox}
\caption[Figure 1]{Upper panel: histogram of a source-free sample simulated from the tail of a Gaussian with mean $55$ and width $15$ (green solid line). The candidate background distribution is given by the best fit of a second-degree polynomial (red dashed line), and it is updated using the source-free data by means of \eqref{fbhat} (purple dot-dashed line). 
\black{The Kernel density estimator of $f_b$ is also displayed for comparison (orange dot-dashed line).}
Bottom panel: comparison density estimate (blue solid line) plotted on the $x$-scale and respective standard errors (light blue area).  }
\label{Fig1}
\end{figure}
Finally, in virtue of the skew-G density model in \eqref{skewG} we can estimate $f(x)$ as
\begin{equation}
\label{fhat}
\widehat{f}(x)=g(x)\widehat{d}(G(x);G,F).
\end{equation}
\black{
Since each $Leg_j(u)$ is a polynomial function of the random variable $U$, 
each $\widehat{LP}_j$ estimate can be expressed as  a linear combination of the first $j$ sample moments of $U$,  e.g.,
\[\widehat{LP}_2=\frac{1}{n}\sum_{i=1}^n Leg_2(u_i)=\sqrt{5}\Bigl(6\widehat{\mu}_2-6\widehat{\mu}_1+1\Bigl)\]
where $\widehat{\mu}_2=\frac{1}{n}\sum_{i=1}^nu_i^2$, $\widehat{\mu}_1=\frac{1}{n}\sum_{i=1}^nu_i$. Therefore,
the truncation point $M$ can be interpreted as the order of the highest moment considered to characterize the distribution of $U$.  (The reader is directed to Section \ref{chooseMsec} for a discussion on the choice of $M$.)}

 
\black{
\subsection{The bias variance trade-off}
\label{biasvariance}
In order to understand how good \eqref{dhat} is in estimating $d(u;G,F)$ we consider the Mean Integrated Squared Error (MISE) of  $\widehat{d}(u;G,F)$, i.e.,
\begin{align}
\label{MISE}
MISE&=E\biggl[\int_0^1 \bigl(\widehat{d}(u;G,F)-d(u;G,F)\bigl)^2\partial u\biggl]\\
\label{MISE2}
&=\sum_{j=1}^M\frac{\sigma^2_j}{n}+\sum_{j>M}LP^2_j
\end{align}
where the first term in \eqref{MISE2} corresponds to the integral of the \eqref{variancedhat} over $[0,1]$; whereas the second term corresponds to the Integrated Squared Bias (IBS), i.e, 
\begin{equation}
\label{intBias}
IBS=\bigintsss_0^1 \biggl(E\bigl[\widehat{d}(u;G,F)\bigl]-d(u;G,F)\biggl)^2 \partial u.
\end{equation}
Interestingly, the latter can also be specified as
\begin{equation}
\label{intBias2}
IBS=\bigintsss_0^1\biggl(\frac{f(x)-g(x)}{g(x)}\biggl)^2\partial u-\sum_{j=1}^MLP^2_j
\end{equation}
(see derivations in Appendix \ref{appB}). The first term on the right hand side of \eqref{intBias2} is particularly important in understanding the role played by $g$ in obtaining a reliable estimate of $f$.
Specifically, the closer $g$ is to $f$ the lower the bias of $\widehat{d}(G(x);G,F)$ and $\widehat{f}(x)$ in \eqref{fhat}.
}

\black{
\noindent\emph{\underline{Practical remarks.}} 
Equation \eqref{MISE2} implies that larger values of $M$ do not necessarily lead to better estimates of $d(u;G,F)$. Specifically, when $n\rightarrow\infty$, the first term in \eqref{MISE2} tends to zero. However, for large values of $M$, more and more terms to contribute to it and thus increasing $M$ may lead to a substantial inflation of the variance in \eqref{variancedhat}.  Conversely, the bias is not affected by sample size and it can be controlled by either choosing  $g$ sufficiently close to $f$ (see \eqref{intBias2}) and/or  increasing $M$ while preserving a good bias-variance trade-off.  }

\black{
\noindent\emph{\underline{Further remarks.}} Equation \eqref{ecdf} implies that the estimator in \eqref{dhat} relies on the empirical distribution of the sample observed  by means of the $\widehat{LP}_j$ estimates. Therefore, an estimator of the comparison density based entirely on the empirical cdf can be expressed by setting $M=n-1$ in \eqref{dhat}. However, as discussed in this section, while this would reduce the bias, it would also icrease the variance drastically. Therefore, for $M<n-1$, the estimator in \eqref{dhat}, not only leads to a reduction of the variance but, in virtue of \eqref{intBias2}, its bias is mitigated when the postulated model $g$ is sufficiently close to the true pdf of the data $f$.
}

\section{Data-driven corrections for misspecified background models}
\label{cali}

\black{
Let $\bm{x}_{\text{B}}=(x_{1},\dots,x_{N})$ be a  sample of observations   from control regions or the result of Monte Carlo simulations,  where we expect  no signal to be present. 
Hereafter, we  refer to  $\bm{x}_{\text{B}}$ as the \emph{source-free sample}.  Therefore,
 $\bm{x}_{\text{B}}$  can be used to ``learn'' the unknown pdf of the background, namely $f_b(x)$, and obtain an estimate for it via \eqref{fhat}. }
 
 Despite the true background model being unknown,  suppose that a candidate pdf, namely $g_b(x)$, is available. The candidate model $g_b(x)$ can be specified from previous experiments or theoretical results  or can be obtained by  fitting specific functions (e.g., polynomial, exponential, etc.) to $\bm{x}_{\text{B}}$. If $g_b(x)$ does not provide an accurate description of $f_b(x)$, the sensitivity of the experiment can be strongly affected. 

Consider, for instance, a source-free sample of $N=5000$ observations whose true (unknown) distribution corresponds to the tail of a Gaussian with mean $55$ and width $15$ over the range $[0,50]$, i.e.,
\begin{equation}
\label{fb}
f_b(x)= \frac{e^{-\frac{1}{2}\bigl(\frac{x-55}{15}\bigl)^2}}{k_{fb}}
\end{equation}
with $k_{fb}=\int_{0}^{50}e^{-\frac{1}{2}\bigl(\frac{x-55}{15}\bigl)^2}\partial x$.  Suppose that a candidate model for the background is obtained by fitting a second-degree polynomial on the source-free sample and adequately normalizing it in order to obtain a proper pdf, i.e.,
\begin{equation}
\label{gb}
g_b(x)=\frac{9.52 -2.22x+  0.15x^2}{k_{gb}}
\end{equation}
with $k_{gb}=\int_{0}^{50}[9.52 -2.22x+  0.15x^2]\partial x$.  For illustrative purposes, assume that  the distribution of the signal is  a Gaussian centered at $25$, with width $4.5$ and pdf 
\begin{equation}
\label{fs}
f_s(x)= \frac{e^{-\frac{1}{2}\bigl(\frac{x-25}{4.5}\bigl)^2}}{k_{fs}}
\end{equation}
with $k_{fs}=\int_{0}^{50}e^{-\frac{1}{2}\bigl(\frac{x-25}{4.5}\bigl)^2}\partial x$. 
The histogram of the source-free sample along with 
\eqref{fb}-\eqref{fs} is shown in Fig. \ref{Fig1}.  
 At the  higher end of the spectrum, the postulated background  (red dashed line) underestimates the true background distribution (green solid line). As a result, using \eqref{gb} as background model increases the chance of false discoveries in this region. Conversely, at the lower end of the spectrum, $g_b(x)$ underestimates $f_b(x)$, reducing  the sensitivity of the analysis. \black{For the sake of comparison, a Kernel density estimate  (orange dot-dashed line) has been computed by selecting the bandwidth parameter as recommended in \cite{sheater}. The latter exhibits substantial bias at the boundary and appears to overfit the data sample.}

It is important to point out that, the discrepancy of $f_b(x)$ from $g_b(x)$ is typically due to the fact that the specific functional form  imposed (in our example, a second-degree polynomial) is not adequate for the data. Thus, changing the values of the fitted parameters (or assigning priors to them) is unlikely to solve the problem. 
However, it is possible to ``repair'' $g_b(x)$ and obtain a suitable estimate of $f_b(x)$ by means of  \eqref{fhat}. Specifically,   $f_b(x)$ can be estimated via
\begin{equation}
\label{fbhat}
\widehat{f_b}(x)=g_b(x)\widehat{d}(G_b(x);G_b,F_b)
\end{equation}
where $\widehat{d}(G_b(x);G_b,F_b)$ is the comparison density  estimated  via \eqref{dhat} on the sample $G_b(x_{1}),\dots,G_b(x_{N})$, whereas $F_b$ and $G_b$ are the true and the postulated background distributions, with pdfs as in \eqref{fb} and \eqref{gb}, respectively. 

In our example, choosing $M=2$ (see Section \ref{chooseMsec}), we obtain
 {\fontsize{3.5mm}{3.5mm}\selectfont{
\begin{equation}
\label{dbhat}
\begin{split}
\widehat{d}(G_b(x);G_b,F_b)=&1+0.063Leg_1[G_b(x)]-\\
&0.082Leg_2[G_b(x)],\\
\end{split}
\end{equation}}}
where  $Leg_1[G_b(x)]$ and $Leg_2[G_b(x)]$  are the first and second normalized shifted Legendre polynomials evaluated at $G_b(x)$.
Notice that, by combining  \eqref{fbhat} and \eqref{dbhat}, we can easily write the background model using  of a  series of shifted Legendre polynomials. This may be especially useful when dealing with complicated likelihoods and for which a functional form is difficult to specify.  

The upper panel of Fig. \ref{Fig1} shows that the ``calibrated'' background model in \eqref{fbhat} as a purple dot-dashed line and matches almost exactly the true background density in \eqref{fb} (green solid line).  
The plot of $\widehat{d}(G_b(x);G_b,F_b)$ in the bottom panel of Fig. \ref{Fig1} 
provides important insights on the deficiencies of \eqref{gb}
as a candidate background model. Specifically, the magnitude and the direction of the departure of 
$\widehat{d}(G_b(x);G_b,F_b)$  from one corresponds to the estimated departure of   $f_b(x)$  from  $g_b(x)$ for each value of $x$. 
Therefore, if $\widehat{d}(G_b(x);G_b,F_b)$ is below one in the region where we  expect the signal to occur, using $\widehat{f_b}(x)$ in place of $g_b(x)$ increases the sensitivity of the analysis. Conversely, if $\widehat{d}(G_b(x);G_b,F_b)$ is above one outside the signal region, the use of $\widehat{f_b}(x)$ instead of $g_b(x)$ prevents from false discoveries.

Notice that in this article we only consider  continuous data. In this respect, the goal is to learn the model of the background considered as a continuum and no binning is applied. Therefore, the histograms presented  here  are only a graphical tool used to display the data distribution and are not intended to represent an actually binning of the data. 

\section{LP-based inference}
\label{inference}
 When discussing the skew-G density model in \eqref{skewG}, we have witnessed that  $f\equiv g$ if $d(u;G,F)=1$ for all $u \in [0,1]$. Additionally, the graph of $\widehat{d}(u;G,F)$ provides an exploratory tool to understand the nature of the deviation of $f(x)$ from $g(x)$.  This section introduces a novel inferential framework to test 
the significance of the departure of $f(x)$ from $g(x)$. Specifically, our goal is to test the hypotheses
\begin{equation}
\label{hp1}
\begin{split}
H_0:d(u;G,F)=1 &\text{ for all $u \in [0,1]$}\\
&vs\\
H_1:d(u;G,F)\neq1 &\text{ for some $u \in [0,1]$.}\\
\end{split}
\end{equation}
 First,  an overall test, namely the \emph{deviance test}, is presented. The deviance test assesses \underline{if}  $f(x)$ deviates significantly from $g(x)$  anywhere over the range of $x$ considered. Second, adequate confidence bands are constructed in order to  assess \underline{where} significant departures occur. 

\subsection{The deviance test} 
\label{dev}\black{
Recall that the $LP_j$ coefficients in \eqref{cd} specify as $LP_j=\int^1_0Leg_u(d)d(u;G,F)\partial u$. Consequently, by orthogonality of the $\{Leg_j(u)\}_j>0$ polynomials and $Leg_0(u)=1$, when
$H_0$ in \eqref{hp1} is true all the $LP_j$ coefficients are equal to zero, including the first $M$ of them.  We can then quantify  the departure of $\widehat{d}(u;G,F)$ from one by means of the \emph{deviance} statistics  \cite{LPFdr} which specifies as $\sum_{j=1}^{M}\widehat{LP}^2_{j}$.}
If the deviance is equal to zero, we may expect that $g$ is approximately equivalent to $f$; hence, we test
\begin{equation}
\label{Dtest}
H_0:\sum_{j=1}^{M}LP^2_{j}=0 \qquad \text{vs\qquad}H_1:\sum_{j=1}^{M}LP^2_{j}>0
\end{equation}
by means of the test statistic
\begin{equation}
\label{D}
D_M=n \sum_{j=1}^{M}\widehat{LP}^2_{j}.
\end{equation}
\black{It can be shown \cite{LPmode} that, as $n\rightarrow\infty$
\begin{equation}
\label{lpH0}
\sqrt{n}\widehat{LP}_j\xrightarrow{d} N(0, 1),
\end{equation} 
where $\xrightarrow{d}$ denotes convergence in distribution, and thus, } 
under $H_0$, $D_M$ is asymptotically $\chi^2_M$-distributed. Hence, an asymptotic p-value for \eqref{Dtest} is given by
\begin{equation}
\label{D_distr}
P(D_M>d_M)\xrightarrow[{n\rightarrow\infty}]{} P(\chi^2_M> d_M),
\end{equation}
where $d_M$ is the value of $D_M$ observed on the data. \\

\noindent\emph{\underline{Practical remarks.}}  
Notice that 
$H_1$ in \eqref{Dtest} implies $H_1$ in \eqref{hp1}. Similarly, $H_0$ in \eqref{hp1} implies $H_0$ in \eqref{Dtest}; however, the opposite is not true in general since there may be some non-zero $LP_j$ coefficients for $j>M$. Therefore, even when  choosing $M$ small may lead to conservative, but yet valid, inference.  

\subsection{Confidence bands}
\label{bands}
\black{The estimator in  \eqref{dhat} only accounts for the first $M+1$ terms  of the polynomial series in \eqref{cd}. Therefore,  $\widehat{d}(u;G,F)$ is a biased estimator of  $d(u;G,F)$. Specifically, as discussed in Section \ref{biasvariance}, the integrated  bias is given by $\sum_{j>M}LP^2_j$, whereas, as show in Appendix \ref{appB}, the bias at a given point $u$ is given by $\sum_{j>M}LP_jLeg(u)$. }
It follows that, when the bias is large,  confidence bands based on $\widehat{d}(u;G,F)$ are shifted away from the true density $d(u;G,F)$.

Despite the bias cannot be easily quantified in the general setting, it follows from \eqref{momentsH0} that, \black{when $H_0$ in \eqref{hp1} (and consequently $H_0$ in \eqref{Dtest})}  is true, both the bias at a point $u$ and the integrated bias are equal to zero. Thus, we can exploit this property   to construct reliable confidence bands under the null. Specifically, the goal is to identify $c_{\alpha}$, such that
\begin{equation}
\label{significance}
\begin{split}
1-\alpha&=P(-c_{\alpha}\leq \widehat{d}(u;G,F)-1\leq c_{\alpha},\text{ for all $u\in[0,1]$}|H_0)\\
     &=P(\max_{u} |\widehat{d}(u;G,F)-1|\leq c_{\alpha}|H_0)\\
\end{split}
\end{equation}
where $\alpha$ is the desired significance level.\footnote{In astrophysics, the statistical significance $\alpha$ is often expressed in terms of number of $\sigma$-deviations from the mean of a standard normal, namely $\sigma$. 
For instance, a 2$\sigma$ significance corresponds to $\alpha=1-\Phi(2)=0.0227$, where $\Phi(\cdot)$ denotes the cdf of a standard normal. }

If the bias determines where the confidence bands are centered, the  distribution and the variance of $\widehat{d}(u;G,F)$ determine their width. \black{
As discussed in Section \ref{LPestimate} (see \eqref{momentsH0}), under $H_0$ in \eqref{hp1}, the $\widehat{LP}_j$ estimates have mean zero, variance $\frac{1}{n}$ and they
are uncorrelated one another. Therefore, when $f\equiv g$, the standard error of $\widehat{d}(u;G,F)$, corresponds to the square root of \eqref{variancedhat} with $\sigma^2_j=1$ and $\sigma^2_{jk}=0$  i.e.,
\begin{equation}
\label{SEdhat0}
SE\Bigl[\widehat{d}(u;G,F)|H_0\Bigl]=\sqrt{\sum_{j=1}^{M} \frac{1}{n} Leg^2_{j}(u)}.
\end{equation}}
Additionally,   \eqref{lpH0} implies  that  $\widehat{d}(u;G,F)$ is  asymptotically normally distributed, hence
\begin{equation}
\label{pivot1}
\frac{\widehat{d}(u;G,F)-1}{\sqrt{\sum_{j=1}^M\frac{1}{n}Leg_j^2(u)}}\xrightarrow{d}N(0,1).
\end{equation}
as $n\rightarrow\infty$, for all $u\in[0,1]$, under $H_0$.

We can then construct approximate confidence bands  under $H_0$ which satisfy \eqref{significance}  by means of tube formulae (see \cite[Ch.5]{larry} and \cite{PL05}), i.e.,
\begin{equation}
\label{CIband}
\Biggl[1-c_\alpha\sqrt{\sum_{j=1}^M\frac{1}{n}Leg_j^2(u)},1+c_\alpha\sqrt{\sum_{j=1}^M\frac{1}{n}Leg_j^2(u)}\Biggl],
\end{equation}
where $c_\alpha$ is the solutions of 
\begin{equation}
\label{CIband2}
2(1-\Phi(c_\alpha))+\frac{k_0}{\pi}e^{-0.5c^2_{\alpha}}=\alpha,
\end{equation}
with $k_0=\sqrt{\sum^M_{j=1}[\frac{\partial}{\partial u}Leg_j(u)]^2}$.  If   $\widehat{d}(u;G,F)$ is within the bands in \eqref{CIband} over the entire range $[0,1]$, we conclude that there is no evidence that $f$ deviates significantly from $g$ anywhere over the range considered \black{and at confidence level $1-\alpha$}. Conversely, we expect  significant departures to occur in  regions  where $\widehat{d}(u;G,F)$ lies outside the confidence bands.\\

\noindent\emph{\underline{Practical remarks.}} Notice that, under $H_0$ in \eqref{Dtest}, the  $\widehat{d}(u;G,F)$ is an unbiased estimator of ${d}(u;G,F)$, regardless of the choice of $M$. This implies that 
the confidence bands in \eqref{CIband} are  only affected by the variance and asymptotic distribution of $\widehat{d}(u;G,F)$ under $H_0$.

\subsection{Choice of $M$}
\label{chooseMsec}
The number of $\widehat{LP}_j$ estimates considered determines the \black{level of ``smoothness''} \footnote{ \black{As an anonymous referee correctly pointed out, $\widehat{d}(u;G,F)$ is always smooth as it is constructed as a series of infinitely differentiable functions. In statistics, however, the word  ``smoothness'' is often used to indicate the flexibility of the estimator considered or, in other words,  its degrees of freedom. Often, this is quantified in terms of magnitude of the second derivative of the function considered. Despite the abuse of terminology, throughout the manuscript we will refer to the latter definition of smoothness.}} of $\widehat{d}(u;G,F)$, with smaller  values of $M$ leading to smoother estimates.
The deviance test can be used to select the value $M$   which maximizes the sensitivity of the analysis according to the following scheme:
\begin{enumerate}
\item[i.] Choose a sufficiently large value $M_{\max}$.
\item[ii.] Obtain the estimates $\widehat{LP}_1,\dots, \widehat{LP}_{M_{\max}}$ as in \eqref{LPest}.
\item[iii.]  For $m=1,\dots,M_{\max}$:
\begin{enumerate}
\item[ ] calculate the deviance test p-value as in \eqref{D_distr}, i.e.,  
\begin{equation}
\label{pvalm}
p(m)=P\Bigl(\chi^2_m>  d_m\Bigl)\end{equation}
with $d_m=n\sum_{j=1}^{m}\widehat{LP}^2_{j}$.
\end{enumerate}
\item[iv.]  Choose $M$ such that
\begin{equation}
\label{chooseM}
M=\underset{m}{\mathrm{argmin}}\{ p(m)\}.
\end{equation}
\end{enumerate}

\subsubsection{Adjusting for post-selection}
As any data-driven selection process, the  scheme presented above affects the distribution of \eqref{dhat}  and can yield to  overly optimistic inference \cite{xiaotong,potscher}.  Despite this aspect being often ignored in practical applications, correct coverage can only be guaranteed  if adequate corrections are implemented. 

\black{The issues arising in the context of post-selection inference can be interpreted in terms of looks-elsewhere effect \cite{gv10,meJINST} where one has to adjust the inference for the fact that, in practice, many different models have been considered and, consequently, many different tests have been conducted for the sake of assessing the goodness of fit.}

In our setting, the number of models under comparison is typically small ($M_{\max}\leq20$); therefore,  post-selection inference  can be easily adjusted by means of Bonferroni's correction \cite{bonferroni35}. Specifically, the adjusted deviance p-value is given by
\begin{equation}
\label{bonf}
M_{\max}\cdot P(\chi^2_M> d_M),
\end{equation}
\begin{algorithm}[!h]
\label{algo}
\caption{a data-scientific signal search} 
\vspace{0.1cm}
\textbf{INPUTS:}  \black{source-free sample ${\bm{x}}_{\text{B}}$;}\\
\hspace{1.5cm}postulated background  distribution $g_b(x)$;\\
\hspace{1.5cm}physics sample $\bm{x}$.\\
\hspace{1.5cm}\emph{If available}: signal distribution, $f_s(x,\bm{\theta}_s)$.\\
\vspace{0.3cm}
\textbf{PHASE A: background calibration}
\begin{enumerate}
\item[\emph{Step 1:}]\black{ Estimate  $\widehat{d}(u;G_b,F_b)$ on ${\bm{u}}_{\text{B}}=G_b({\bm{x}}_{\text{B}})$ and test \eqref{hp1} via deviance test and CD plot.}
\item[\emph{Step 2:}] \textbf{if} $F_b\not \equiv G_b$, set $\widehat{f}_b(x)=g_b(x)\widehat{d}(u;G_b,F_b)$; \\
 \textbf{else} set $\widehat{f}_b(x)=g_b(x)$.
\end{enumerate}
\vspace{0.3cm}
\textbf{PHASE B: signal search}\\
\vspace{0.15cm}
\textbf{Stage 1: nonparametric signal detection}
\begin{itemize}
\item[\emph{Step 3:}] set $g(x)=\widehat{f}_b(x)$.
\item[\emph{Step 4:}]  estimate  $\widehat{d}(u;G,F)$ on $\bm{u}=G(\bm{x})$ and test \eqref{hp1} via deviance test and CD plot.
\item[\emph{Step 5:}]  \textbf{if} $G\not \equiv F$, claim   \underline{evidence in favor of the signal} and go to Step 6; \\ 
 \textbf{else} set $\widehat{f}(x)=g(x)$, claim that
\underline{no signal is} \underline{present} and stop. 
\end{itemize}
\vspace{0.15cm}
\textbf{Stage 2: semiparametric signal characterization}
\begin{enumerate}
\item[\emph{Step 6:}]  \textbf{if} $f_s(x,\bm{\theta}_s)$ given, fit $g_{bs}(x)$ in \eqref{gbs};\\
\textbf{else} use the CD plot of $\widehat{d}(u;G,F)$ and the theory available to specify/fit 
a suitable model for $f_s(x,\bm{\theta}_s)$  and fit $g_{bs}(x)$ in \eqref{gbs}.
\item[\emph{Step 7:}] estimate  $\widehat{d}(u;G_{bs},F)$ on $\bm{u}=G_{bs}(\bm{x})$ and test \eqref{hp1} via deviance test and CD plot.
\item[\emph{Step 8:}] \textbf{if} $G_{bs}\not \equiv F$,  claim  \underline{evidence of  unexpected signal}  and use the CD plot of $\widehat{d}(u;G_{bs},F)$ and the theory available to further investigate the nature the deviation from $G_{bs}$;\\
\textbf{else} go to Step 9.
\item[\emph{Step 9:}] compute $\widehat{\widehat{d}(}u;G,F)$ as in \eqref{semipard} and use it to refine $\widehat{f}_b(x)$ or $f_s(x,\widehat{\bm{\theta}_s})$ as in \eqref{refine}. Go back to Step 3.
\end{enumerate}
\end{algorithm}
where $M$ is the value selected  via \eqref{chooseM}, whereas confidence bands can be adjusted by substituting  $c_\alpha$ in \eqref{CIband}, with $c_{\alpha,M_{\max}}$ satisfying
\begin{equation}
\label{CIband3}
2(1-\Phi(c_{\alpha,M_{\max}}))+\frac{k_0}{\pi}e^{-0.5c^2_{\alpha,M_{\max}}}=\frac{\alpha}{M_{\max}}.
\end{equation}

\noindent\emph{\underline{Practical remarks.}} 
As noted in Section \ref{LPmodelling}, the estimate \eqref{dhat} involves the first $M$ sample moments of $U$; therefore, $M_{\max}$ can be interpreted as the order of the highest moment which we expect to contribute in discriminating the distribution of $U$ from uniformity. \black{Notice that, in addition to the inflation of the variance of \eqref{dhat}, when $M$ is large, the computation of normalized shifted Legendre of higher order may face numerical instability (see Section \ref{AICtest}). } Therefore, 
as a rule of thumb, $M_{\max}$ is typically chosen $\leq20$. Finally, Steps i-iv aim to select the approximant based on the first most significant $M$ moments, while excluding powers of higher order. A further note on model-denoising is given in Section \ref{denoising}.
\section{A data-scientific approach to signal searches}
\label{DS}
The tools presented in Sections \ref{LPmodelling} and \ref{inference} provide a natural framework to simultaneously
\begin{itemize}
\item[(a)] assess the validity of the postulated background model and, if necessary, update it using the data (Section \ref{cali});
\item[(b)] perform signal detection on the physics sample;
\item[(c)] characterize the signal when a model for it is not available.
\end{itemize}

Furthermore, if the model for the signal is known (up to some free parameters), it is possible to
\begin{itemize}
\item[(d)] further refine the background or signal distribution;
\item[(e)] detect hidden signals from new unexpected sources.
\end{itemize}

Notice that, since Bonferroni's correction leads to an upper bound for the overall significance, the resulting coverage will be higher than the nominal one. Alternatively,  approximate post-selection confidence bands and inference can be constructed using Monte Carlo and/or resampling methods  and repeating the selection process at each replicate.

Tasks (a)-(e) can be tackled in two main phases. In the first phase, the postulated background model is ``calibrated'' on a source-free sample in order to improve the sensitivity of the analysis and reduce the risk of false discoveries. The second phase focuses on searching for the signal of interest and involves both  a nonparametric signal detection stage and  a semiparametric stage for signal characterization.  Both phases and respective steps are described in details below and summarized in Algorithm 1.

\begin{figure}[htb]
\centering
 \begin{adjustbox}{center}
\includegraphics[width=90mm]{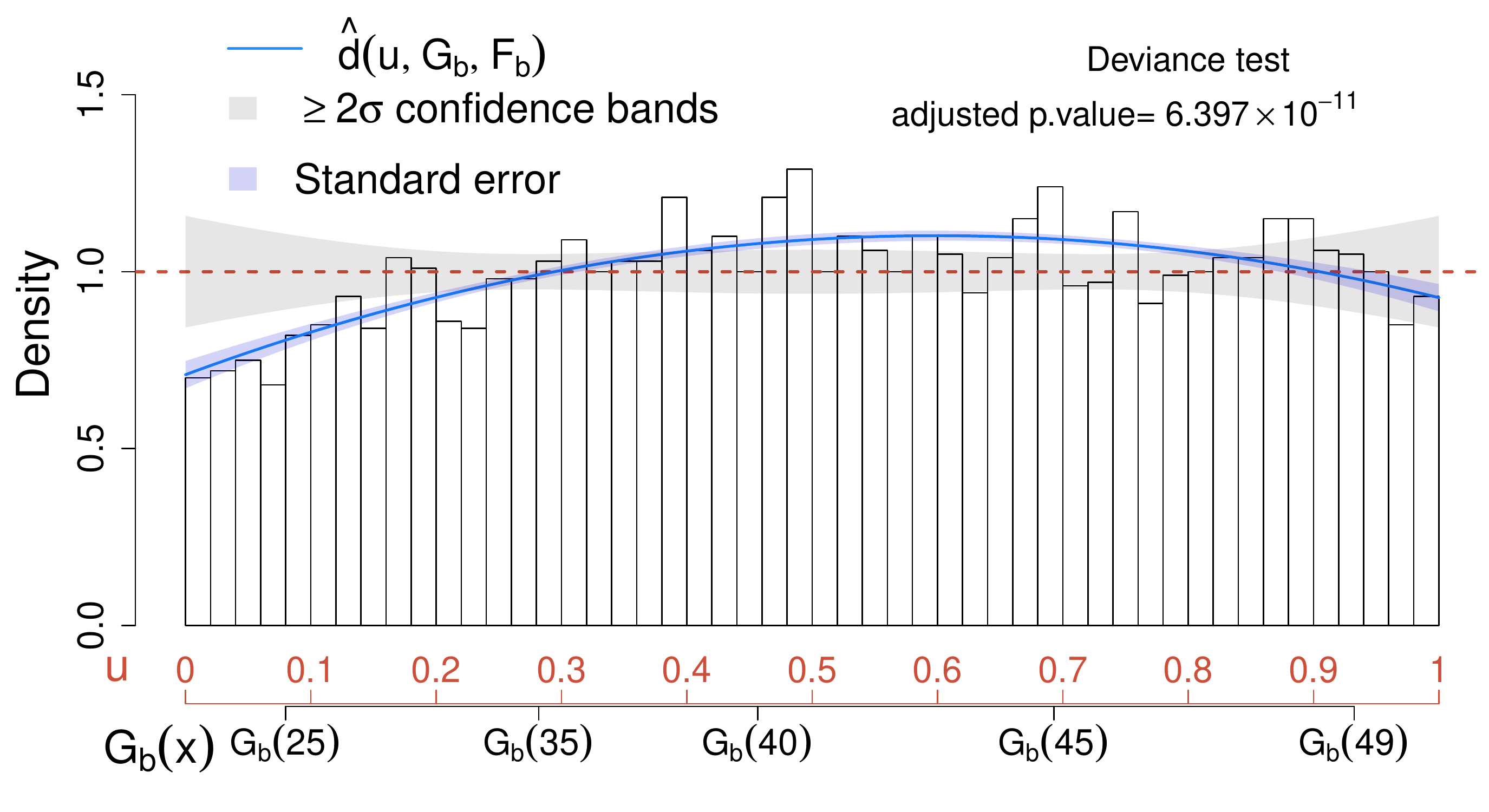} 
 \end{adjustbox}
\caption[Figure 2]{Deviance test and CD plot for the source-free sample. The comparison density is estimated via \eqref{dbhat} (solid blue line), \black{ whereas its standard error (light blue area) is computed as the squared root of the estimate of the variance in \eqref{variancedhat}.} Finally, confidence bands have been constructed around one (grey areas) via \eqref{CIband} with $c_\alpha$ replaced by $c_{\alpha,M_{\max}}$ in \eqref{CIband3}. The notation $\geq 2\sigma$ is used to highlight  that Bonferroni's correction has been applied to adjust for post-selection inference, leading to an increase of the nominal coverage. }
\label{Fig2}
\end{figure}

\subsection{Background calibration}
\label{bkgcali}
As discussed in Section \ref{cali}, deviations of  $\widehat{d}(G_b(x);G_b,F_b)$   from one suggest that a further refinement of the candidate background model  $g_b$  is needed. However, as $M$ increases,  the deviations of  $\widehat{d}(G_b(x);G_b,F_b)$  from one may become more and more prominent while the variance inflates. Thus, it is  important to assess if such deviations are indeed significant. In order address this task,   the analysis of Section \ref{cali} can be further refined  in light of the inferential tools introduced in Section \ref{inference}.

For the toy example discussed in Section \ref{cali}, we have seen that   $g_b$ overestimates $f_b$ in the signal region  and underestimates it at the higher end of the range considered (Fig. \ref{Fig1}). We can now assess if any of these deviations are significant by implementing the deviance test in \eqref{D}-\eqref{D_distr}, whereas,  to identify where the most significant departures occur, we construct confidence bands under the null model  as in \eqref{CIband}, i.e.,  assuming that no ``update'' of  $g_b$ is necessary.

The results are collected in the \emph{comparison density plot} or \emph{CD plot} presented in Fig. \ref{Fig2}. First, a value $M=2$ has been selected as in \eqref{chooseM}, and the respective deviance test (adequately adjusted via Bonferroni) indicates that the deviation of $f_b$ from $g_b$ is significant at a $6.430\sigma$ significance level (adjusted p-value of $6.397\cdot10^{-11}$). Additionally, the estimated comparison density in \eqref{dbhat}  lies outside the $2\sigma$ confidence bands in the region $[0,50]$   where the signal specified in \eqref{fs} is expected to occur. Hence, using \eqref{fbhat}  instead of \eqref{gb} is recommended in order to improve the sensibility of the analysis in the signal region.\\

\noindent\underline{\emph{Important remarks on the CD plot.}} When comparing different models for the background  or when assessing if the data distribution deviates from the model  expected  when no signal is present,  it is common practice to visualize the results of the analysis by superimposing the models under comparison to the histogram of the data observed on the original scale (e.g.,  upper panel of  Fig. \ref{Fig1}).
This corresponds to a data visualization in the density domain. Conversely,   the 
CD plot (e.g.,    Fig. \ref{Fig2}) provides a representation of the data in the quantile domain,   which offers the advantage of connecting the true  density of the data with the quantiles of the postulated model (see \eqref{cd1}-\eqref{cd2}). Consequently, the most substantial departures of the data distribution    from the expected model are magnified, and  those due to random fluctuations are smoothed out (see, also, Section \ref{upperlimits}). 
Furthermore, the deviance tests and the CD plot together provide a powerful goodness-of-fit tool and exploratory which, conversely from classical methods such as Anderson-Darling \cite{anderson} and  Kolmogorov-Smirnov \cite{darling},  not only allow to test \underline{if}  the distributions under comparison differ, but they also allow to assess \underline{how} and \underline{where} they differ. As a result, the CD plot can be used to characterize the unknown signal distribution (see Section \ref{signalcar}) and to identify exclusion regions (e.g., Case I in Section \ref{nonpar}). 

\black{As an additional advantage, the deviance test appears to enjoy  higher detection power than classical approaches. This aspect is highlighted in Table \ref{GOF} where several  methods for goodness of fit or two-samples comparisons are implemented, along with the deviance test, for all the cases discussed in Section \ref{DS}. }\\

\noindent\underline{\emph{Reliability of the calibrated background model.}} The size $N$ of the source-free sample plays a fundamental role in the validity of $\widehat{f}_b(x)$ as a reliable background model. Specifically, the randomness involved in \eqref{fbhat} only depends on the $\widehat{LP}_j$ estimates. If $N$ is sufficiently large, by the strong law of large numbers, 
\[P\bigl(\underset{{N\rightarrow\infty}}{\lim}\widehat{LP}_j=LP_j\bigl)=1.\]
Therefore, despite the variance of $\widehat{f}_b(x)$ becoming negligible as $N\rightarrow\infty$, one has to account for the fact that $\widehat{f}_b(x)$ leads to a biased estimate of $f_b(x)$ when $f_b\not\equiv g_b$ (see Section \ref{biasvariance}). For sufficiently smooth densities, a visual inspection is often sufficient to assess if $\widehat{d}(u;G_b,F_b)$ (and, consequently, $\widehat{f}_b(x)$) provides a satisfactory fit for the data, whereas, for more complex distributions the effect of the bias can be  mitigated considering larger values of $M$ and model-denoising (see Section \ref{denoiseAIC}). 

{\fontsize{3mm}{3mm}\selectfont{
\begin{table*}
\begin{tabular}{|c|ccc|cc|}
\hline
  &         &       & & &\\[-1.5ex]
&   \multicolumn{3}{c|}{Goodness-of-fit test p-values} & \multicolumn{2}{c|}{Two-samples test p-values }   \\[-1.5ex]
  &         &       & & &\\[-1.5ex]

  &         &       & & &\\[-1.5ex]
         \multirow{1}{*}{\bf Sample}&  \textbf{Anderson-Darling}& \textbf{Cramer-von Mises} & \textbf{Deviance (adjusted)} & \textbf{Kolmogorov-Smirnov} & \textbf{Wilcoxon Rank Sum }\\[-1ex]
  &         &       & & &\\[-1.5ex]
                                                  \hline
  &         &       & & &\\[-1.5ex]
  \multirow{1}{*}{Calibration} &  $1.2\cdot 10^{-7}$    &   $4.2\cdot10^{-7}$      &     $3.2\cdot 10^{-12}$ ($6.4\cdot 10^{-11}$) &- & -  \\[-1ex]
  &         &       & & &\\[-1.5ex]
                                                  \hline 
  &         &       & & &\\[-1.5ex]
         \multirow{1}{*}{Case I}      &   $0.7776$    &   $0.7711$      &     $0.2657$ ($>1$) &0.9248 &0.5487  \\[-1ex]
  &         &       & & &\\[-1.5ex]
                                                  \hline 
  &         &       & & &\\[-1.5ex]
Case II&    $4.6\cdot10^{-7}$       &    $8.2\cdot10^{-11}$     &     $9.0\cdot10^{-33}$ ($1.8\cdot10^{-31}$) &   $1.9\cdot10^{-13}$ & $4.5\cdot10^{-12}$\\[-1ex]
  &         &       & & &\\[-1.5ex]
                                                  \hline 
  &         &       & & &\\[-1.5ex]
Case III& $4.6\cdot10^{-7}$ & $2.6\cdot10^{-10}$ &        $2.6\cdot10^{-28}$ ($5.2\cdot10^{-27}$) &   $ 2.1\cdot10^{-15}$ & $2.2\cdot10^{-16}$\\[-2ex]
  &         &       & & &\\
\hline 
 \end{tabular}
\caption[Table 1]{\black{Comparison of deviance test and classical inferential tools. The first two columns report the p-values of Anderson-Darling \cite{anderson} and Cramer-von Mises \cite{darling} goodness-of-fit tests obtained assuming as theoretical distribution the same $G$ indicated in Sections \ref{bkgcali} and \ref{signalsearch} for the the calibration phase, case I, II and III, respectively. The raw deviance p-values
and their post-selection adjusted counterparts are reported in the third columns. Finally, the fouth and fifth column report, respectively, the Kolmogorov-Smirnov \cite{darling} and Wilcoxon rank sum \cite{wilcoxon} tests used to compare directly the physics samples in Case I, II and III with the source-free sample used in Section \ref{bkgcali}.}   }
\label{GOF}
\end{table*} 
}}
\begin{figure*}[htb]
\begin{tabular*}{\textwidth}{@{\extracolsep{\fill}}@{}c@{}c@{}}
      \includegraphics[width=90mm]{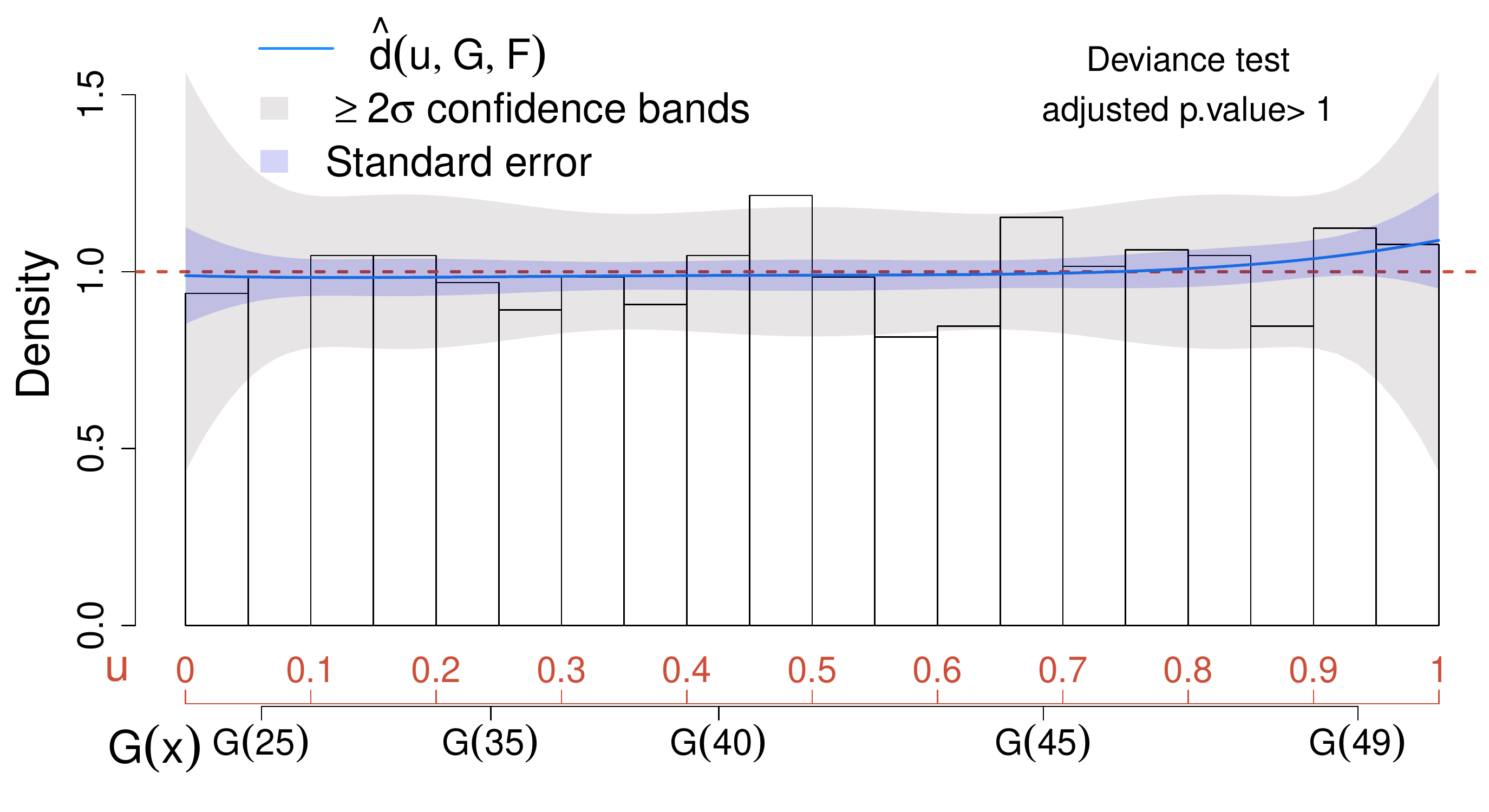} & \includegraphics[width=90mm]{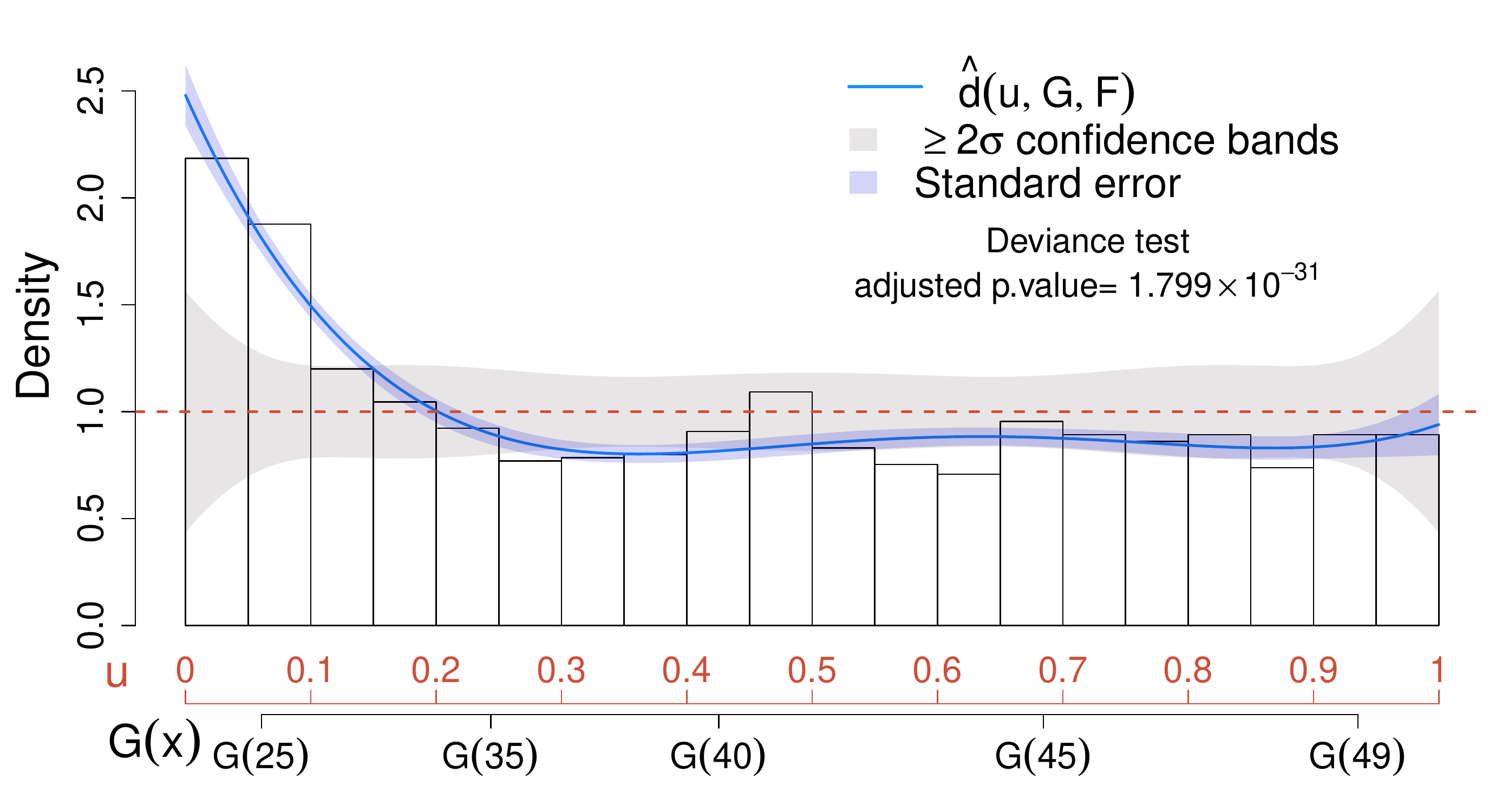}\\
\end{tabular*}
\caption[Figure 3]{Deviance test and CD plots for Case~I where no signal is present (left panel) and  Case~II where the signal is present (right panel).  In both cases, the postulated distribution $G$ corresponds to the cdf of the calibrated background model in \eqref{gest}. For the sake of comparison, $d(u;G,F)$ has been estimated via \eqref{dhat} with $M=4$  for both samples. }
\label{Fig3}
\end{figure*}
\begin{figure}[htb]
\centering
 \begin{adjustbox}{center}
\includegraphics[width=0.9\columnwidth]{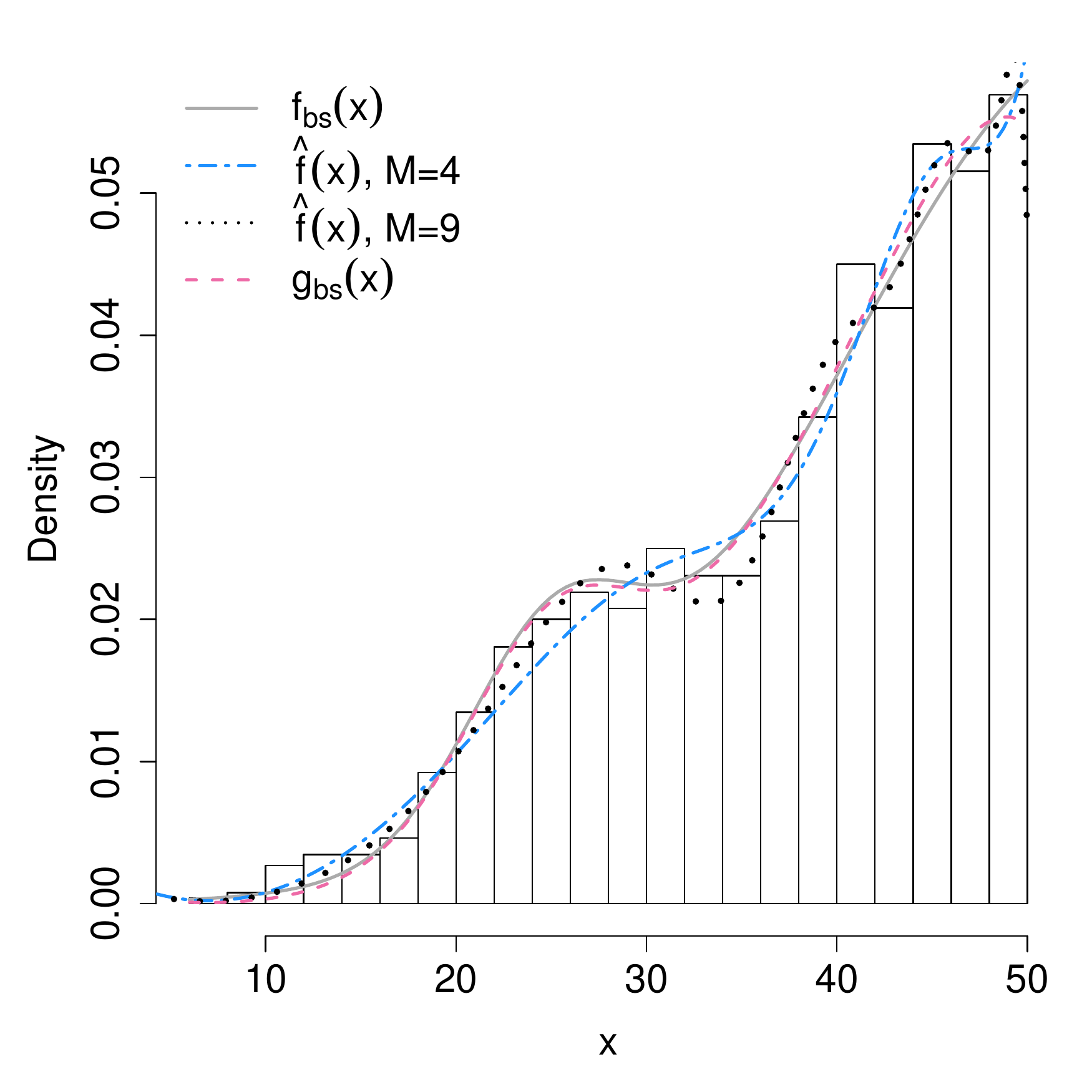} 
 \end{adjustbox}
\caption[Figure 4]{Histogram of a physics sample of $n=1300$ observations from both background and signal and with pdf as  in \eqref{fbs} (grey solid line). The true density has been estimated   semiparametrically, as in \eqref{gbs} (pink dashed line), whereas the  nonparametric estimates of $f(x)$ have been computed as in \eqref{fhat}, by plugging-in the $\widehat{d}(G(x);G,F)$ estimates obtained with  $M=4$ (blue dot-dashed line) and $M=9$ (black dotted line). }
\label{Fig4}
\end{figure}
\subsection{Signal search}
\label{signalsearch}
\subsubsection{Nonparametric signal detection}
\label{nonpar}
The background calibration phase allows  the specification of a well tailored model for the background, namely $\widehat{f}_b(x)$,  which simultaneously integrates the initial guess, $g_b$, and the information carried by the source-free data sample. Hereafter, we disregard the source-free data sample and    focus on analyzing the physics sample.

Under the assumption that the source-free sample  has no significant source contamination,  we expect that, if the signal is absent, both the source-free and the physics sample follow the same distribution. Therefore,   the calibrated  background model, $\widehat{f}_b(x)$, plays  the role of the postulated distribution for the physics sample, i.e.,  the model that we expect the data to follow when no signal is present; hence, we set $g(x)=\widehat{f}_b(x)$.

Let  $f(x)$ be the (unknown) true pdf of the physics sample which may or may not carry evidence in favor of the source of interest.
When no model for the signal is specified, it is reasonable to consider any significant deviation of $f$ from $g$ as an indication that a signal of unknown nature may be present. 
In this setting, similarly to the background calibration phase, we can construct deviance tests and CD plots to assess  if and where significant departures of $f$ from $g$ occur.  Two possible scenarios are considered -- a physics sample which collects only background data (Case I) and a physics sample of  observations from both background and signal (Case II). 

\emph{\textbf{Case I: background-only.}} Let $\bm{x}$ be a physics sample of $n=1300$ observations whose true (unknown) pdf $f(x)$  is equivalent to $f_b(x)$ in \eqref{fb}. We set  
\begin{equation}
\label{gest}
g(x)=\widehat{f}_b(x)=g_b(x)\widehat{d}(G_b(x);G_b,F_b)
\end{equation}
 where $g_b(x)$ and $\widehat{d}(G_b(x);G_b,F_b)$ are defined as in \eqref{gb} and \eqref{dbhat}, respectively.
The resulting  CD plot and deviance test are reported in the  left panel of  Fig. \ref{Fig3}.

When applying the scheme in Section \ref{chooseMsec} with $M_{\max}=20$, none of the values of $M$ considered leads to significant results; therefore, for the sake of comparison with Case II below, we choose $M=4$.
Not surprisingly, the estimated comparison density approaches one over the entire range and lies entirely within the  confidence bands. This suggests that the true distribution of the data does not differ significantly from the model which accounts only for the background.  Similarly, the deviance test leads to very low significance (adjusted p-value $>1$);
 hence, we conclude that our physics sample does not provide evidence in favor of the new source.

\emph{\textbf{Case II: background + signal.}} Let $\bm{x}$ be a physics sample of $n=1300$ observations whose true (unknown) pdf $f(x)$  is equal to $f_{bs}(x)$ in \eqref{fbs}
\begin{equation}
\label{fbs}
f_{bs}(x)=(1-\eta)f_b(x)+\eta f_s(x)
\end{equation}
with $f_b(x)$ and $f_s(x)$ defined as in \eqref{fb} and \eqref{fs} respectively, and  $\eta=0.15$. The histogram of the data and the graph of $f_{bs}(x)$ are plotted in Fig. \ref{Fig4}.
As in Case I, we set $g(x)$ as in \eqref{gest}.  

\begin{figure*}[htb]
\begin{tabular*}{\textwidth}{@{\extracolsep{\fill}}@{}c@{}c@{}}
      \includegraphics[width=90mm]{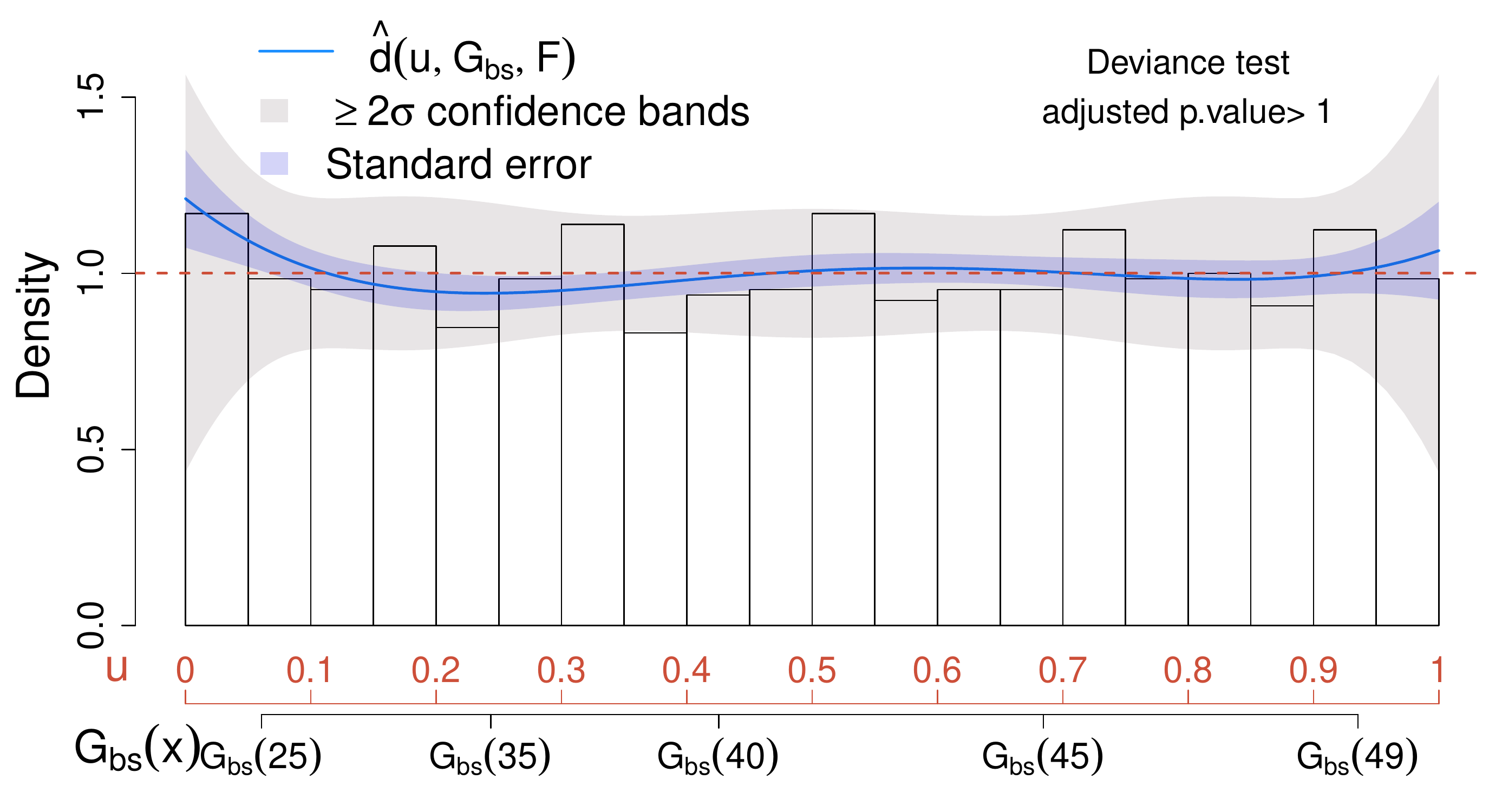} & \includegraphics[width=90mm]{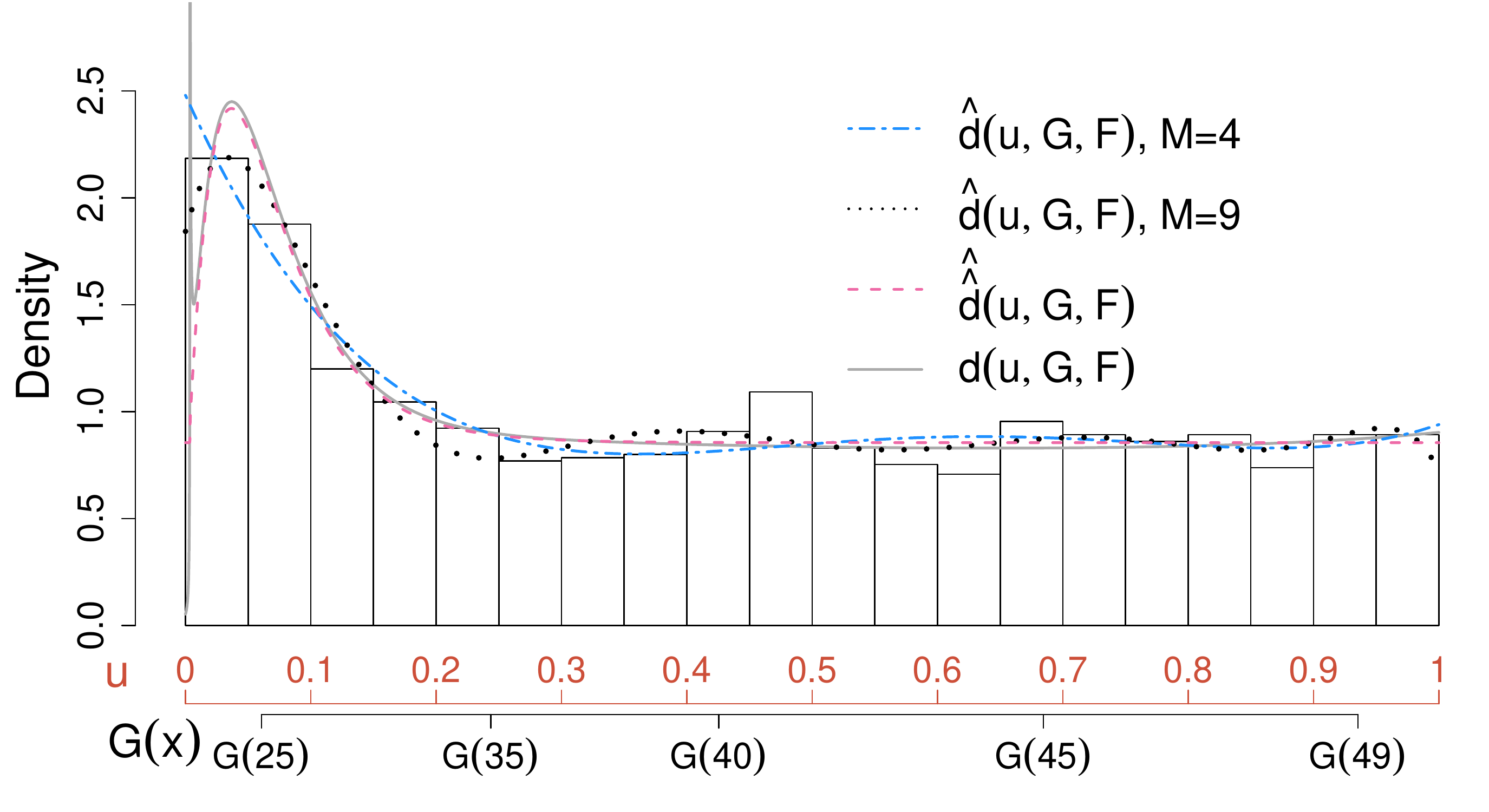}\\
     \includegraphics[width=90mm]{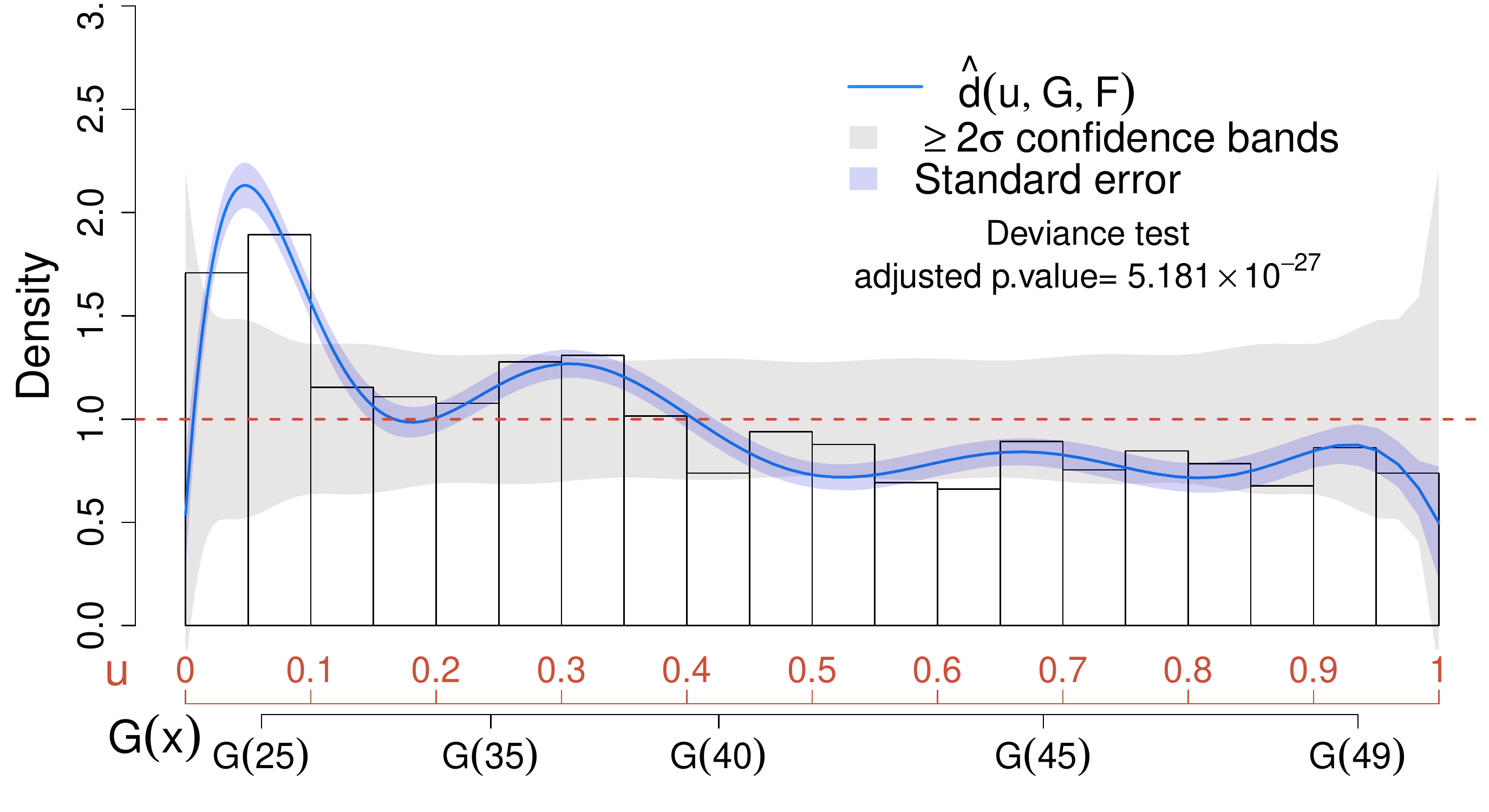} & \includegraphics[width=90mm]{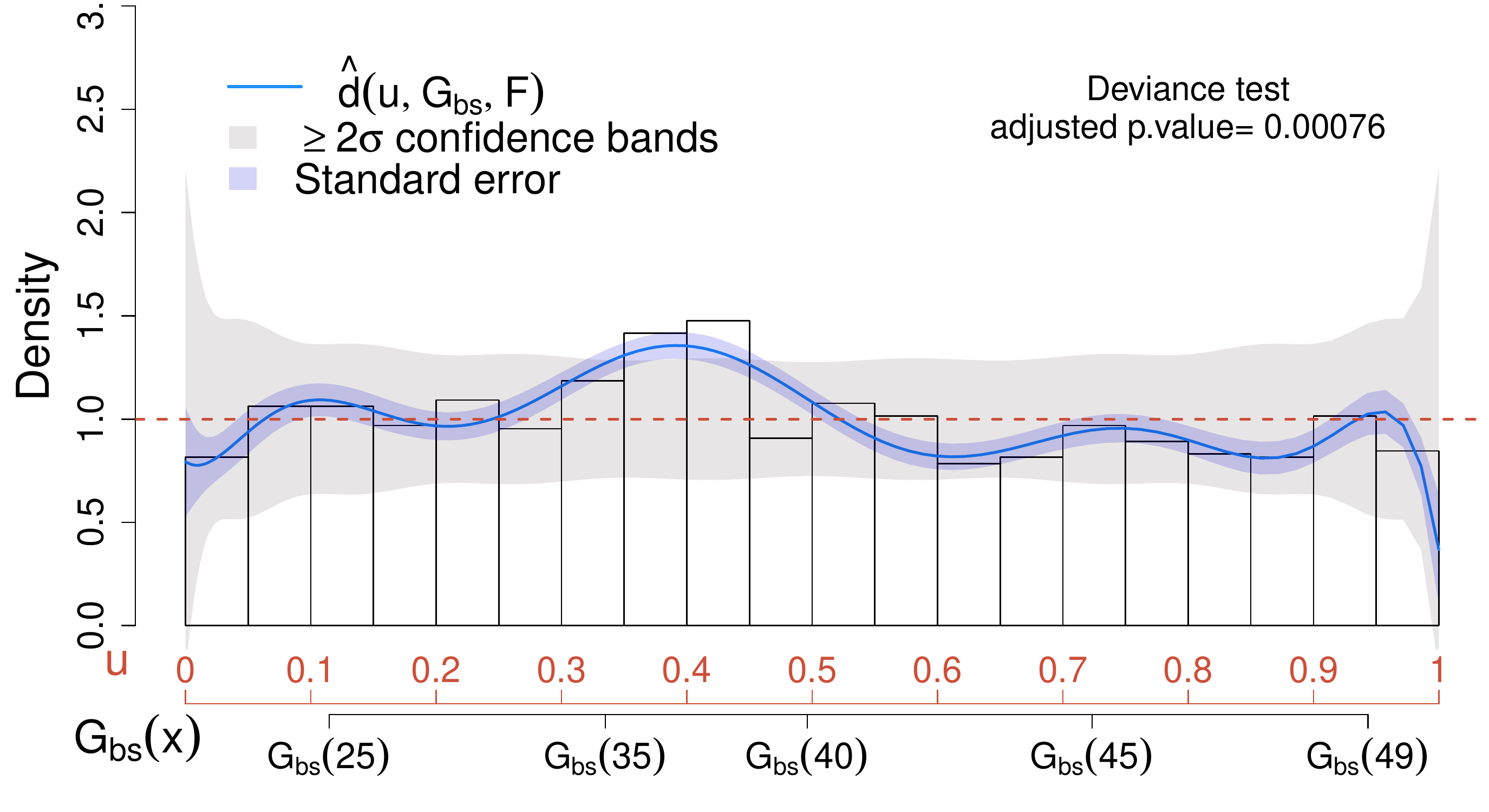}\\
\end{tabular*}
\caption[Figure 5]{Upper panels: deviance test and CD plots for Case~IIa where the signal is present (right panel), and  the postulated distribution$G_{bs}$ corresponds to the cdf of the estimated background+signal model in \eqref{gbs} with $\widehat{\eta}=0.146$. The comparison density estimate has been obtained considering $M=3$.
Bottom panels: Deviance test and CD plots for Case~III  where, in addition to the signal of interest, an additional resonance is present. The data are first analyzed considering the background-only pdf in \eqref{gb} as the postulated model (left panel). The analysis is then repeated by assuming the fitted background + signal model in \eqref{gbs} as the postulated distribution (right panel). Both estimates of the comparison density  in the left and right panels have been computed as in \eqref{dhat} with $M=9$. }
\label{Fig5}
\end{figure*} 
The CD plot and deviance test in the  right panel of  Fig. \ref{Fig3} show a significant departure of the data distribution  from the background-only model in \eqref{gest}. The maximum significance of the deviance is achieved at $M=4$, leading to a rejection of the null hypothesis at a $11.611\sigma$ significance level (adjusted p-value$=1.799\cdot10^{-31}$). The CD plot shows  a prominent  peak at the lower end of the spectrum; hence, we conclude that there is evidence in favor of the signal, and we proceed to characterize its distribution as described in Section \ref{signalcar}.

\subsubsection{Semiparametric signal characterization}
\label{signalcar}

The signal detection strategy proposed in Section \ref{nonpar} does not require the specification of a distribution for the signal. However, if a model for the signal is known (up to some free parameters), the analysis can be further refined by providing a parametric estimate of the comparison density and  assessing if additional signals from new unexpected sources are present.

\textbf{\emph{Case IIa: background + (known) signal.} } Assume that a model for the signal, $f_s(x,\bm{\theta}_s)$, is given, with $\bm{\theta}_s$ being a vector of unknown parameters. Since the CD plot in the right panel of Fig. \ref{Fig3} provides evidence in favor of the signal, we expect the data to be distributed according to the pdf 
\begin{equation}
\label{fhatbs}
\widehat{f}_{bs}(x)=(1-\eta)\widehat{f}_b(x)+\eta f_s(x,\bm{\theta}_s), \qquad 0\leq\eta\leq 1,
\end{equation}
where $\widehat{f}_b(x)$ is the calibrated background distribution in \eqref{gest}  and  $\eta$ and $\bm{\theta}_s$ can be estimated via Maximum Likelihood (ML).
Letting $\widehat{\eta}$ and $\widehat{\bm{\theta}}_s$ be the ML estimates of $\eta$ and $\bm{\theta}_s$ respectively, we specify 
\begin{equation}
\label{gbs}
g_{bs}(x)=(1-\widehat{\eta})\widehat{f}_b(x)+\widehat{\eta} f_s(x,\widehat{\bm{\theta}}_s)
\end{equation}
as postulated model.
For simplicity,  let $f_s$ to be fully specified as in \eqref{fs};  we construct the deviance test and the CD plot  to assess if \eqref{gbs} deviates significantly from the true distribution of the data. The scheme in Section \ref{chooseMsec} has been implemented with $M_{\max}=20$, and none of the values of $M$ considered led to significant results. 
The CD plot and deviance test for $M=4$ are reported in the upper left panel of Fig. \ref{Fig5}. Both the large p-value of the deviance test (adjusted p-value$> 1$) and the CD plot  suggest that no significant deviations occur; thus, \eqref{gbs} is a reliable model for the physics sample. 

Moreover, we can use \eqref{gbs} to further refine our $\widehat{f}_b(x)$ or $f_s(x,\widehat{\bm{\theta}}_s)$ distributions. Specifically, we first construct  a  semiparametric estimate of $d(G(x);G,F)$, i.e., 
\begin{equation}
\label{semipard}
\widehat{\widehat{d}(}G(x);G,F)=(1-\widehat{\eta})\frac{\widehat{f}_b(x)}{f_s(x,\widehat{\bm{\theta}}_s)}+\widehat{\eta},
\end{equation}
and   rewrite
\begin{equation}
\label{refine}
\begin{split}
\widehat{\widehat{f}_b}(x)&=\frac{\widehat{f}_b(x)\widehat{\widehat{d}(}G(x);G,F)-\widehat{\eta}f_s(x,\widehat{\bm{\theta}}_s)}{(1-\widehat{\eta})}\\
\widehat{\widehat{f}_s}(x)&=\frac{\widehat{f}_b(x)\widehat{\widehat{d}(}G(x);G,F)-(1-\widehat{\eta})\widehat{f}_b(x)}{\widehat{\eta}}.\\
\end{split}
\end{equation}

In the upper right panel of Fig. \ref{Fig5},  the true comparison density (grey dashed line) of our physics sample is compared with its semiparametric estimate computed as  in \eqref{semipard} (pink dashed line) with $f_s(x,\widehat{\bm{\theta}}_s)=f_s(x)$ in \eqref{fs}. The graphs of two nonparametric estimates of $d(u;G,F)$ computed via \eqref{dhat}  with $M=4$ and $M=9$ (blue dot-dashed line and black dotted line), respectively, are added to the same plot. Not surprisingly, incorporating the information available on the signal distribution drastically improves the accuracy of the analysis. The semiparametric estimate matches $d(u;G,F)$ almost exactly, whereas both nonparametric estimates show some discrepancies from the true comparison density. All the estimates suggest that there is only one prominent peak in correspondence of the signal region.

When moving from the comparison density domain to the 
density domain in Fig. \ref{Fig4}, the discrepancies between the  nonparametric estimates and the true density $f(x)$ are substantially magnified. 
Specifically, when computing \eqref{dhat} and \eqref{fhat} with $M=4$  (blue dot-dashed line),  the height signal peak is underestimated whereas, when choosing $M=9$, the $\widehat{f}(x)$ exhibits high bias at the boundaries\footnote{Boundary bias is a common problem among nonparametric density estimation procedures \cite[e.g.,][Ch.5, Ch.8]{larry}. When aiming for a non-parametric estimate of the data density $f(x)$, solutions exists to mitigate this problem \cite[e.g.,][]{efromovich}. } (dotted black line).

\textbf{\emph{Case IIb: background + (unknown) signal.} } When the signal distribution is unknown,  the CD plot of $\widehat{d}(u;G,F)$ can be used to guide the scientist in navigating across the  different theories  on the astrophysical phenomenon under study and specify a suitable model for the signal, i.e., $f_{s}$.
 The model proposed can then be validated, as in Case IIa, by fitting \eqref{gbs}  and constructing deviance tests and CD plots. 

At this stage, the scientist has the possibility to  iteratively query the data  and explore the distribution of the signal 
 by assuming different  models. A viable signal characterization is achieved when no significant  deviations of $\widehat{d}(u,G_{bs},F)$ from one are observed (e.g., see upper left panel of Fig. \ref{Fig5}). Notice that a similar approach can be followed also in the background calibration stage (Section \ref{bkgcali}) to provide a parametric characterization of the background distribution. 

\textbf{\emph{Case III: background + (known) signal +  unexpected source.} } The tools proposed so far can also be used to detect   signals 
from unexpected sources whose pdfs are, by design, unknown. 

\begin{figure*}
\begin{tabular*}{\textwidth}{@{\extracolsep{\fill}}@{}c@{}c@{}}
      \includegraphics[width=90mm]{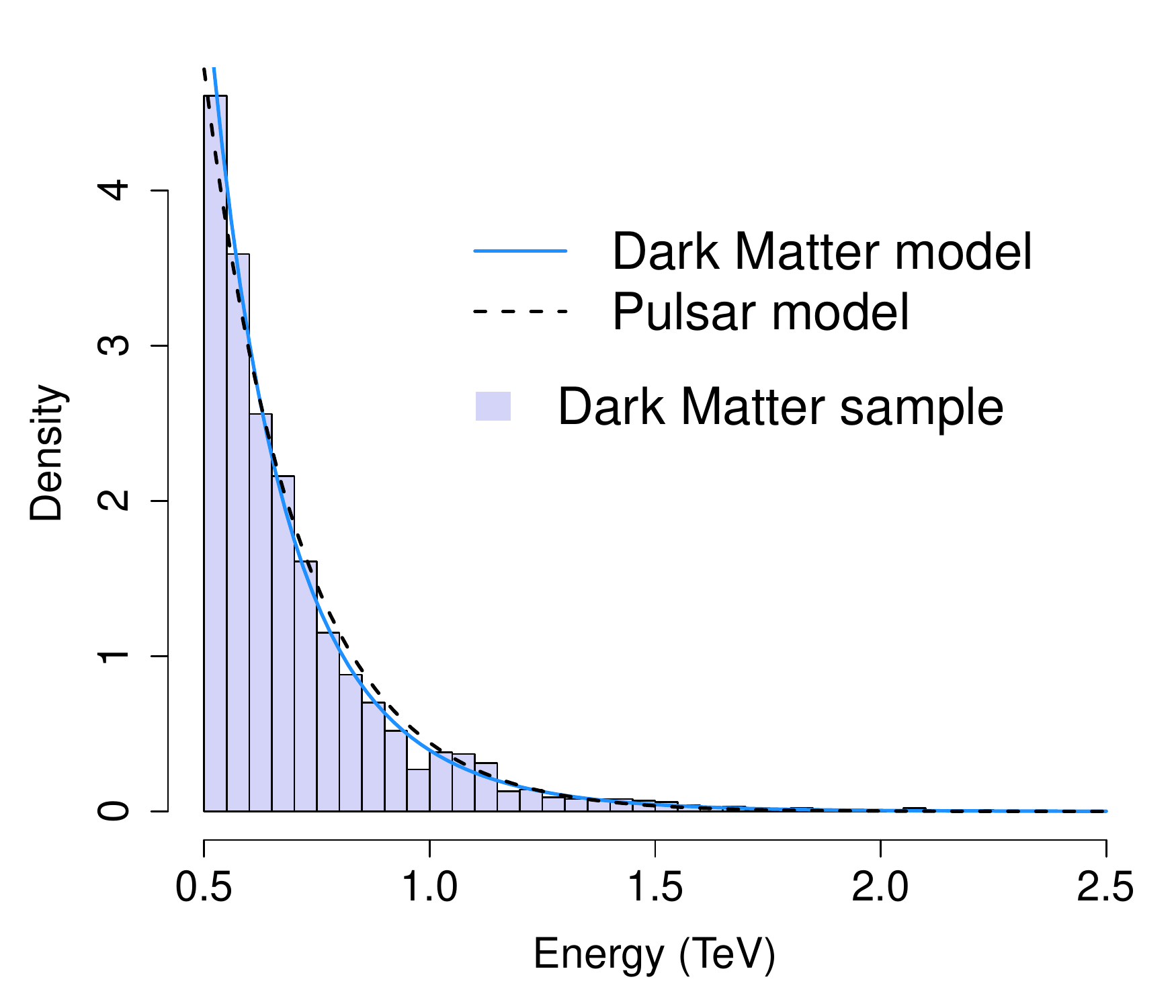} & \includegraphics[width=90mm]{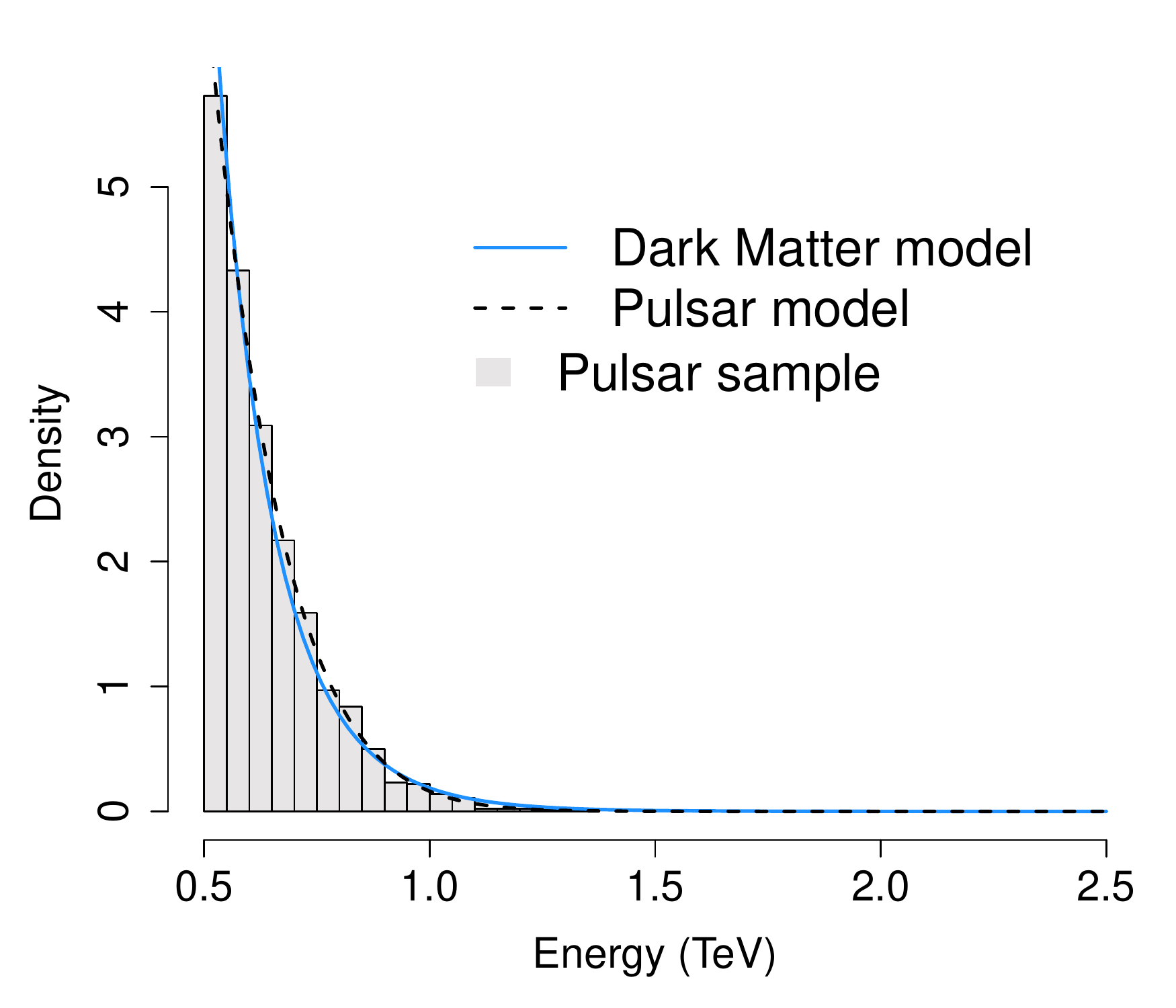}\\
\end{tabular*}
\caption[Figure 6]{\black{Dark matter and pulsar samples. The left panel corresponds to the histogram of a sample of 2000 observations simulated from the model in \eqref{DM} with $M_\chi=2.5$.
The right panel corresponds to the histogram of a sample of 2000 observations simulated from the model in \eqref{PS} with $\tau=2$. The best fit of \eqref{DM} and \eqref{PS} are also reported as blue solid line and black dashed lines, respectively, on top of each histogram.}}
\label{hists}
\end{figure*}

Suppose that the physics sample  $\bm{x}$ contains  $n=1300$ observations whose true (unknown) pdf $f(x)$  is equal to $f_{bsh}(x)$ 
\begin{equation}
\label{fbsh}
f_{bsh}(x)=(1-\eta_1-\eta_2)f_b(x)+\eta_1 f_s(x)+\eta_2  f_h(x)
\end{equation}
where $f_h(x)$ is the pdf of  the unexpected signal and assume its distribution to be normal with center at 37 and width 1.8. Let $f_b(x)$ and $f_s(x)$ be defined as in \eqref{fb} and \eqref{fs}, respectively, and let  $\eta_1=0.15$ and $\eta_2=0.1$. 

We can start with a nonparametric signal detection stage by setting $g(x)=\widehat{g}_{bs}(x)$   in \eqref{gest}, with $f_s$ defined as in \eqref{fs} and $\widehat{\eta}$ estimated via MLE. The respective CD plot and deviance tests are reported in the bottom left panel of  Fig. \ref{Fig5}.

Choosing $M=9$, as in   \eqref{chooseM}, both the CD plot and deviance test indicate a significant departure from the expected background-only model and a prominent peak is observed in correspondence of the signal of interest  centered around 25. A second but weaker peak appears to be right on the edge of our confidence bands, suggesting the possibility of an additional source.
At this stage, if  $f_s$ was unknown, we could proceed with a semiparametric signal characterization as in Case IIb. Whereas  assuming that the  distribution of the signal of interest is known and given by   \eqref{fs}, we fit \eqref{gbs},  aiming to  capture a significant deviation in correspondence of the second bump. 
This is precisely what we observe in the bottom right panel of Fig. \ref{Fig5}. Here the estimated comparison density deviates  from \eqref{gest} around 35, providing evidence in favor of an additional signal in this region.  We can then proceed as in Case IIb by exploring the theories available and/or collecting more data to further investigate the nature and the cause of the unanticipated bump.

\black{
\section{Signal detection without calibration sample and model selection}
\label{PSDMsec}
There are situations where a source-free sample is simply not available and thus the calibration phase in Section \ref{bkgcali} cannot be implemented. The tools described in Sections \ref{LPmodelling}
and \ref{inference} can, however, still be applied in order to perform signal detection and goodness-of-fit when a model for the signal is known, up to some free parameters. In this framework, we expect the data to either come only from the signal (with at most some negligible background contamination) or only from the background.}

\black{
In order to illustrate how to proceed in this setting, we consider a dark matter search where the postulated model for dark matter $\gamma$-ray emissions is the one of \cite[][Eq. 29]{bergstrom}, i.e.,
\begin{equation}
\label{DM}
g_{DM}(y)=\frac{0.73 M_{\chi}^{1.5}}{yk_{M_{\chi}}}\exp\biggl\{-7.8\frac{y}{M_{\chi}}\biggl\}
\end{equation}
with $y\in [0.5, 5]$ Teraelectron Volt (TeV), $M_{\chi}\in [0.5, 5]$ TeV and $k_{M_{\chi}}$ is a normalizing constant. The goal is to show that, when considering a background-only sample, the method proposed correctly rejects  \eqref{DM} as suitable model for the data; whereas, when considering  a dark matter sample,   the dark matter model in \eqref{DM} is ``accepted''.}
 
 \black{
To further increase the complexity of the problem, we consider a situation where the background  sample corresponds to $\gamma$-ray emissions due to a pulsar, with distribution  
\begin{equation}
\label{PS}
g_{PS}(y)= \frac{1}{y k_\tau}\exp\biggl\{-\biggl(\frac{y}{y_0}\biggl)^{\tau}\biggl\},
\end{equation}
with $y_0=0.5$, $y\in [0.5, 5]$ TeV, $\tau>0$ and and $k_{\tau}$ is a normalizing constant. Notice that, as discussed in \cite{baltz}, distinguishing $\gamma$-ray emissions due to  pulsars from those due to dark matter is a particularly challenging task. The histograms of the two datasets considered are shown in Figure \ref{hists}; the overlapping curves correspond to the best fit of the models in \eqref{DM} and \eqref{PS} on each sample. Interestingly, for both samples,  \eqref{DM} and \eqref{PS} provide a very similar fit to the data; hence the importance of correctly selecting the most adequate model or, excluding the dark matter hypothesis when observing emissions due to pulsars.}

\begin{figure*}
\begin{tabular*}{\textwidth}{@{\extracolsep{\fill}}@{}c@{}c@{}}
      \includegraphics[width=90mm]{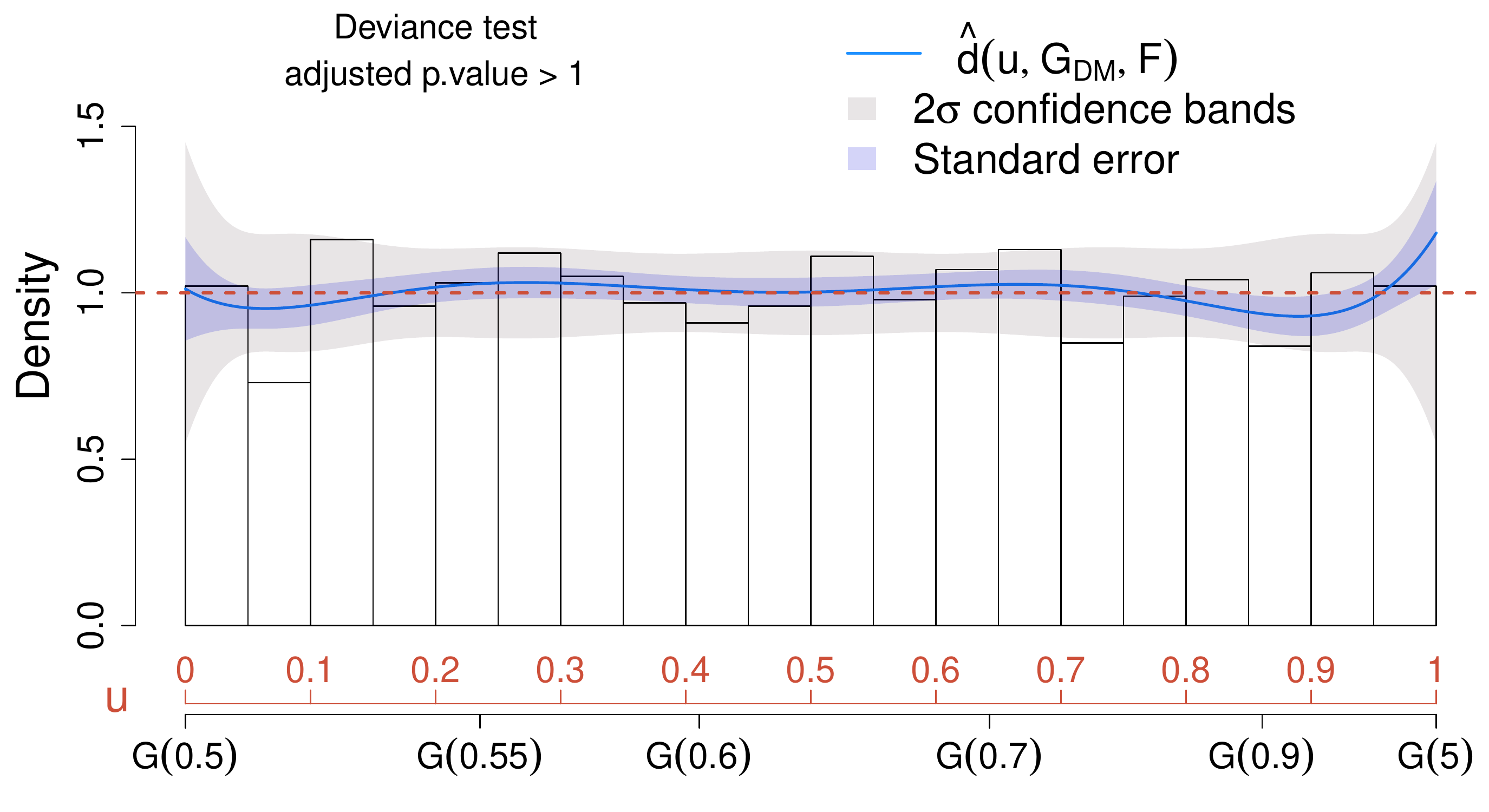} & \includegraphics[width=90mm]{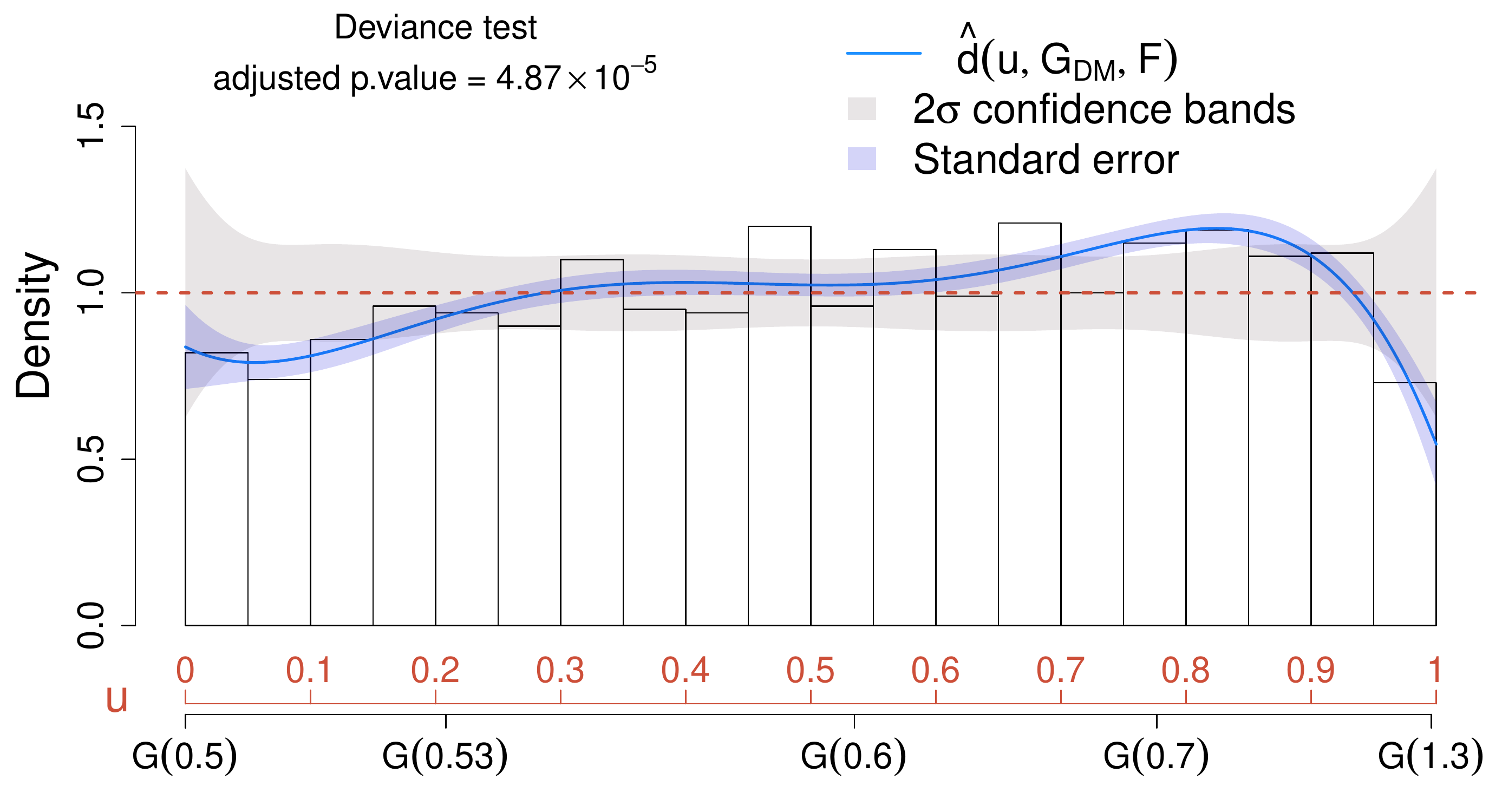}\\
            \includegraphics[width=90mm]{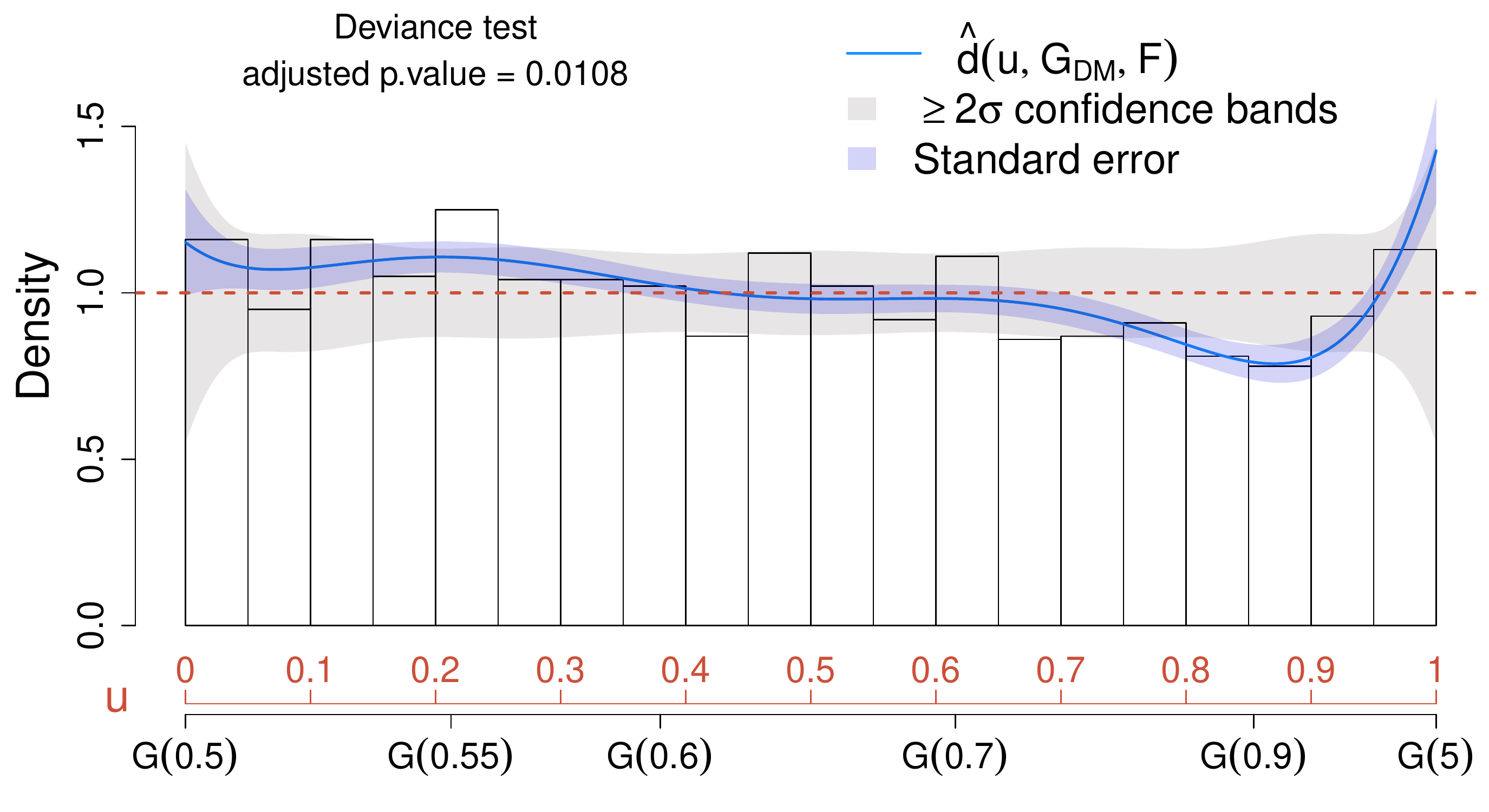} & \includegraphics[width=90mm]{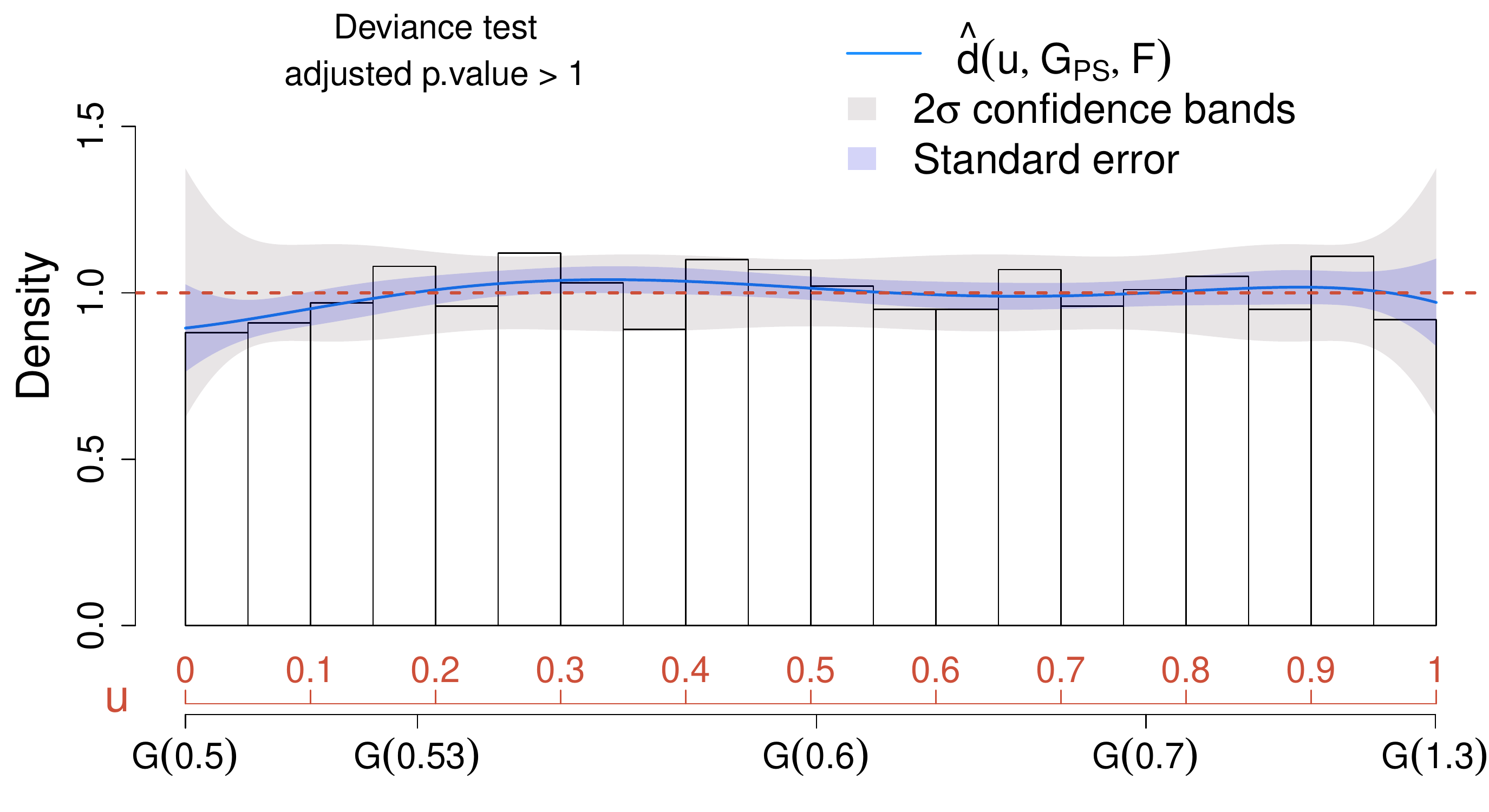}\\
\end{tabular*}
\caption[Figure 7]{
\black{
CD plots and deviance test for dark matter and pulsar samples. The upper left panel displays CD plot and deviance test for the dark matter sample with $g=g_{DM}$ in \eqref{DM}. The upper right panel compares the distribution of the pulsar sample with the model in  \eqref{DM}. 
 The lower left panel displays CD plot and deviance test for the dark matter sample with $g=g_{PS}$ in \eqref{PS}. The lower right panel displays CD plot and deviance test for the pulsar sample with $g=g_{PS}$ in \eqref{PS}. For the plots on the left,
 the size of the basis selected is $M=6$, whereas, for the plots on the right, $M=5$.}}
\label{cdPSDM}
\end{figure*}

\black{
The upper panels of Figure \ref{cdPSDM} display the CD plots obtained by setting $g=g_{DM}$ in \eqref{DM} as postulated model and comparing it with the distribution of the  dark matter sample (upper left panel) and of the background pulsar sample (upper right panel). Remarkably, the CD plots and the adjusted deviance tests correctly  lead to the conclusion that the distribution of the dark matter sample does not deviates significantly from \eqref{DM}, whereas the distribution of the pulsar sample does deviate substantially from \eqref{DM} and the deviance test (adequately adjusted for post-selection inference)  rejects the dark matter model with  $3.897\sigma$ significance (adjusted p-value of $4.870\cdot 10^{-5}$).
Notice that, in both cases, we are ignoring the information regarding the pulsar distribution and the only inputs considered are the data and the signal model in \eqref{DM}.}

\black{
Finally, when incorporating the knowledge of the  pulsar distribution  in \eqref{PS} into the analysis, one can select between the models in \eqref{DM} and \eqref{PS} by constructing additional  CD plots and deviance test for both samples and  setting $g=g_{PS}$ in \eqref{PS}. The results are shown in lower panels of  Figure \ref{cdPSDM}. As expected, the dark matter model is rejected (lower left panel)   with $2.297\sigma$ significance (adjusted p-value of $0.0108$) whereas the  pulsar model is ``accepted'' (lower right panel).}

\black{
\section{Background mismodelling due to instrumental noise and upper limits constructions}
\label{instrument}}
When  conducting real data analyses one has to take into account that the data generating process is affected by both statistical  and non-random uncertainty due to the instrumental noise. As a result, even when a model for the background is known, the data distribution may substantially deviate from it due to the smearing introduced by the detector \cite[e.g.,][]{lyonsPHY}. In order to account for the instrumental error affecting the data, it is common practice to consider folded distributions where the errors due to the detector are often modelled assuming a normal distribution or estimated via non-parametric methods \cite[e.g.,][]{PHY,PHY2}. \black{In Section \ref{modinstr}, it is shown how the same approach described in Sections \ref{bkgcali} and \ref{nonpar} can be  used to assess if the instrumental error is negligible and, when not, how to update the postulated background model in order to incorporate the instrumental noise. Section \ref{upperlimits} discusses upper limits constructions by means of comparison distributions}.

\black{
\subsection{Modelling the instrumental error}
\label{modinstr}}
The data  considered come from a simulated observation by the Fermi Large Area Telescope \cite{atwood} with realistic representations of the effects of the detector and present backgrounds  \cite{meJINST,meMNRAS}. The Fermi-LAT is a pair-conversion $\gamma$-ray telescope on board the earth-orbiting Fermi satellite. It measures energies and images $\gamma$-rays between about a 100 MeV and several TeV. The goal of the analysis is to assess if the data could result from the self-annihilation of a dark matter particle. 
\begin{figure*}
\begin{tabular*}{\textwidth}{@{\extracolsep{\fill}}@{}c@{}c@{}}
      \includegraphics[width=90mm]{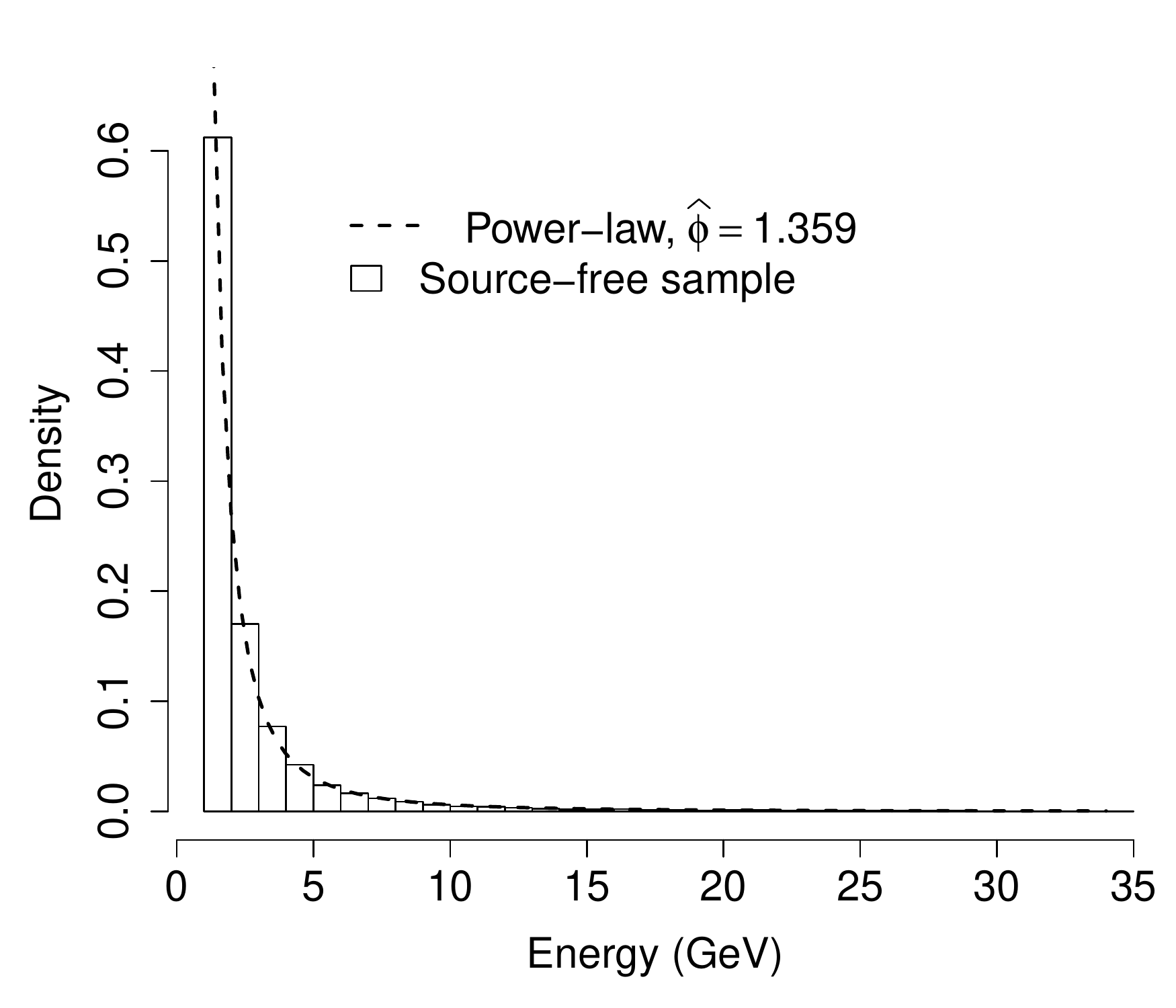} & \includegraphics[width=90mm]{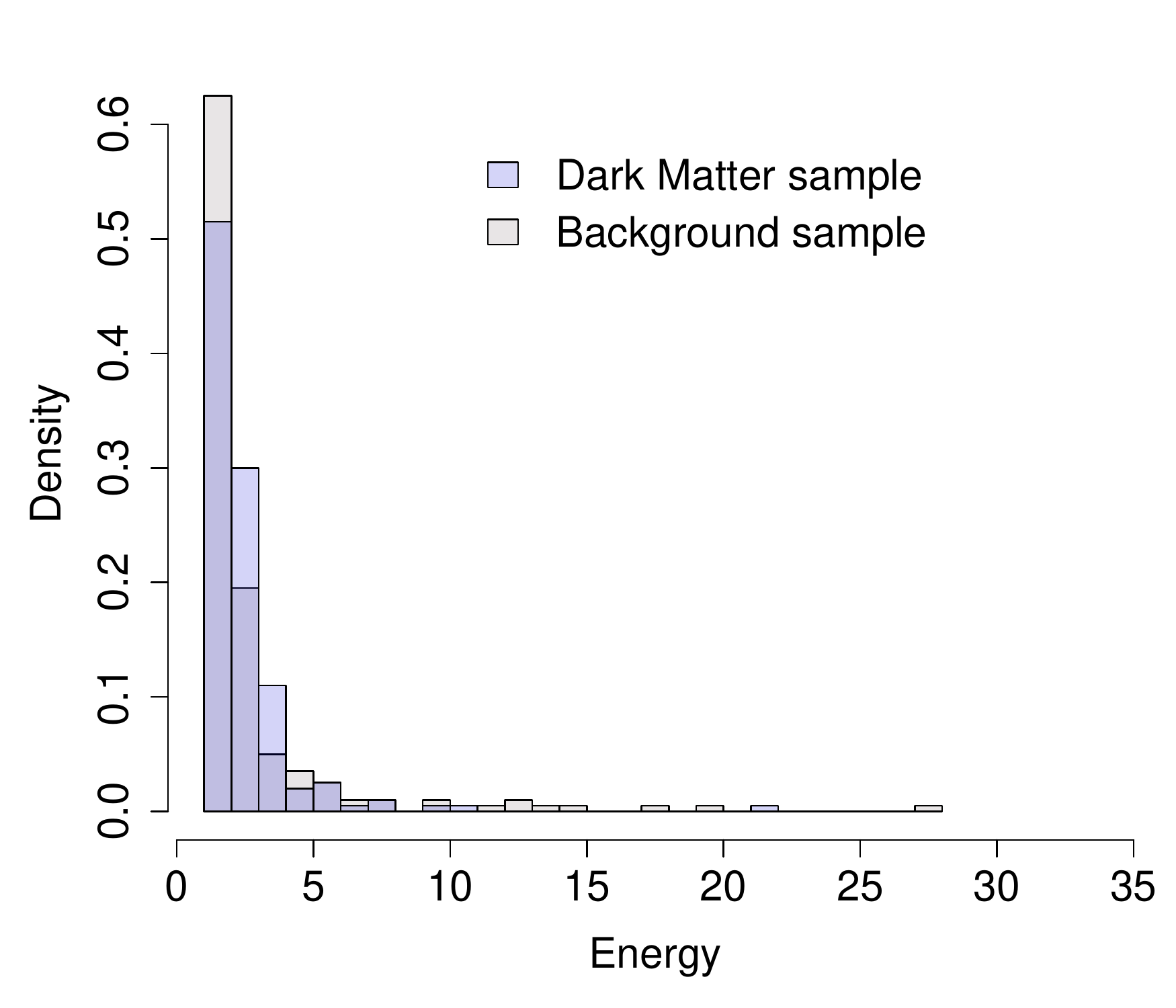}\\
\end{tabular*}
\caption[Figure 8]{Histograms of simulated Fermi-LAT  samples. \black{The left panel displays the histogram  of a source-free simulated Fermi-LAT   sample of $N=35,157$ observations, whereas the black dashed line  corresponds to the best fit of the power-law model in \eqref{gb2}. The right panel shows} the histogram of two simulated Fermi-LAT physics  samples of $n=200$ observations. The  grey histogram  corresponds to the background-only sample, whereas the blue histogram corresponds to the dark matter signal sample.}
\label{Fig8}
\end{figure*}
\begin{figure*}
\begin{tabular*}{\textwidth}{@{\extracolsep{\fill}}@{}c@{}c@{}}
\includegraphics[width=90mm]{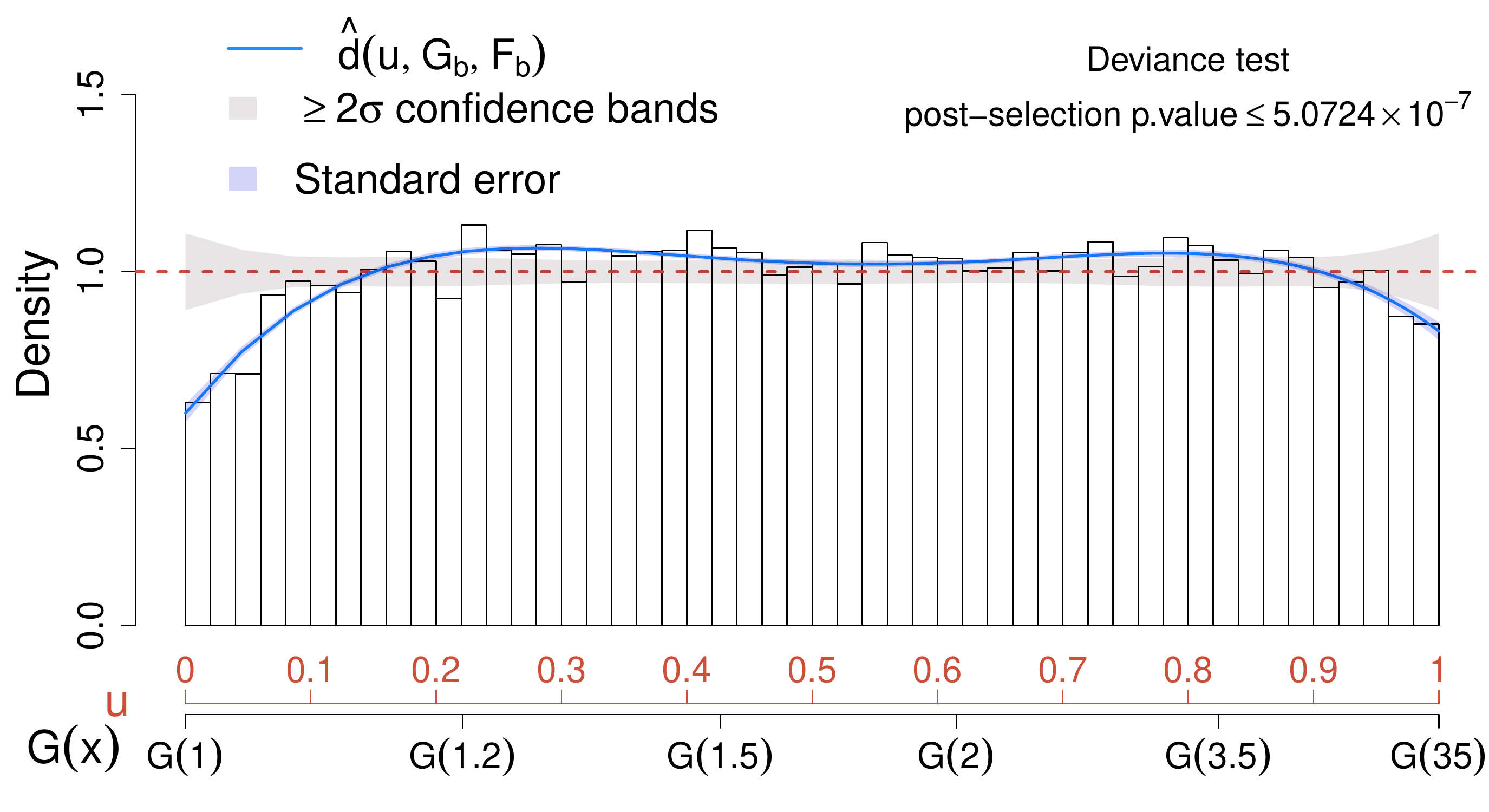} & \includegraphics[width=90mm]{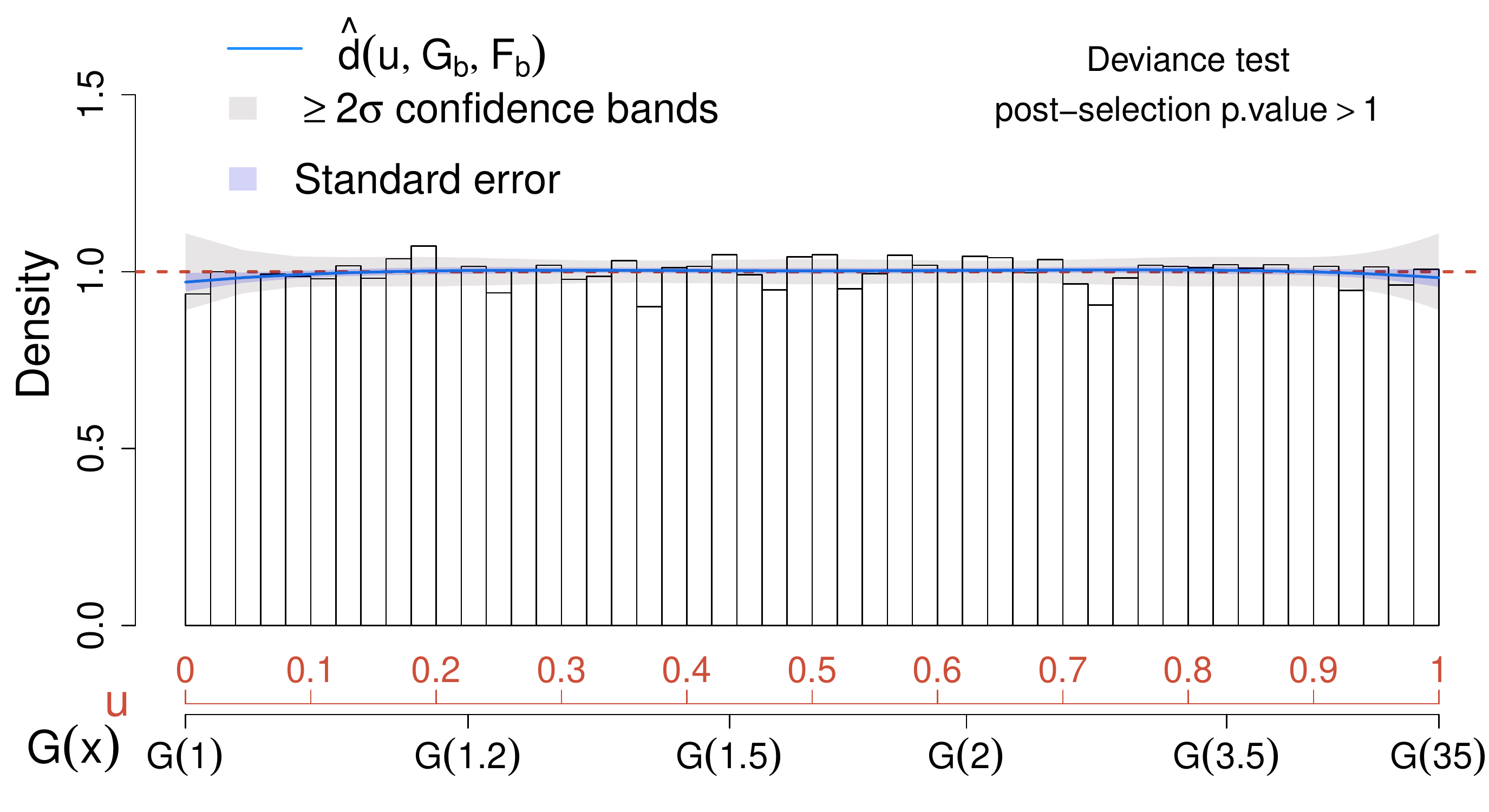}\\
\end{tabular*}
\caption[Figure 6]{\black{Deviance test and CD plot for the source-free simulated Fermi-LAT sample including the instrumental error (left panel) and simulated sample with no instrumental error (right panel). In both cases $M=4$.}  }
\label{Fig6}
\end{figure*}
Let the distribution of the astrophysical background  be \black{a power-law, i.e.,}
\begin{equation}
\label{gb2}
g_b(x)=\frac{1}{k_{\phi}x^{\phi+1}}
\end{equation} 
where $k_{\phi}$ is a normalizing constant and $x\in [1,35]$ Giga electron Volt (GeV). \black{Equation \eqref{gb2} corresponds to the distribution we would expect the background to follow if there was no smearing of the detector. The left panel of Figure \ref{Fig8} shows the histogram of a  source-free sample of 35,157 i.i.d. observations  from a power-law distributed background source with index 2.4 (i.e., $\phi=1.4$ in \eqref{gb2})  and contaminated by  instrumental  errors of unknown distribution.}

\black{In order to assess if \eqref{gb2} is a suitable distribution for these data, we  proceed by fitting  \eqref{gb2} via maximum likelihood and setting it as postulated background distribution.  The best fit of \eqref{gb2} is displayed on the left panel of Figure \ref{Fig8} as a black dashed line.}

\black{We proceed estimating $d(G_b(x);G_b,F_b)$ and $f_b$ as in \eqref{dhat} and  \eqref{fbhat} respectively, with $M=4$ (chosen as in Section \ref{chooseMsec}). The deviance test and CD plot are reported  in the left panel of Figure \ref{Fig6} and suggest that significant departures from the fitted power-law model occur. This implies that the instrumental error is not negligible and thus, in order to account for it, we consider \eqref{fbhat2} as ``calibrated'' background density the model }
\begin{equation}
\label{fbhat2}
\begin{split}
\widehat{f}_b(x)&=\frac{1}{k_{\hat{\phi}}x^{\hat{\phi}+1}}\Bigl(1+0.027Leg_1[G_b(x)]-0.067Leg_2[G_b(x)]\\
& + 0.026Leg_3[G_b(x)]-0.045Leg_4[G_b(x)]\Bigl),\\
\end{split}
\end{equation} 
where $G_b(x)$ is the cdf of \eqref{gb2} and $\hat{\phi}=1.359$ is the ML estimate of $\phi$ in \eqref{gb2}. 

\black{For the sake of comparison, the same analysis has been repeated considering $35,157$ i.i.d. observations  from a power-law  background source with index 2.4, without instrumental error. The respective CD plot and deviance test are shown on the right panel of  Figure \ref{Fig6} and indicate that the power-law model in \eqref{gb2}, with $\phi$ replace by its MLE (i.e., $\widehat{\phi}=1.391$), provides a good fit for the data, i.e., the instrumental error is, in this case, absent or negligible. }

\black{
\subsection{Signal detection and upper limit construction}
\label{upperlimits}}
Once obtained a calibrated background distribution, we proceed with the signal detection phase by setting $g(x)=\widehat{f}_b(x)$ in \eqref{fbhat2}. Similarly to Section \ref{nonpar},   two physics samples are given; one containing 200 observations from the background source distributed, as in \eqref{gb2},  and the other  containing 200 observations from a dark matter emission. The signal distribution from which the data have been simulated is the pdf of $\gamma$-ray dark matter energies in \cite[][Eq. 28]{bergstrom} \black{with $M_{\chi}=3.5$}. Both physics samples include the contamination due to the instrumental noise \black{with unknown distribution}. The respective histograms are shown in the right panel of  Fig. \ref{Fig8}. 

\begin{figure*}
\begin{tabular*}{\textwidth}{@{\extracolsep{\fill}}@{}c@{}c@{}}
      \includegraphics[width=90mm]{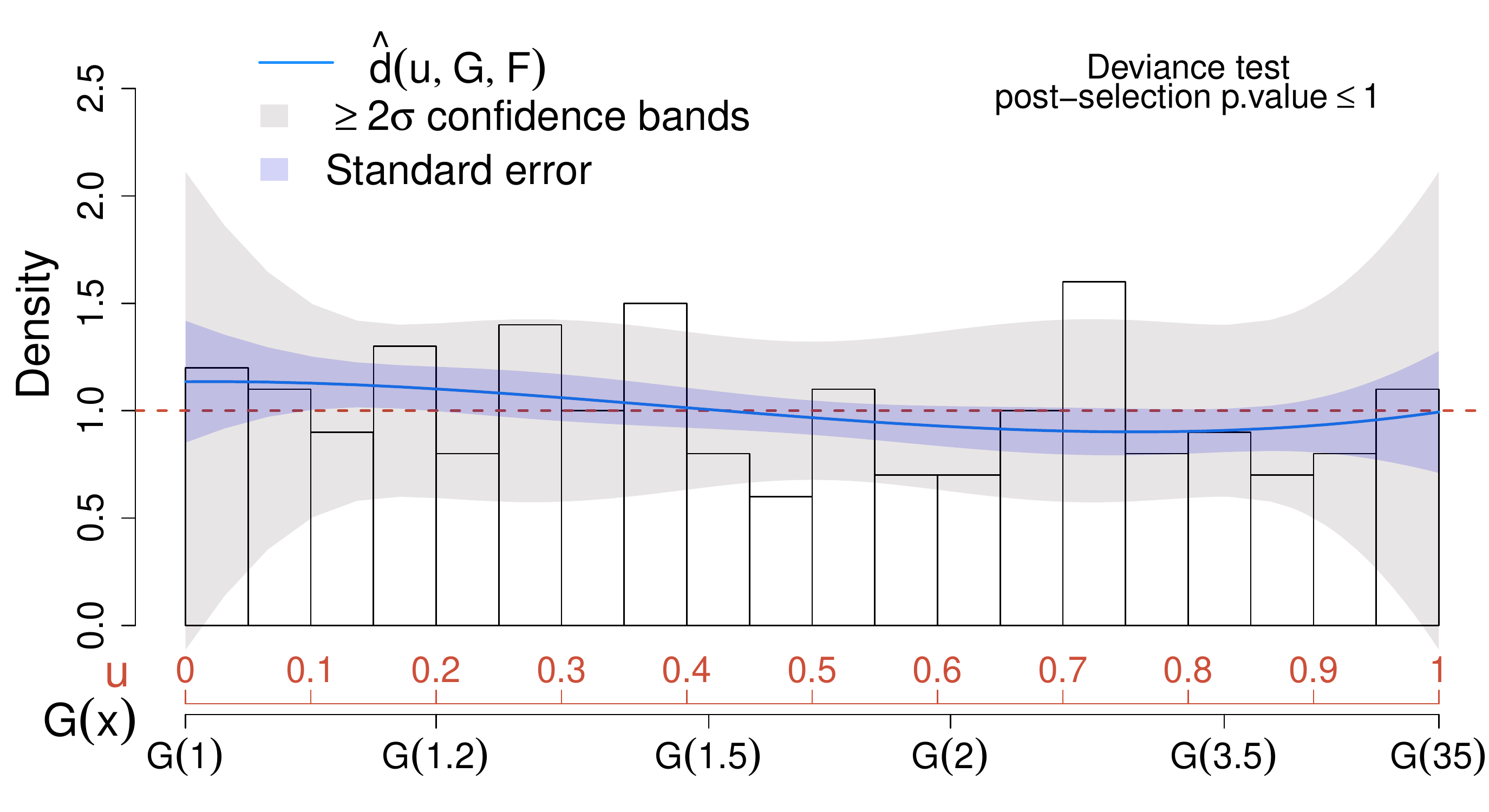} & \includegraphics[width=90mm]{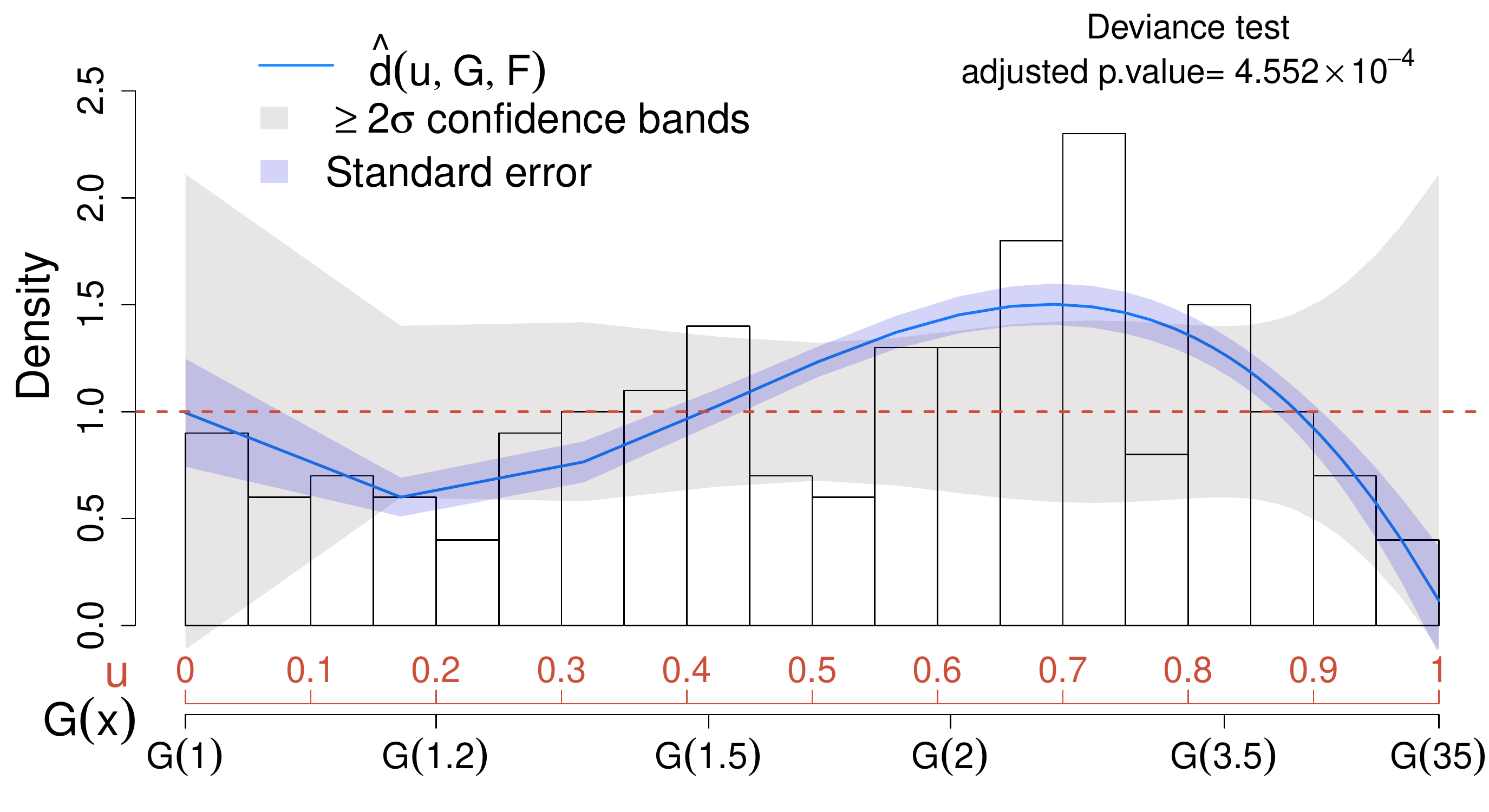}\\
      \includegraphics[width=90mm]{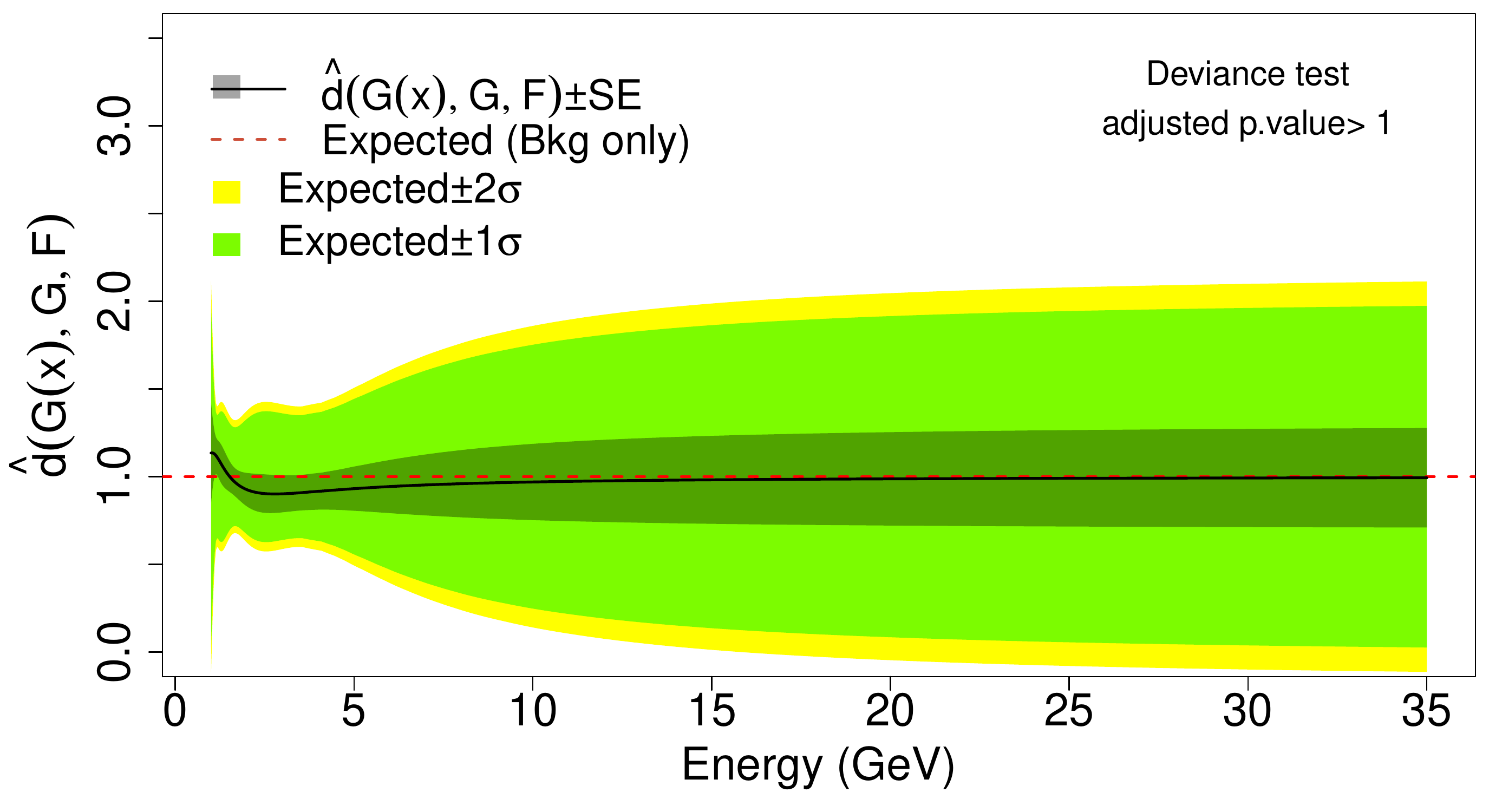} & \includegraphics[width=90mm]{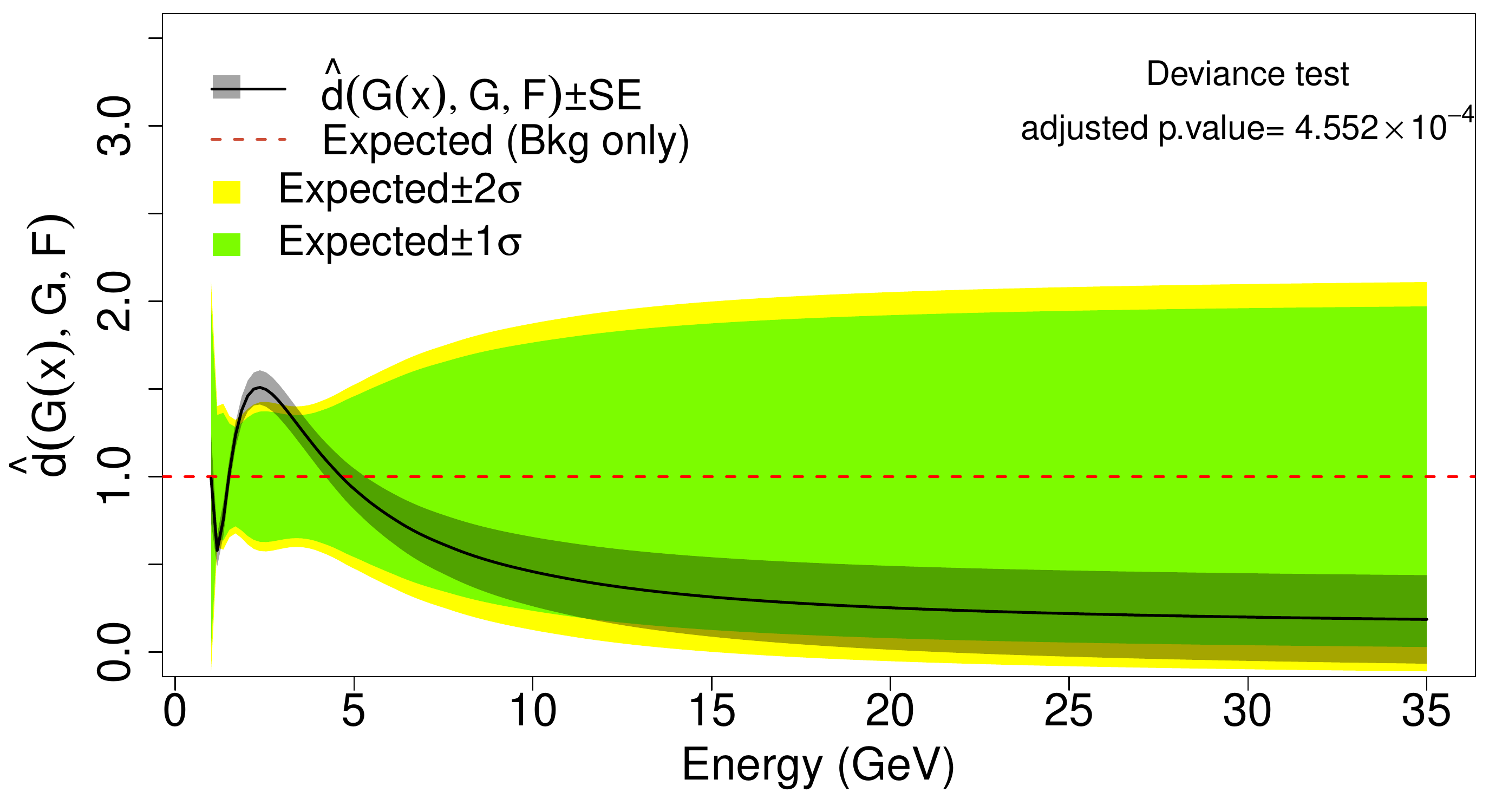}\\
\end{tabular*}
\caption[Figure 7]{\black{Deviance test, CD plots and Brazil plots for the simulated Fermi-LAT  background-only   sample of size 200 (left panels) and the  simulated Fermi-LAT   dark matter signal   sample of size 200 (right panels). In both cases, the postulated distribution $G$ corresponds to the cdf of the calibrated background model in \eqref{fbhat2}. For the sake of comparison, $d(u;G,F)$ has been estimated via \eqref{dhat}, with $M=3$  in both cases. }}
\label{Fig7}
\end{figure*}

The selection scheme in Section \ref{chooseMsec} suggests that  no significant departure from \eqref{fbhat2} occurs on the background-only physics sample, whereas, for the signal sample, the strongest significance is observed at $M=3$; therefore, for the sake of comparison, we choose $M=3$ in both cases. The respective deviance tests and CD plots  are reported  in   Fig. \ref{Fig7}. As expected, the upper left panel of Fig. \ref{Fig7} shows a flat estimate of the comparison density on the background-only sample. Conversely, the upper right panel of Fig. \ref{Fig7} suggests that an extra bump is present over the $[2,3.5]$  region with $3.318\sigma$ significance (adjusted p-value = $4.552\cdot 10^{-4}$). As in \eqref{semipard}, it is possible to proceed with the signal characterization stage (see Section \ref{signalcar}); however, in this setting, one has to account for the fact that also the signal distribution must include the smearing effect of the detector.

\black{
As an anonymous referee pointed out, it is important to discuss how upper  limits and Brazil plot can be constructed via LP modelling and how they relate to the constructs discussed so far in this manuscript. Indeed, the confidence bands reported in the CD plots are themselves upper limits. Specifically, in the signal detection framework of Section \ref{nonpar},  the confidence bands in \eqref{CIband} are constructed assuming that there is no signal in the data. Specifically, they correspond to the regions where the comparison density estimator is expected to lie, at $1-\alpha$ confidence level,  if the data includes background-only events. Conversely, any deviation from the confidence bands characterizes the quantiles of the distribution where the data distribution does not conform with the one postulated under the assumption that no signal is present. }

\black{When the interest is in identifying areas of the search region where deviations from the background model occur, one can exploit the fact that   $u=G(x)$, and thus upper limits and classical ``Brazil plots'' based on the comparison density can be obtained by plotting \eqref{dbhat} and the respective confidence bands in \eqref{CIband} as a function of $x$. This is shown, for our Fermi-LAT example in the bottom panels of Figure \ref{Fig7}. Indeed, the upper and bottom panels in Figure \ref{Fig7} carry essentially the same information in two different domains. Specifically, the CD plots display the departure of $f$ from $g$ in the quantile domain whereas the Brazil plots show the same differences in the frequency domain. For signal detection purpose, the bottom panels may be preferred to identify the location where substantial deviations among the background and signal model occur. Whereas, the CD plots are more suitable for goodness-of-fit purposes as they provide a simulataneous visualization of  the differences occurring at each quantile of the distribution.  }

\section{model-denoising}
\label{denoising}

\black{As discussed in Section \ref{biasvariance}, the choice of $M$ affects the resulting estimator of $d(u;G,F)$ in terms of both bias and variance. 
When dealing with complex background distributions,  a large value of $M$ may be necessary to reduce the bias of the estimated comparison density. At the same times, however, a large value of $M$ leads to an inflation of the variance. In other words, considering a basis of $M$ shifted Legendre polynomials may lead to overfitting.}

\black{Practically speaking, overfitting leads to wiggly (i.e., non-smooth) estimates and thus one may overcome this limitation by attempting to denoise the estimator in \eqref{dhat}}. Section \ref{denoiseAIC}  reviews the model-denoising approach proposed by \cite{LPapproach,LPmode}, whereas  Section \ref{AICtest}  briefly discusses inference and model selection in this setting. Finally, Section \ref{comparison} compares the results obtained with a full and a denoised solution on the examples of Section \ref{DS}.

\subsection{AIC denoising}
\label{denoiseAIC}
Let $\widehat{LP}_1,\dots, \widehat{LP}_M$ be the estimate of the first $M$  coefficients of the expansion  in \eqref{cd}.  The most ``significant'' ${LP}_j$ coefficients are selected by  sorting the respective $\widehat{LP}_j$ estimates  so that
\[\widehat{LP}^2_{(1)}\geq \widehat{LP}^2_{(2)}\geq \dots \geq \widehat{LP}^2_{(M)}\]
and choosing the value $k=1,\dots,M$ for which $AIC(k)$ in \eqref{AIC} is maximum
\begin{equation}
\label{AIC}
AIC(k)=\sum_{j=1}^{k} \widehat{LP}^2_{(j)}-\frac{2k}{n}.
\end{equation}
The AIC-denoised estimator  of $d(u;G,F)$ is given by
\begin{equation}
\label{dhat2}
\widehat{d}^*(u;G,F)=1+\sum_{j=1}^{k^*_M} \widehat{LP}_{(j)} Leg_{(j)}(u)
\end{equation}
where $\widehat{LP}_{(j)}$ is the estimate whose square is the $j^{\text{th}}$ largest among $\widehat{LP}^2_{1},\dots,\widehat{LP}^2_{M}$, $Leg_{(j)}(u)$ is the respective shifted Legendre polynomial and 
\begin{equation}
\label{mstar}
k^*_M=\underset{k}{\mathrm{argmax}} \{AIC(1),\dots, AIC(M)\}.
\end{equation}

\noindent\emph{\underline{Practical remarks.}} Recall that the first $M$ coefficients $LP_j$  can be expressed as a linear combination of the first $M$ moments of $U$. Thus, the 
AIC-denoising approach selects the  $LP_j$ coefficients which carry all the ``sufficient''  information on  the first $M$ moments of the distribution.

\subsection{Inference after denoising}
\label{AICtest}
The deviance   test can be used, as in Section \ref{chooseMsec}, to choose the size of the initial basis of $M$ polynomials among $M_{\max}$ possible models. 
Finally, the $k^\star_M$ largest coefficients are chosen  by maximizing \eqref{AIC}. This two-step procedure selects  $\widehat{d}^*(u;G,F)$ in \eqref{dhat2} from a pool of $M_{tot}=M_{\max}+\frac{M(M-1)}{2}$ possible estimators. Therefore, the Bonferroni-adjusted p-value of the deviance test is given by
\begin{equation}
\label{adjk}
M_{tot}\cdot P(\chi^2_{k_M^*}>d_{k_M^*})
\end{equation}
withe $d_{k_M^*}=\sum_{j=1}^{k_M^*} \widehat{LP}^2_{(j)}$. Similarly, confidence bands can be constructed as
\begin{equation}
\label{CIband4}
\Biggl[1-c_{\alpha,M_{tot}}\sqrt{\sum_{j=1}^{k_M^*}\frac{1}{n}Leg_{(j)}^2(u)},1+c_\alpha\sqrt{\sum_{j=1}^{k_M^*}\frac{1}{n}Leg_{(j)}^2(u)}\Biggl]
\end{equation}
where $c_{\alpha,M_{tot}}$ is the solutions of 
\begin{equation}
\label{CIband3b}
2(1-\Phi(c_{\alpha,M_{tot}}))+\frac{k_0}{\pi}e^{-0.5c^2_{\alpha,M_{tot}}}=\frac{\alpha}{M_{tot}}.
\end{equation}

\black{\emph{\underline{Practical remarks.}} Given the possibility of denoising our solution, one may legitimately wonder why not to consider a large value of 
$M_{\max}$, e.g., $M_{\max}=100$ and then select $k^\star_{M_{\max}}$ directly. In other words, why should we first implement the procedure in Section \ref{chooseMsec} 
and, only after, refine our estimator as in Section \ref{denoiseAIC} and not vice-versa?
There are two main reasons why such approach is discouraged. }

\black{
First of all, one has to take into account that ignoring the selection stage proposed in Section \ref{chooseMsec},
there is no guarantee that the resulting $k^\star_{M_{\max}}$ would include all the $\widehat{LP}_j$ terms that provide the strongest evidence in favor of $H_1$ in \eqref{Dtest}. Therefore, the resulting p-value  can in principle be lower than the one in \eqref{adjk}. Indeed the AIC criterion in \eqref{AIC}, aims to improve the fit of the estimator to the data, whereas the deviance selection criteria in \eqref{chooseM} aim to maximize the power of the inferential procedure.}

\black{ Second,
choosing $M_{\max}=100$ is computationally unfeasible with most of the standard programming languages such as \texttt{R}  and \texttt{Python}, and  the numerical computation  of \eqref{dhat} may easily lead to divergent or inaccurate results.   }

 \begin{figure*}[htb]
\begin{tabular*}{\textwidth}{@{\extracolsep{\fill}}@{}c@{}c@{}c@{}c@{}}
\includegraphics[width=45mm]{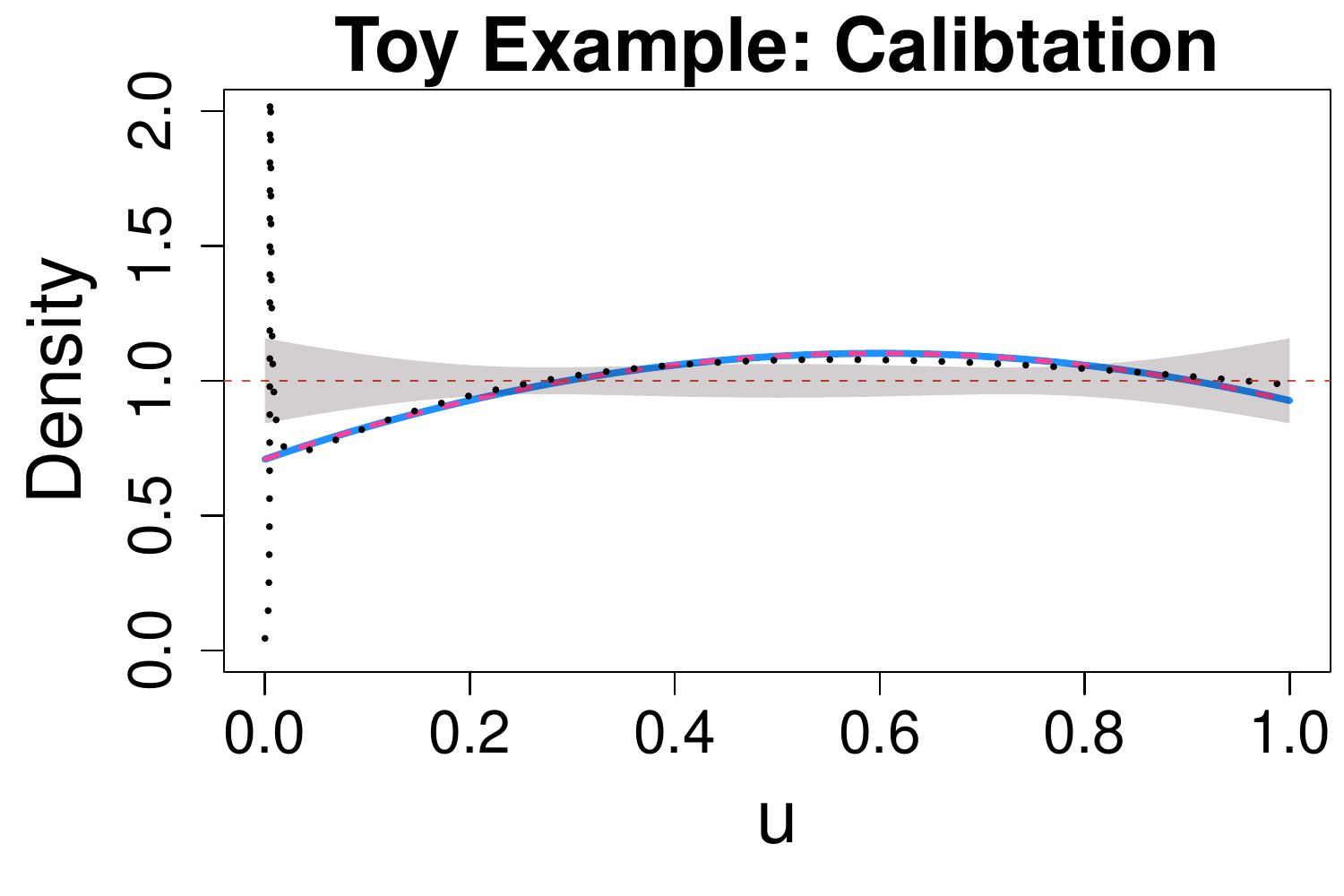} &\includegraphics[width=45mm]{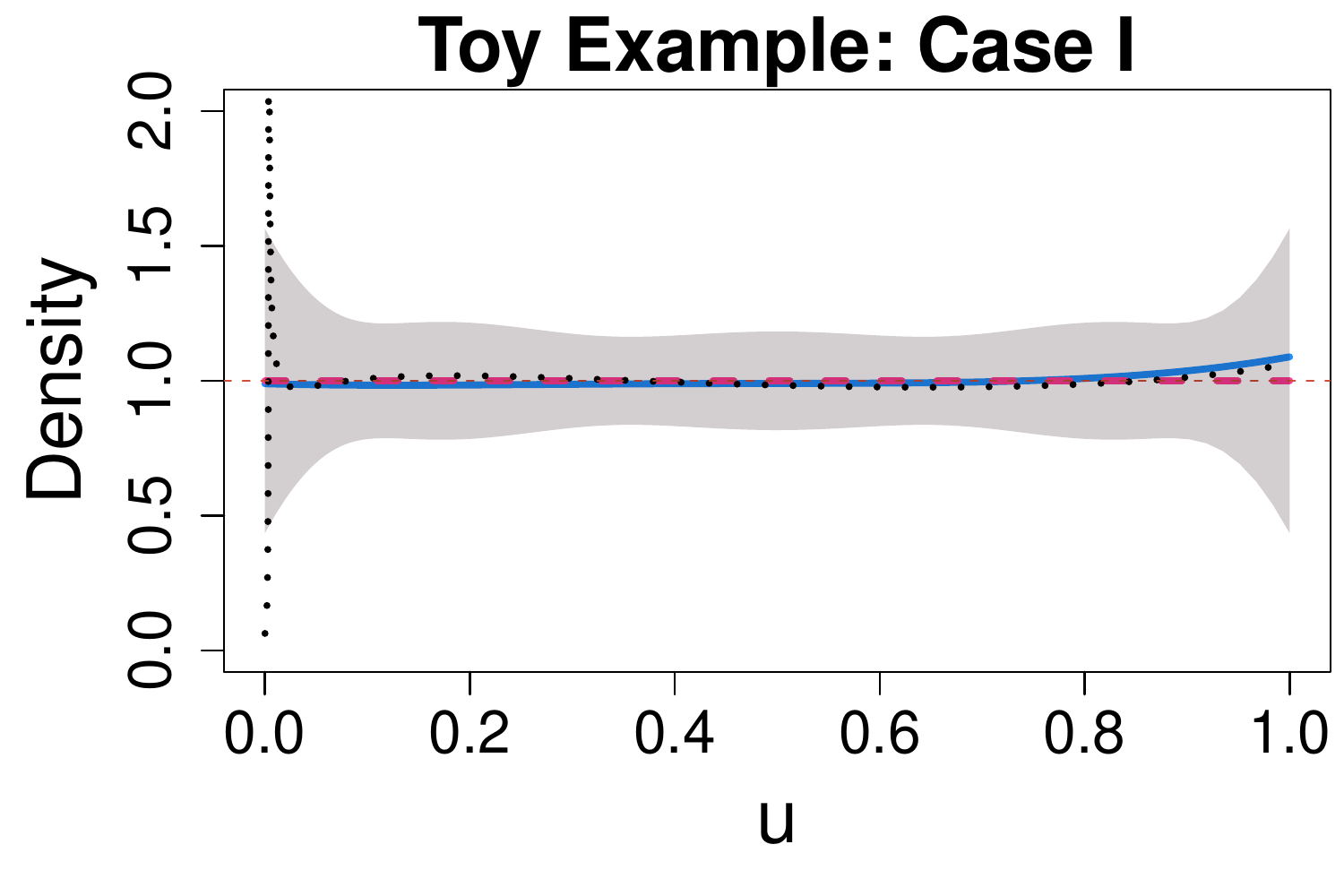}&\includegraphics[width=45mm]{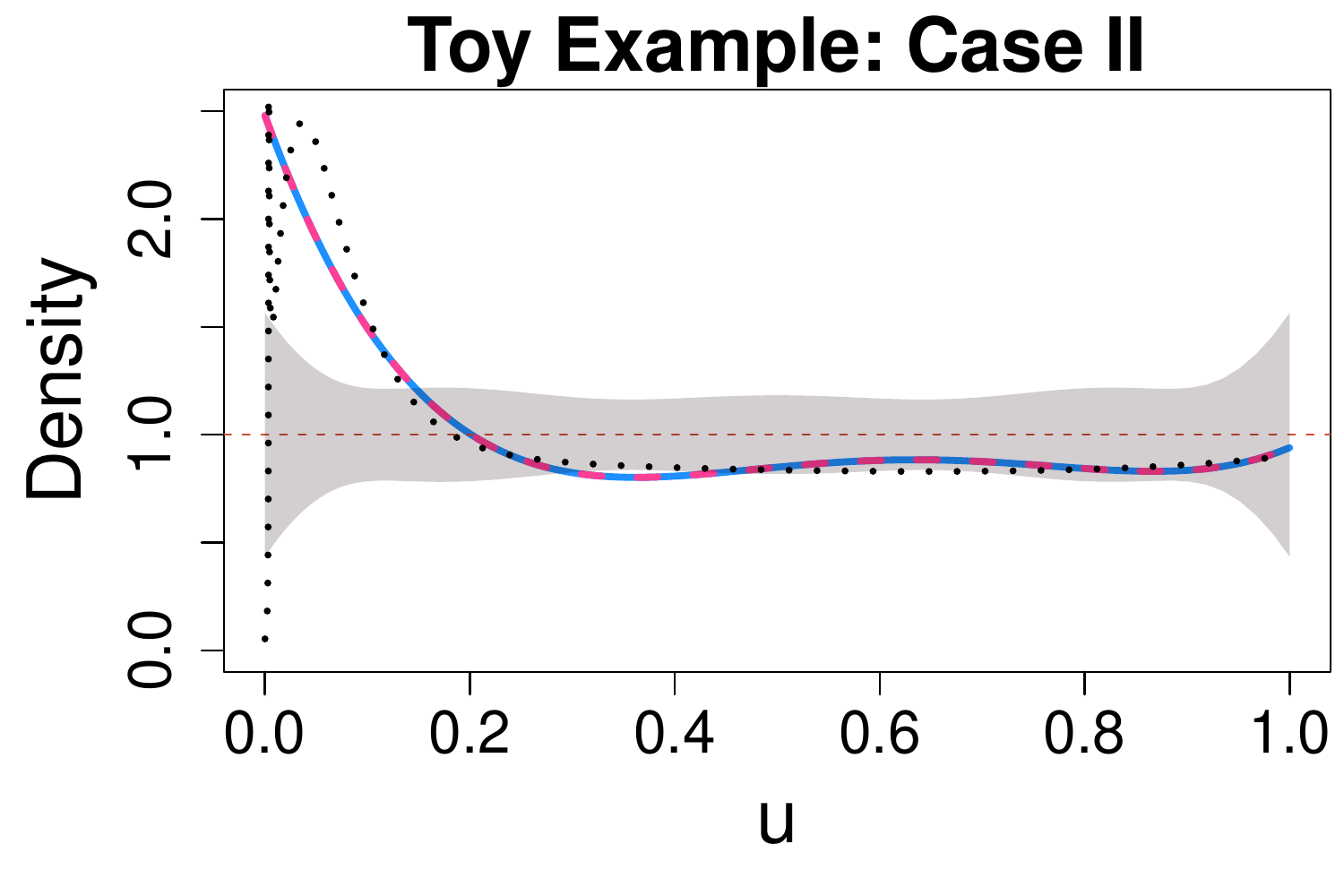}&
\includegraphics[width=45mm]{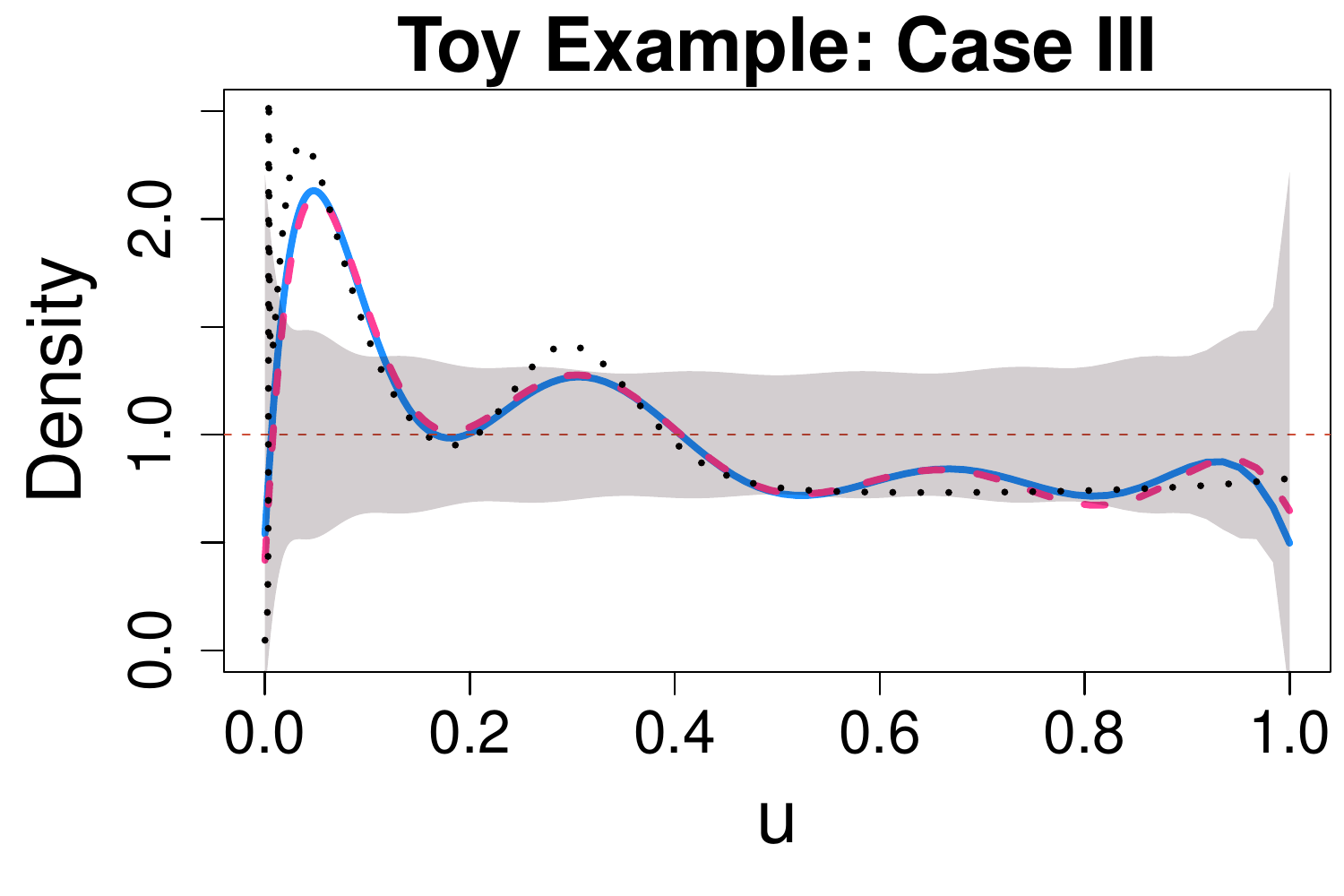}\\[-3ex]
 \multicolumn{4}{c}{\includegraphics[width=180mm]{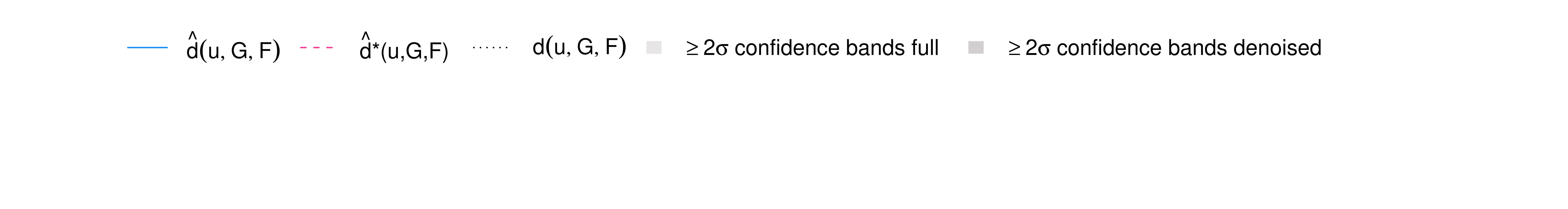}} \\[-12ex]
\end{tabular*}
\caption[Figure 8]{\black{Comparison of the estimators $\widehat{d}(u;G,F)$ (blue solid lines) and $\widehat{d}^*(u;G,F)$ (pink dashed lines) for the  toy examples in Section \ref{DS}. The true comparison densities  $d(u;G,F)$ are displayed as  dotted black curves. The light gray areas corresponds to the confidence bands of the full solution whereas dark gray areas refer to the confidence bands of the denoised solution. In all the examples proposed, the latter is almost entirely overlapping with the former. }}
\label{denoisefig}
\end{figure*}
{\fontsize{3mm}{3mm}\selectfont{
\begin{table*}
\begin{tabular}{|l|c|cc|c|}
\hline
                & &&&\\
                                 &\textbf{$M,k_M^*$}& \textbf{Method} &\textbf{Deviance } & \textbf{Adjusted  }\\
                                 &\textbf{selected }&                               &\textbf{ p-values} & \textbf{ p-values}\\
         & &       &&\\[-1.5ex]
                                 \hline
                    &       &       &&     \\[-1.5ex]
  \textbf{Toy example}& $M=2$&Full&  $p(2)=3.199\cdot 10^{-12}$ &   $20\cdot p(2)=6.397\cdot 10^{-11}$\\
   \textbf{Calibration} & $k^*_2=2$ &Denoised& $p(2)=3.199\cdot 10^{-12}$ &   $21\cdot p(2)=6.717\cdot 10^{-11}$\\[-1ex]   
                    &     &               &&    \\[-1.5ex]
   \hline
                       &        &         &&    \\[-1.5ex]
   \textbf{Toy example}&M=18 &Full & $p(18)=0.2657$&$20\cdot p(18)>1$\\
   \textbf{Case I} &$k^*_{18}=2$ &Denoised& $p(2)=5.096\cdot 10^{-4}$& $21\cdot p(2)=0.0882$ \\[-1ex] 
                       &          &       &&     \\[-1.5ex]
   \hline
                        &            &    &&     \\[-1.5ex]
   \textbf{Toy example} &$M=4$&Full& $p(4)=8.994\cdot 10^{-33}$ &   $20\cdot p(4)=1.799\cdot 10^{-31}$\\
   \textbf{Case II}  &$k^*_4=4$&Denoised& $p(4)=8.994\cdot 10^{-33}$&   $21\cdot p(4)=2.338\cdot 10^{-31}$ \\[-1ex]
                                   &                &                          &&      \\[-1.5ex]
\hline
                                        &              &                       &&     \\[-1.5ex]
   \textbf{Toy example} & $M=9$&Full&$p(9)=2.590\cdot 10^{-28}$&   $20\cdot p(9)=5.181\cdot 10^{-27}$\\
   \textbf{Case III}  & $k^*_9=6$&Denoised& $p(6)=4.457\cdot 10^{-30}$ &   $35\cdot p(1)=2.496\cdot 10^{-28}$\\ [-1ex] 
                                                         &          &          && \\
\hline 
 \end{tabular}
\caption[Table 2]{Model selection and inference for the  toy example in Section \ref{DS}. The second column reports the M and $k^*_M$ values selected as in \eqref{chooseM} and \eqref{mstar}, respectively. The third column collects the unadjusted deviance p-values for the full and denoised solutions. The Bonferroni-adjusted p-values, computed as in \eqref{bonf} and \eqref{adjk} are reported in the fourth column. The correction terms applied correspond to $M_{\max}=20$ for the full solution and $M_{tot}=M_{\max}+\frac{M(M-1)}{2}$ for the denoised solution.    }
\label{AICdev}
\end{table*} }}

\subsection{Comparing full and denoised solution}
\label{comparison}

Fig. \ref{denoisefig} compares the fit of the estimators   $\widehat{d}(u;G,F)$ and $\widehat{d}^*(u;G,F)$ for  the examples   in Section \ref{DS}. For all the cases considered, $M$ and $k^*_M$ have been selected  as in \eqref{chooseM} and \eqref{mstar} (see second column of Table \ref{AICdev}). When no significance was achieved for any of the values of $M$  considered, a small basis of $M=3$ or $M=4$ polynomials was chosen for the full estimator $\widehat{d}(u;G,F)$, which was then further denoised in order to obtain  $\widehat{d}^*(u;G,F)$. 
Table \ref{AICdev} shows the results of the deviance tests of the full and the denoised solution for the examples in Section \ref{DS}. The unadjusted p-values  and the Bonferroni-adjusted p-values are reported in the second and third columns, respectively.
In \black{half of the} cases, $k^*_M=M$ and the estimators $\widehat{d}(u;G,F)$ and $\widehat{d}^*(u;G,F)$ overlap over the entire range $[0,1]$. The inferential results were also approximately equivalent in the majority of the situations considered.   

The main differences are observed in the analysis of the background-only physics sample (Case I). In this case, the deviance-selection procedure leads to non-significant results for all the values of $M$ considered; the minimum p-value is observed at $M=18$ (unadjusted p-value = $0.2657$). In this setting, the denoising process leads to
$k^*_{18}=2$  and the respective unadjusted p-value is $5.096\cdot10^{-4}$. This   further emphasizes the importance of adjusting for model selection  in order to avoid false discoveries.
For modelling purposes and for the sake of comparison with the case where a signal is present, a  basis of $M=4$  was selected.
Since the true distribution of the data is the same as the postulated one, the denoising process sets all the coefficients equal to zero ($k^*_M=0$). 

For Case III,  only $k^*_M=6$ out of $M=9$ coefficients are selected when denoising  (see Table \ref{AICdev}). 
Despite the  right panel of Fig. \ref{denoisefig} shows that the full and the denoised solution are almost overlapping, the latter leads to an increased sensitivity (adjusted p-value=$2.496\cdot10^{-28}$) compared to the full solution (adjusted p-value=$5.181\cdot10^{-27}$).

\begin{figure*}
\begin{tabular*}{\textwidth}{@{\extracolsep{\fill}}@{}c@{}c@{}}
\includegraphics[width=1\columnwidth]{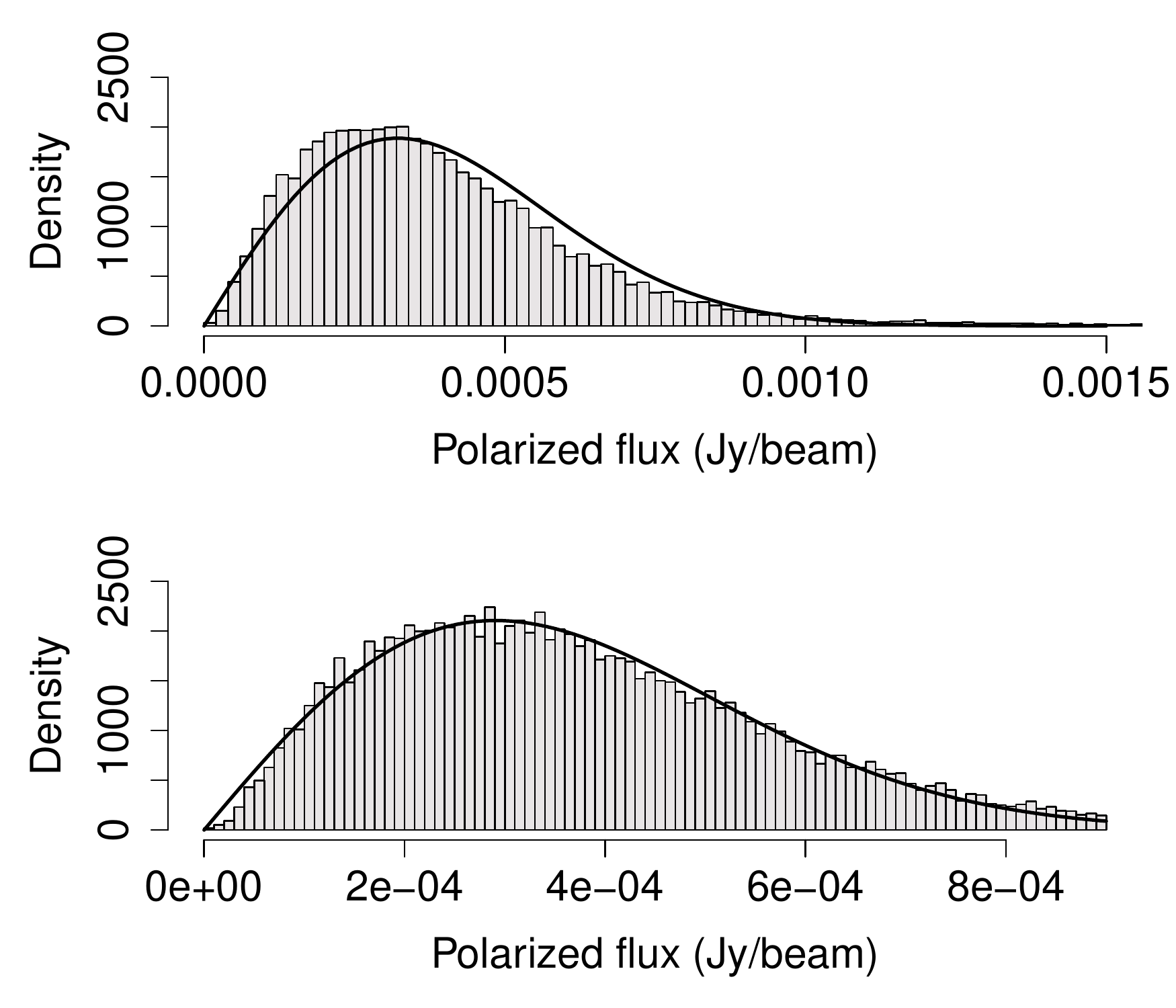}&\includegraphics[width=1\columnwidth]{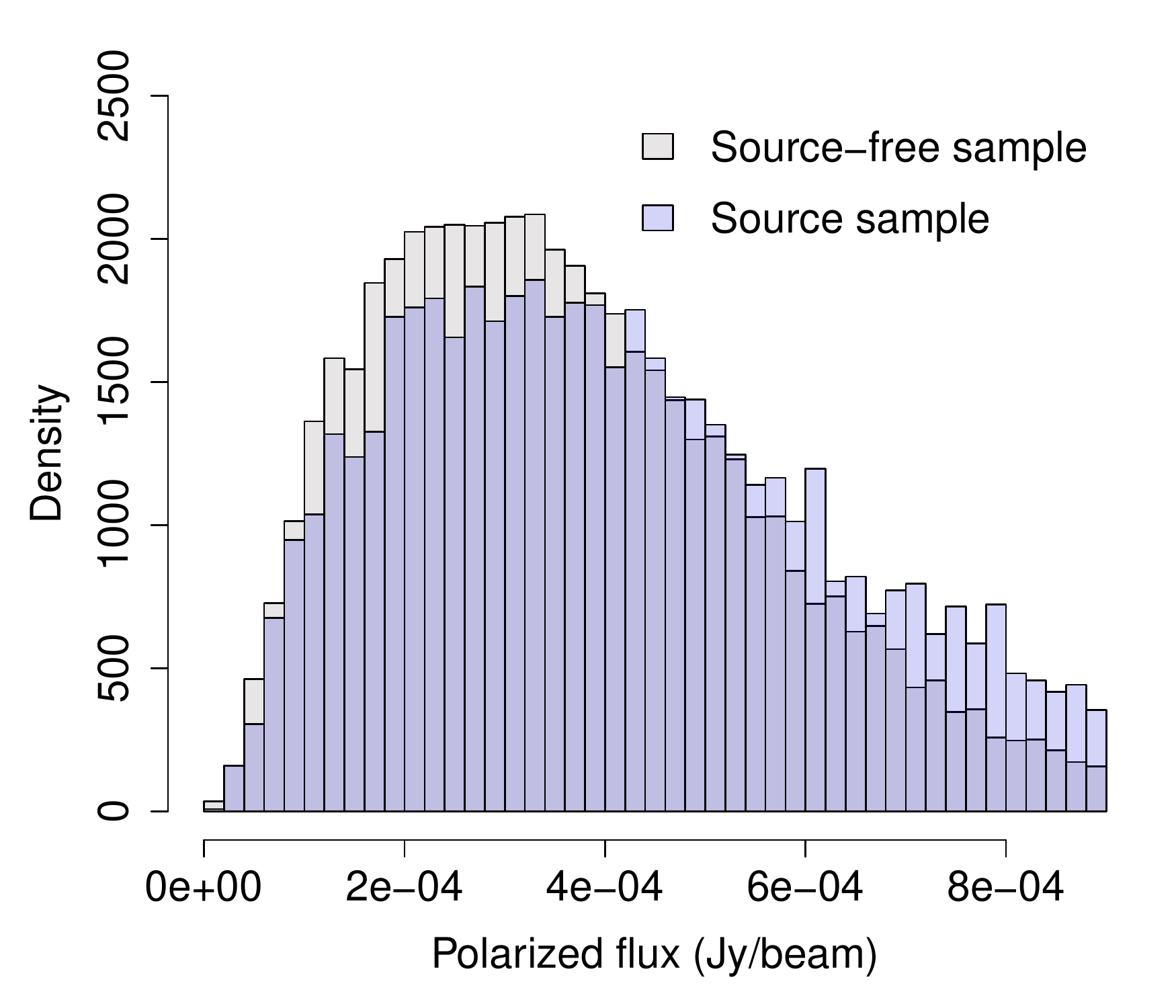}\\
 \end{tabular*}
\caption[Figure 9]{Histograms of the NVSS samples. \black{ The left panels show the source-free sample with and without outliers (upper left and lower left panels, respectively).
In both cases, the best fit of the Rayleigh model in \eqref{gb3} is displayed as a black curve. The right panel compares the source-free sample without outliers (grey histogram) with the 
source sample (blue histogram)  truncated over the search area considered. }}
\label{NVSSfig}
\end{figure*}
\begin{figure}[htb]
\centering
 \begin{adjustbox}{center}
\includegraphics[width=1\columnwidth]{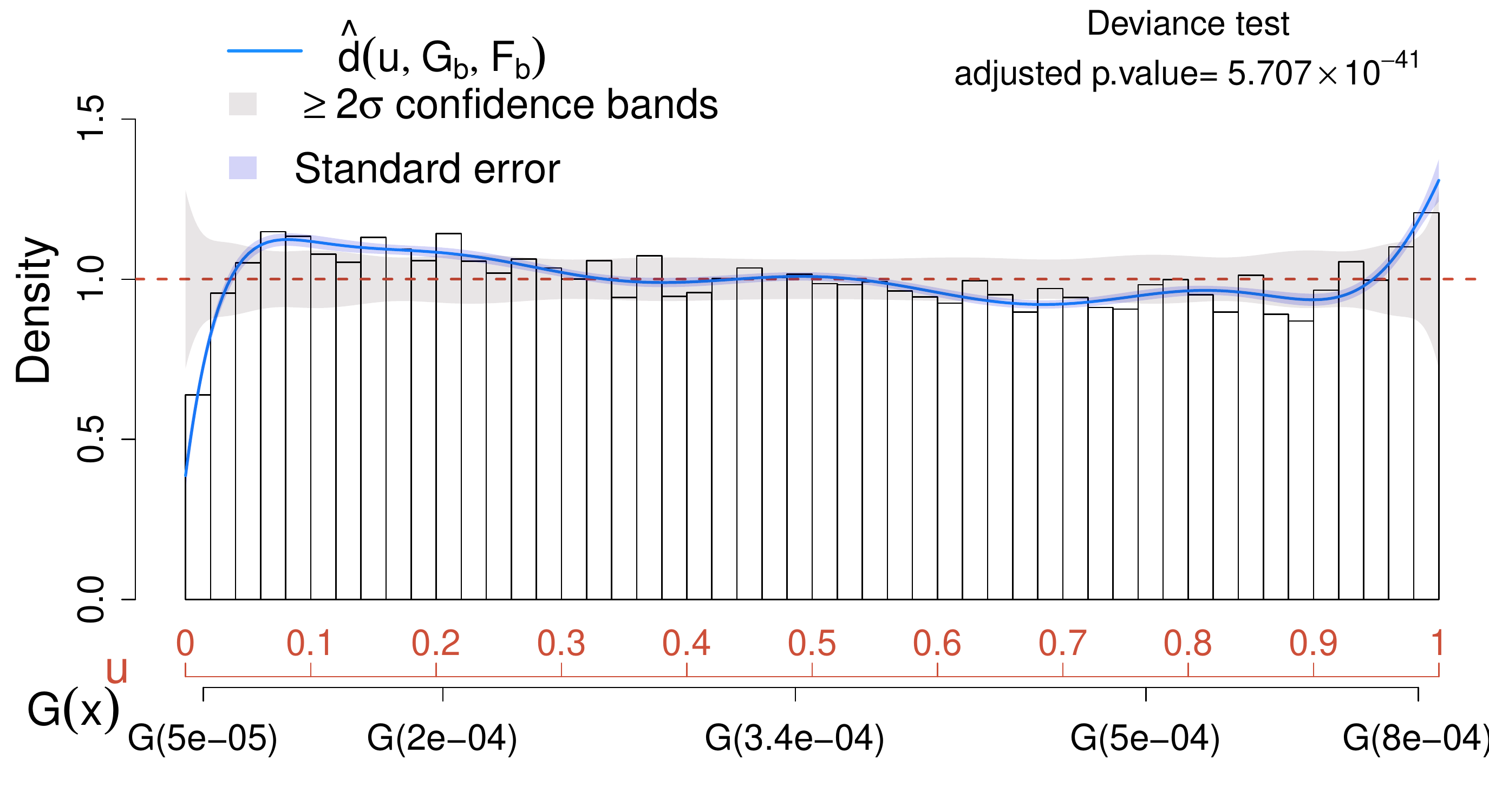}
 \end{adjustbox}
\caption[Figure 10]{ Deviance test and CD plot for the NVSS  source-free sample of size $28,739$  compared to the Rayleigh distribution in \eqref{gb3}. }
\label{Fig10}
\end{figure}
These results suggest that the denoising approach can easily adapt to situations where a sparse solution is preferable (i.e., when only few of the $M$ coefficients $LP_j$  are non-zero)  without enforcing sparsity when many of the $M$ coefficients considered are needed to adequately fit the data  (e.g., bottom right panel of Fig. \ref{denoisefig}). 
From an inferential perspective, denoising can improve the sensitivity of the analysis; however, in order to avoid false discoveries,  extra care needs to be taken  when the deviance selection procedure  leads to large p-values for all the $M_{\max}$ models considered.

\section{An application to stacking experiments}
\label{stacking}
In  radio astronomical surveys, stacking techniques are often used to combine noisy images or ``stacks'' in order to increase the signal-to-noise ratio and improve the sensitivity of the analysis in detecting  faint sources \cite[e.g.,][]{lawrence,white,jeroen}. 
In polarized signal searches, for instance, a faint population of sources is considered  when the median  polarized intensity observed over control regions differs significantly from the median of the region where the sources are expected to be present. In this context, \black{under simplifying assumptions}, the distribution of the  intensity of the source polarization is often  assumed to \black{to have Rice distribution i.e., 
\begin{equation}
\label{rice}
f(x)=\frac{xe^{-\frac{x^2+\nu^2}{2\sigma^2}}}{k_{\nu\sigma^2}}\text{Bessel}\bigl(\frac{x\nu}{\sigma^2}\bigl)
\end{equation}
where $\text{Bessel}(\cdot)$ denotes the Bessel function of first kind of order zero and $k_{\nu\sigma^2}$ is a normalizing constant. Furthermore, \eqref{rice} reduces to a Rayleigh pdf when no signal is present \cite{simmons}, i.e, when $\nu=0$}.  Below, it is shown how the  methods described in Sections \ref{bkgcali} and \ref{nonpar} can be used to  assess whether the Rayleigh distribution is a reliable model for the background and, \black{when too simplistic}, investigate the impact of incorrectly assuming a Rayleigh distribution on the \black{reliability} of the analysis.

\begin{figure*}
\begin{tabular*}{\textwidth}{@{\extracolsep{\fill}}@{}c@{}c@{}}
      \includegraphics[width=90mm]{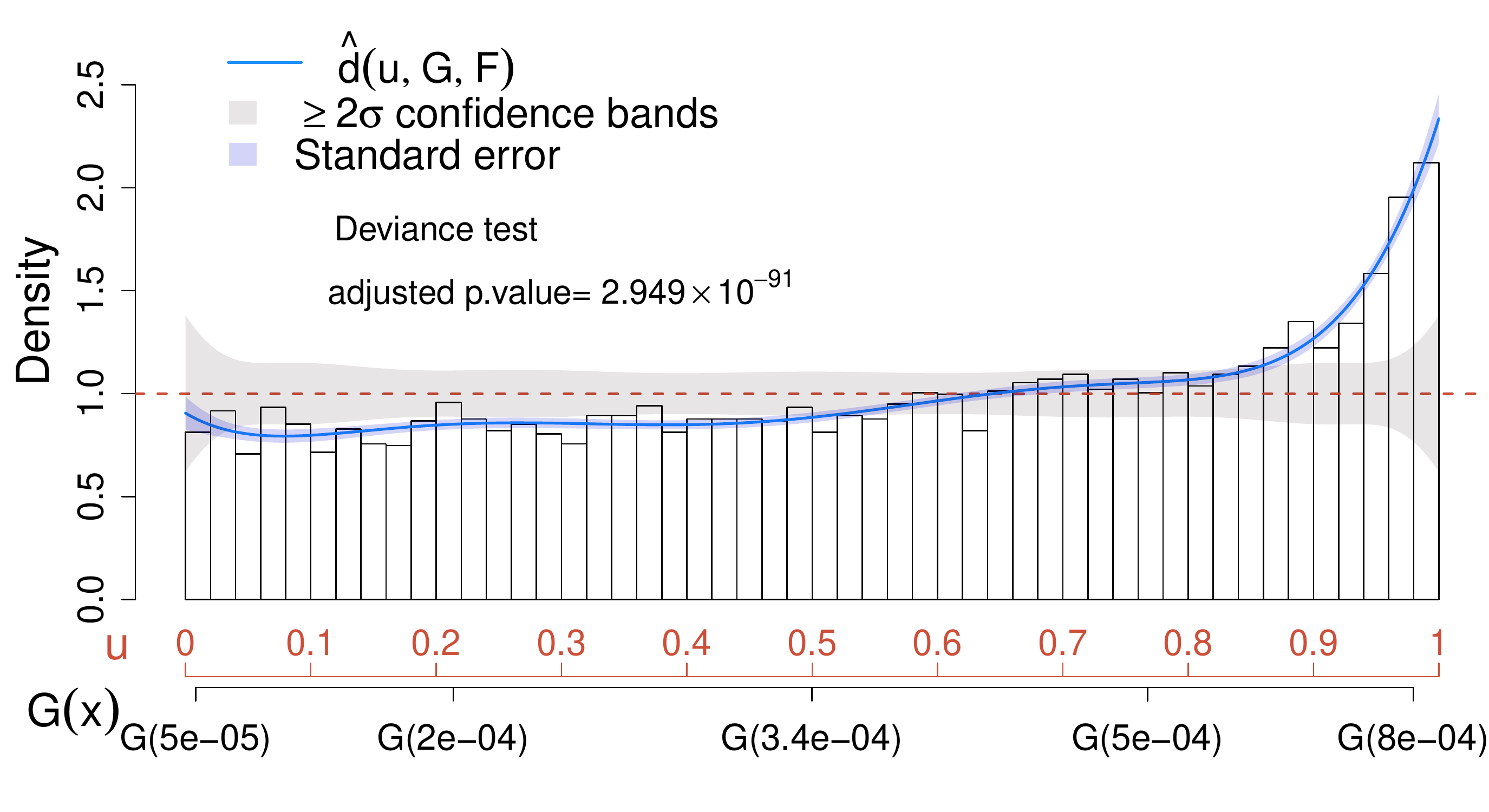} & \includegraphics[width=90mm]{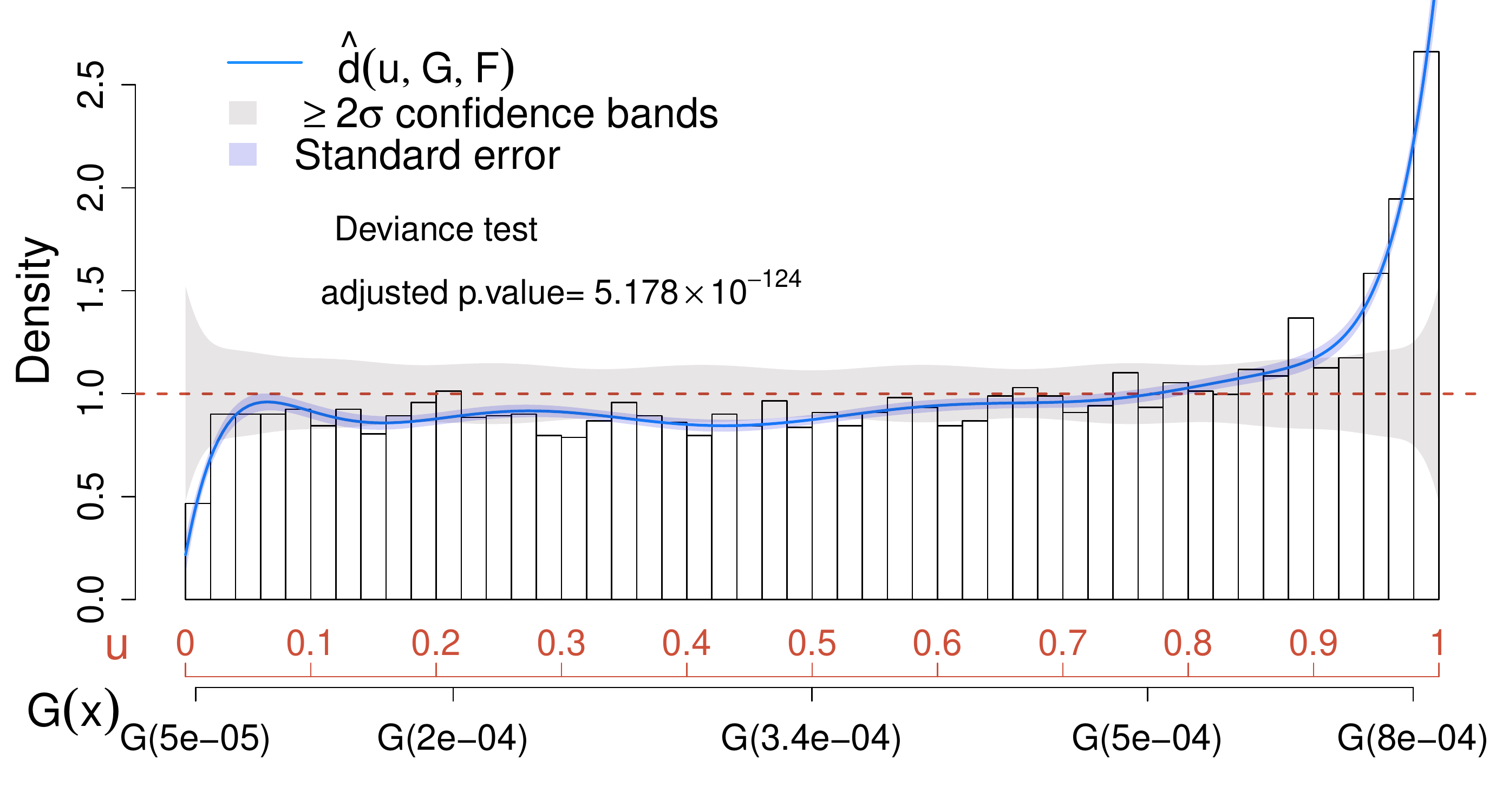}\\
\end{tabular*}
\caption[Figure 11]{Deviance tests and CD plots for the NVSS source sample assuming $g(x)$ to be the calibrated background model in \eqref{fbhat3} (left panel) and 
when letting $g(x)$ be the pdf in of the truncated Rayleigh distribution in \eqref{gb3}. In both cases the estimator of the comparison density has been denoised as described in Section \ref{denoiseAIC}. The values of $M$ and $k^*$ considered are $M=k^*=6$  and  $M=9$, $k^*=8$ for the  estimators on the left and right panels, respectively.}
\label{compare}
\end{figure*}
The data  considered  comes from the  NRAO VLA Sky Survey (NVSS) \cite{NVSS}. The NVSS is an astronomical survey of the Northern hemisphere carried out by the Very Large Array of the National Radio Astronomy Observatory. The NVSS has detected 1.8 million sources in total intensity, but only $14\%$ of these have  reported a \black{polarized signal peak} greater than $3\sigma$ \cite{jeroen}. \black{The original source-free sample contained $29,915$ observations  collected from four different control regions for each source with a brightness in total intensity between 0 and 0.0093 Jy/beam (see upper panel of Figure \ref{NVSSfig}). However, such sample appears to contain several outliers which affect the data distribution, making it far from Rayleigh.
A better Rayleigh fit  is obtained when removing the outliers,\footnote{In statistics, an  observation $x_i$ is considered an outlier if $x_i<Q_{0.25}-1.5[Q_{0.75}-Q_{0.25}]$ or $x_i>Q_{0.75}+1.5[Q_{0.75}-Q_{0.25}]$ where $Q_{0.25}$ and $Q_{0.75}$ are the first and the third sample quartiles.}  (see bottom left panel of Figure \ref{NVSSfig}). Since understanding the cause of these anomalous observations is beyond the scope of this manuscript, we proceed excluding  them from the analysis and  we focus on assessing the validity of the Rayleigh assumption on the remaining $28,739$ observations on  the region
$[0 , 0.0009]$ Jy/beam. It has to be noted that the nominal noise in NVSS polarization is 0.00029 Jy/beam and we may expect as reasonable threshold for the detection of one individual source to be three times the noise. Hence,  a source sample of $6,220$ observations has been selected from positions where compact radio sources with a brightness in total intensity between 0 and 0.0009 Jy/beam are known to be present. Both source-free and source samples are assumed to be i.i.d.  The histograms of the source-free and signal samples considered are shown in the right panel of Fig. \ref{NVSSfig}.}

As first step, we fit a Rayleigh distribution (adequately truncated over the range $[0,0.0009]$) on the source-free sample, i.e.,  
\begin{equation}
\label{gb3}
g_b(x)=\frac{x e^-\frac{x^2}{2\widehat{\sigma}^2}}{k_{\widehat{\sigma}^2}}
\end{equation} 
 where $k_{\widehat{\sigma}^2}$ is a normalizing constant, $\widehat{\sigma}=0.0003$ is the ML estimate of the unknown parameter $\sigma$, and  $x\in[0,0.0009]$ Jy/beam. In order to assess if \eqref{gb3} provides a good fit for the data, we 
  estimate the comparison density $d(G_b(x);G_b,F_b)$  by, first, selecting $M$ as in \eqref{chooseM} and then applying the AIC-based denoising approach described in Section \ref{denoiseAIC}.
In this case, the denoised solution selects $k^*=9$ out of $M=10$ polynomial terms. The  deviance tests and the CD plot  in Fig. \ref{Fig10} suggest that, despite the fact that the median of the data coincides with the one of  the Rayleigh model, overall, the latter does not provide a good fit for the distribution of the source-free sample. Specifically, the data distribution shows a higher right tail than one expected under the Rayleigh assumption, whereas the first quantiles are overestimated by the Rayleigh.
Therefore, the researcher can either decide to use a more refined parametric model for the background or consider the calibrated background distribution of the form in \eqref{fbhat},   which in our setting specifies as
\begin{equation}
\label{fbhat3}
\begin{split}
&\widehat{f}_b(x)=\frac{x e^-\frac{x^2}{2\widehat{\sigma}^2}}{k_{\widehat{\sigma}^2}}\Bigl(1-0.018Leg_1[G_b(x)]+0.012Leg_2[G_b(x)]\\
&+0.052Leg_3[G_b(x)]-0.014Leg_4[G_b(x)]+0.047Leg_5[G_b(x)]\\
&-0.018Leg_6[G_b(x)]+0.031Leg_7[G_b(x)]+0.016Leg_9[G_b(x)]\\
&-0.015Leg_{10}[G_b(x)]\Bigl),\\
\end{split}
\end{equation} 
where $G_b(x)$ is the cdf of \eqref{gb3}. 

The strategy described in Section \ref{nonpar} allows us to identify  where significant differences between the control and source sample occur.  In order to assess the effect of incorrectly assuming a Rayleigh background, we  compare the distribution of the physics sample with both the Rayleigh and the calibrated background distribution in \eqref{fbhat3}.
Figure \ref{compare} reports deviance tests and CD plots obtained on the physics sample when setting $g(x)=\widehat{f}_b(x)$ in \eqref{gb3} (left panel) and  $g(x)=\widehat{f}_b(x)$ in \eqref{fbhat3} (right panel).
Both analyses provide strong evidence that the distribution of the physics sample differs significantly  from the postulated models $\widehat{f}_b(x)$  and $g_b(x)$, and the most substantial discrepancies occur on the right tail of the distribution. 
However, since the Rayleigh model underestimates the right tail of the background distribution (see Fig. \ref{Fig10}), it leads to an artificially enhanced sensitivity in this region.
The differences between the two CD plots are less prominent around the median expected under $\widehat{f}_b(x)$  and $g_b(x)$ (i.e., in correspondence of  $u=0.5$ in both plots). 

\black{Fig \ref{compare}} suggests that, for these data,  assuming a background Rayleigh distribution  would not substantially affect the results  of a comparison \black{between the source-free and signal sample} based on the median. However,  focusing solely on the median can strongly \black{limit} the overall sensitivity of the analysis since the major differences occur at the higher quantiles of the distribution. \black{On the other hand, assuming a Rayleigh distribution for the background would artificially inflate the evidence in favor of the source. Specifically, the sigma significance of the deviance test obtained under the Rayleigh background assumption is $23.655\sigma$ (adjusted p-value = $5.178\cdot 10^{-124}$), whereas the one obtained using \eqref{fbhat3} is $20.225\sigma$ (adjusted p-value = $2.948\cdot 10^{-91}$).  }

Conversely, the calibrated background model in \eqref{fbhat3} allows us to safely compare  the entire distribution of the polarized intensity in the source and control regions via CD plots and deviance tests without \black{affecting} the sensitivity of the analysis.

\section{Discussion}
\label{discussion}
This article proposes a unified framework  for signal detection and characterization  under background mismodelling. 
From a methodological perspective,  the methods presented here extend     LP modelling to the inferential setting.

The solution discussed  is articulated in two main phases: a calibration phase where the background model is ``trained'' on a source-free sample   and a signal search phase conducted on the physics sample collected by the experiment. If a model for the signal is given, the method proposed allows the identification of hidden signals from new unexpected sources and/or the refining of the postulated background or signal distributions. Furthermore, the tools presented in this manuscript can be easily extended to situations where a source-free sample is not available and  the background is unknown (up to some free parameters). \black{As discussed in Section \ref{PSDMsec}, however, in this setting the signal distribution is required to be known, and the physics sample is expected to contain only signal-like events, i.e., the background is almost completely reduced}.  

\black{The theory of Section \ref{biasvariance} and} the analyses  in Section \ref{DS} have highlighted that, despite a fully non-parametric approach provides reliable inference, it may lead to unsatisfactory estimates when the postulated pdf $g$ is substantially different from the true density $f$. In this setting, a semiparametric stage can be performed in order provide a reliable model for the data.

\black{Each individual step in both the nonparametric and the semiparametric stage of Sections  \ref{signalcar} and \ref{bkgcali} provides useful scientific insights on the  signal and  background distribution. Hence, an automatized implementation of the steps of Algorithm 1  based solely on the p-values of the deviance tests  is discouraged as it would lead to a substantial loss of scientific knowledge on the phenomena under study.}

Finally, it is important to point out that, despite this article's  focus on the one-dimensional searches on continuous data,  all the constructs presented in Sections \ref{LPmodelling} and \black{the deviance test in \ref{dev} also apply to the discrete case  when considering i.i.d. events. More work  is needed to extend these results and those of Section \ref{bands} to searches in multiple dimensions and when considering Poisson events with functional mean. 
In the first case the difficulty mainly lies in generalizing the constructs of Section \ref{inference} to account for the dependence structure occuring across multiple dimensions. In the second case, the main challenge lies in identifying the equivalent of \eqref{skewG} to model the mean of the distribution, while incorporating the Poisson error.}

\section*{Code availability}
\black{The \texttt{LPBkg}  Python package \cite{python} and the \texttt{LPBkg} R package \cite{rr}  allow the implementation of the methods proposed   in this manuscript. Detailed tutorials on how to  use the functions provided are also available at \url{http://salgeri.umn.edu/my-research}. }

\section*{Acknowledgments}
The author thanks Jeroen Stil, who provided the NVSS datasets used in Section \ref{stacking}, and 
Lawrence Rudnick, who first recognized the usefulness of the method proposed in the context of stacking experiments. 
Conversations with Subhadeep Mukhopadhyay have been of great help when this work was first conceptualized. Discussions and e-mail exchanges with  Charles Doss  and Chad Shafer  are gratefully acknowledged. \black{Finally, the author thanks an  anonymous referee whose feedback has been  substantial to improve the overall quality of the paper. }

\appendix{

\black{
\section{Moments of the $\widehat{LP}_j$ estimates}
\label{appA}
Consider the general setting where $f\not\equiv g$ and thus $d(u;G,F)\neq1$ over $[0,1]$. It follows that each $u_i$ is independently and identically distributed with pdf $d(u;G,F)$; hence, all the expectations in $E[\widehat{LP}_j], V(\widehat{LP}_j)$ and $Cov(\widehat{LP}_j,\widehat{LP}_k)$ are taken with respect to $d(u;G,F)$.
Specifically,
\begin{equation*}
\begin{split}
E[\widehat{LP}_j]&=E\biggl[\frac{1}{n}\sum_{i=1}^nLeg_j(U_i)\biggl] \\
&=E[Leg_j(U)]\\
&=\int_0^1Leg_j(u)d(u;G,F)\partial{u} =LP_j
\end{split}
\end{equation*}
where the  second equality follows by the fact that each observed value $u_i$ is a realization of a random variable $U_i$ and each $U_1,\dots,U_n$ is identically distributed as the random variable $U$, whose pdf is given by the comparison density $d(u;G,F)$. Notice that  $d(u;G,F)=1$ implies that $\int_0^1Leg_j(u)d(u;G,F)\partial{u}=\int_0^1Leg_j(u)\partial{u}=0$, from which the first equivalence in \eqref{momentsH0} follows. Moreover,
\begin{equation*}
\begin{split}
V(\widehat{LP}_j)&=\frac{1}{n^2}V\biggl(\sum_{i=1}^nLeg_j(U_i)\biggl)\\
&=\frac{1}{n}V\bigl(Leg_j(U)\bigl)=\frac{\sigma^2_j}{n}\\
\end{split}
\end{equation*}
where $V\bigl(Leg_j(U)\bigl)=\int_0^1(Leg_j(u)-LP_j)^2d(u;G,F)\partial{u}=\sigma^2_j$. The second equality holds because of independence and identical distribution of each $u_i$. Notice that   if  $d(u;G,F)=1$, $\sigma^2_j=1$ in virtue of the orthonormality of the $Leg_j(u)$ polynomials. Hence the second equivalence in \eqref{momentsH0} holds. Finally,
\begin{equation*}
\begin{split}
Cov(\widehat{LP}_j,\widehat{LP}_k)&=Cov\biggl(\frac{1}{n}\sum_{i=1}^nLeg_j(U_i),\frac{1}{n}\sum_{i=1}^nLeg_k(U_i)\biggl)\\
&=\frac{1}{n}Cov\bigl(Leg_j(U),Leg_k(U)\bigl)=\frac{\sigma_{jk}}{n}
\end{split}
\end{equation*}
  also in this case, the second equality follows by independence and identical distribution of each $u_i$ and 
   {\fontsize{3mm}{3mm}\selectfont{
 \begin{equation}
 \label{covLPj}
\begin{split}
Cov\bigl(Leg_j(U),Leg_k(U)\bigl)&=\int_0^1(Leg_j(u)-LP_j)(Leg_k(u)-LP_k)d(u;G,F)\partial{u}\\
&=\int_0^1Leg_j(u)Leg_k(u)d(u;G,F)\partial{u}-LP_jLP_k\\&=\sigma_{jk}. \\
\end{split}
\end{equation}}}
Because of the orthogonality of the $Leg_j(u)$, $\sigma_{jk}=0$ when $d(u;G,F)=1$. Hence the third equivalence in \eqref{momentsH0}.}

\black{
\section{Bias, variance and MISE of $\widehat{d}(u;G,F)$}
\label{appB}
Given a point $u$ over $[0,1]$, the bias of \eqref{dhat} at $u$ is
\begin{align}
\label{biasder1}
\text{Bias}\bigl[\widehat{d}(u;G,F)\bigl]&=E\bigl[\widehat{d}(u;G,F)\bigl]-d(u;G,F)\\
\label{biasder2}
 &=\sum_{j=1}^ME[\widehat{LP}_j]Leg_j(u)-\sum_{j>0}{LP}_jLeg_j(u)\\
 \label{biasder3}
 &=\sum_{j>M}{LP}_jLeg_j(u); 
 \end{align}
here \eqref{biasder2} follows from \eqref{cd} and \eqref{dhat}. Whereas, the integrated squared bias is
\begin{align}
\label{ibsder1}
IBS=&\bigintssss_0^1 \biggl(\sum_{j>M}{LP}_jLeg_j(u)\biggl)^2 \partial u\\
\label{ibsder2}
=& \sum_{j>M}{LP}^2_j\int_0^1Leg^2_j(u)\partial u \\
&+2\sum_{M<j<k}{LP}_j{LP}_k\int_0^1Leg_j(u)Leg_k(u) \partial u\\
\label{ibsder3}
=&\sum_{j>M}{LP}^2_j;
 \end{align}
where \eqref{ibsder3} holds because of orthonormality of the  $Leg_j(u)$ polynomials. Notice that
\begin{align}
\label{ibsderB1}
IBS&=\sum_{j>M}{LP}^2_j=\sum_{j>0}{LP}^2-\sum_{j=1}^M{LP}^2\\
\label{ibsderB3}
&=\int_0^1 \bigl(d(u;G,F)-1\bigl)^2 \partial u-\sum_{j=1}^M{LP}^2\\
\label{ibsderB4}
&=\bigintsss_0^1\biggl(\frac{f(G^{-1}(u))-g(G^{-1}(u))}{g(G^{-1}(u))}\biggl)^2\partial u-\sum_{j=1}^MLP^2_j
 \end{align}
where \eqref{ibsderB3} follows by Parseval's identity whereas \eqref{ibsderB4}  follows from \eqref{cd1}.}

\black{
The variance of \eqref{dhat} at a given point $u$ is given by
\begin{align}
\label{varder1}
V\bigl[\widehat{d}(u;G,F)\bigl]=&V\biggl(\sum_{j=1}^M \widehat{LP}_jLeg_j(u)\biggl)  \\
\label{varder2}
=&\sum_{j=1}^{M}Leg^2_j(u)V\bigl(\widehat{LP}_j\bigl)\\
\label{varder3}
&+2\sum_{j<k}Leg_j(u)Leg_k(u)Cov\bigl(\widehat{LP}_j,\widehat{LP}_k\bigl)\\
\label{varder4}
=&\sum_{j=1}^{M}\frac{\sigma^2_j}{n}Leg^2_j(u)+2\sum_{j<k}\frac{\sigma_{jk}}{n}Leg_j(u)Leg_k(u).
\end{align}
By orthonormality of the polynomials $Leg_j(u)$, the integral of \eqref{varder4} over $[0,1]$  is 
\begin{equation}
\label{IVder}
\bigintsss_0^1V\bigl[\widehat{d}(u;G,F)\bigl]\partial u=\sum_{j=1}^{M}\frac{\sigma^2_j}{n}.
\end{equation}
also in this case, equality follows by orthonormality of the  $Leg_j(u)$.
Finally, the MISE is
\begin{align}
\label{MISEder1}
MISE\Bigl[\widehat{d}(u;G,F)\Bigl]=&E\biggl[\int_0^1 \bigl(\widehat{d}(u;G,F)-d(u;G,F)\bigl)^2\partial u\biggl]\\
\label{MISEder2}
=&\int_0^1 E\biggl[\bigl(\widehat{d}(u;G,F)-d(u;G,F)\bigl)^2\biggl] \partial u\\
\label{MISEder3}
=&\int_0^1 V\bigl[\widehat{d}(u;G,F)\bigl]\\
&\qquad+ \bigl(E\bigl[\widehat{d}(u;G,F)\bigl]-d(u;G,F)\bigl)^2 \partial u\\
\label{MISEder4}
=&\sum_{j=1}^{M}\frac{\sigma^2_j}{n}+\sum_{j>M}{LP}^2_j
\end{align}
where \eqref{MISEder2} holds because of Fubini-Tonelli theorem, whereas the last equality follows by \eqref{ibsder3} and \eqref{IVder}.}

\end{document}